\documentclass[aps,prb,floatfix,twocolumn,letterpaper,superscriptaddress,
  showpacs
  ]{revtex4-1}
\usepackage{graphicx,amsmath,amsfonts,amsthm,amssymb}
\usepackage{color}
\usepackage[ colorlinks, linktocpage=true, urlcolor=blue ]{hyperref}
\usepackage[all]{xy}
\usepackage{relsize}
\newcommand{\del}[1]{{\iffalse #1 \fi}}
\newcommand{\add}[1]{#1}

\DeclareMathOperator{\Tr}{Tr}
\DeclareMathOperator{\ev}{ev}

\newcommand\phia{\text{{\relscale{1.5}${\phi}$}}}
\newcommand\nphia{\text{{\relscale{1.2}$\phi$}}}
\newcommand{\xt}{\times}
\newcommand{\ot}{\otimes}
\newcommand{\f}[1]{\vcenter{\hbox{\includegraphics{#1}}}}
\newcommand{\bbm}{\mathbb{M}}
\newcommand{\bbn}{\mathbb{N}}
\newcommand{\bbc}{\mathbb{C}}
\newcommand{\bbz}{\mathbb{Z}}

\newcommand{\cB}{\mathcal{B}}
\newcommand{\cC}{\mathcal{C}}
\newcommand{\cD}{\mathcal{D}}
\newcommand{\cK}{\mathcal{K}}
\newcommand{\cM}{\mathcal{M}}
\newcommand{\cN}{\mathcal{N}}
\newcommand{\cR}{\mathcal{R}}
\newcommand{\cH}{\mathcal{H}}
\newcommand{\cV}{\mathcal{V}}
\newcommand{\vcyl}{\mathcal{V}_\text{cyl}}
\newcommand{\cZ}{\mathcal{Z}}
\newcommand{\mz}[4]{\left( \begin{array}{cc} #1&#2\\#3&#4\end{array} \right)}
\newcommand{\cl}[1]{\lfloor{#1}\rceil}
\newcommand{\ri}{\mathrm{i}}
\newcommand{\ee}{\mathrm{e}}
\newcommand{\one}{\textbf{1}}
\newcommand{\res}[1]{\left\langle{#1}\right\rangle}
\renewcommand{\v}[1]{\boldsymbol{#1}}

\begin{document}
\graphicspath{{figures/}}

\begin{titlepage}
  \title{Topological quasiparticles and the holographic\\
         bulk-edge relation in 2+1D string-net models}

  \author{Tian Lan}
\affiliation{Perimeter Institute for Theoretical Physics, Waterloo, Ontario N2L
2Y5, Canada}
  \affiliation{Department of Physics and Astronomy,
  University of Waterloo, Waterloo, Ontario N2L 3G1, Canada}
\author{Xiao-Gang Wen} 
\affiliation{Perimeter Institute for Theoretical Physics, Waterloo, Ontario N2L
2Y5, Canada}
\affiliation{Department of Physics, Massachusetts Institute of Technology,
Cambridge, Massachusetts 02139, USA}
\affiliation{Collaborative Innovation Center of Quantum Matter, Beijing, China}

\begin{abstract}
String-net models allow us to systematically construct and classify 2+1D
topologically ordered states which can have gapped boundaries. We can use a
simple ideal string-net wavefunction, which is described by a set of F-matrices [or more
precisely, a unitary fusion category (UFC)], to study all the universal
properties of such a topological order.  In this paper, we describe a finite
computational method -- \emph{Q-algebra approach}, that allows us to compute
the non-Abelian statistics of the topological excitations [or more precisely,
the unitary modular tensor category (UMTC)], from the string-net wavefunction
(or the UFC). We discuss several examples, including the topological phases
described by twisted gauge theory ({i.e.,} twisted quantum double
$D^\alpha(G)$).  Our result can also be viewed from an angle of holographic
bulk-boundary relation.  The 1+1D anomalous topological orders, that can appear
as edges of 2+1D topological states, are classified by UFCs which describe the
fusion of quasiparticles in 1+1D.  The 1+1D anomalous edge topological order
uniquely determines the 2+1D bulk topological order (which are classified by
UMTC).  Our method allows us to compute this bulk topological order ({i.e.,}
the UMTC) from the anomalous edge topological order ({i.e.,} the UFC).


\end{abstract}
\maketitle
\end{titlepage}
\tableofcontents

\section{Introduction}
A major problem of physics is to classify phases and phase transitions of
matter.  The problem was once thought to be completely solved by Landau's
theory of symmetry breaking\cite{Lan37}, where the phases can be classified
by their symmetries.  However, the discovery of fractional quantum Hall (FQH)
effect\cite{TSG8259} indicated that Landau's theory is incomplete.  There are
different FQH phases with the same symmetry, and the symmetry breaking theory
failed to distinguish those phases.  FQH states are considered to possess new
topological orders\cite{Wtop,WNtop,Wrig} beyond the symmetry breaking theory.

We know that all the symmetry breaking phases are labeled by two groups
$(G_H,G_\Psi)$, where $G_H$ is the symmetry group of the Hamiltonian and
$G_\Psi$ is the symmetry group of the ground state.  This fact motivates us to
search for the complete ``label'' of topological order.

Here, the ``label'' that labels a topological order corresponds to a
set of universal properties that can fully determine the phase and distinguish
it from other phases.  Such universal properties should always remain the same
as long as there is no phase transition.  In particular, they are invariant
under any small local perturbations.  Such  universal properties are called
topological invariants in mathematics.

In 2+1D, it seems that anyonic quasiparticle statistics, or the modular data
$T,S$ matrices, are the universal properties. The set of universal properties
that describes quasiparticle statistics is also referred to as unitary modular
tensor category (UMTC).  $T,S$ matrices ({i.e.,} UMTC) can fully determine
the topological phases, up to a bosonic $E_8$ FQH
state.\cite{Wrig,KW9327,W1221,ZV1313,CV1308} \add{In Section \ref{qpart} we
will introduce topological quasiparticle excitations and their statistics, {\it
i.e.} fusion and braiding data, in 2+1D topological phases and on 1+1D gapped
edges.}

Since the universal properties do not depend on the local details of the
system, it is possible to calculate them from a simple renormalization
fixed-point model.  In this paper we will concentrate on a class of 2+1D
fixed-point lattice model, the Levin-Wen string-net model\cite{LW05}.  As a
fixed-point model, the building blocks of Levin-Wen models are effective
degrees of freedom with the form of string-nets.  
The fixed-point string-net wavefunction is completely determined
by important data -- the
F-matrices. The F-matrices are also referred to as unitary fusion category (UFC).
\del{
  More precisely,  Levin-Wen models and the ideal string-net wave function are
  labeled by the unitary fusion categories (UFC).

  In Ref. \onlinecite{LW05} the F-matrices are assumed to obey the tetrahedral
  symmetry.  In this paper we generalized the original Levin-Wen model by
  dropping the tetrahedral symmetry assumption of F-matrices.\cite{H0904}
  Therefore, it will be interesting to find an efficient algorithm to calculate
  the $T,S$ matrices of string-net models.
}

\add{
  Therefore, a central question for string-net models is how to calculate the
  $T,S$ matrices from F-matrices (or how to calculate the UMTC from the UFC). In Ref. \onlinecite{LW05} the $T,S$ matrices
  can be calculated by searching for \emph{string operators}. String operators
  are determined by a set of non-linear algebraic equations involving the
  F-matrices. However, this
  algorithm is not an efficient one. The equations determining string operators
  have infinite many solutions and there is no general method to pick up the
  irreducible solutions. In this sense it is even not guarantied that one can
  find all the (irreducible) string operators.

  In this paper we try to fix this weak point.
  Motivated by the work of Kitaev and Kong~\cite{KK12,Kon12},
  we introduce the \emph{Q-algebra} approach to
  compute quasiparticle statistics.  The idea using Q-algebra modules to
  classify quasiparticles is analog to using group representations to classify
  particles.
  It is well known that in a system with certain symmetry the energy
  eigenspaces, including excited states of particles,
  form representations of the symmetry group.
  String-net models are fixed-point models thus renormalization can be viewed
  as generalized ``symmetry''.
  Moreover we show that renormalization in string-net
  models can be exactly described by \emph{evaluation} linear maps. This allows
  us to introduce the Q-algebra, which describes the renormalization of
  quasiparticle states. Quasiparticles are identified as the invariant
  subspaces under the action of the Q-algebra, {i.e.,} Q-algebra modules.

  Roughly speaking, the Q-algebra is the ``renormalization group'' of
  quasiparticles in string-net models, a linearized, weakened version of a
  group. The notions of algebra modules and group representations are
  almost equivalent.
  Modules over the group algebra are in one to one correspondence with group
  representations up to similarity transformations. The only
  difference is that ``module'' emphasizes on the subspace of
  states that is invariant under the action of the group or
  algebra, while ``representation'' emphasizes on how the group or algebra
  acts on the ``module''.

  The specific algorithm to compute the Q-algebra modules is also analog to
  that to compute the group representations. For a group, firstly, we write
  the multiplication rules. Secondly, we take the multiplication rules as the
  ``canonical representation''. Thirdly, we try to simultaneously
  block-diagonalize the canonical representation. Finally, the irreducible
  blocks correspond to irreducible representations, or simple modules over the
  group algebra. The canonical representation of a group contains all types of
  irreducible representations of that group. This is also true for the
  Q-algebra. The multiplication rules of the Q-algebra are fully determined by
  the F-matrices ({i.e.,} the UFC, see (\ref{qalg}) and (\ref{qalgt})).  
  Therefore, following this block-diagonalization process we
  have a finite algorithm to calculate the quasiparticle statistics from
  F-matrices. We are guarantied to find all types of quasiparticles by
  block-diagonalizing the canonical representation of the Q-algebra.
  Simultaneous block-diagonalization is a straightforward algorithm, however,
  it is not a quite efficient way to decompose the Q-algebra. The algorithm
  used in this paper is an alternative one, \emph{idempotent decomposition}.

  The notions of algebra, module and idempotent play an important role in our
  discussion and algorithm.
  On the other hand, we think it a necessary step to proceed from ``groups and
  group representations'' to ``algebras and modules'', since we are trying to
  extend our understanding from ``symmetry breaking phases'' to ``topologically
  ordered phases''.
  We provide a brief introduction in Appendix
  \ref{secalgmod} to these mathematical notions in case the reader is not
  familiar with them.

}

\del{
  We will first drop the tetrahedron-reflectional symmetry of the F-matrices and
  keep the tetrahedron-rotational symmetry assumption, reformulate the string-net
  model, and give an algorithm to calculate MTC that describes the quasiparticle
  statistics (see section \ref{qalgsec}). This approach is motivated and based
  on Kitaev and Kong's work\cite{KK12,Kon12}. They pointed out that the
  quasiparticles can be classified by \emph{module functors}, or equivalently
  \emph{modules} over certain local operator algebra.  In this paper we call such
  algebras as \emph{Q-algebra}.  We show that the Q-algebra actually describe the
  ``renormalization'' of quasiparticles, therefore the quasiparticle types
  (fixed-point quasiparticles, quasiparticles up to local operators) are modules
  over the Q-algebra.  With the tetrahedron-rotational symmetry, the string
  operators that create quasiparticles at their ends are well defined and closely
  related to Q-algebra modules.  One can calculate the quasiparticle statistics
  with Q-algebra modules, as well as with string operators.
}

\add{
  Another weak point of the original version of Levin-Wen model in Ref
  \onlinecite{LW05} is that the F-matrices are assumed to be symmetric under
  certain index permutation. More precisely, the F-matrices have 10 indices
  which can be associated to a tetrahedron, 6 indices to the edges and 4
  indices to the
  vertices. If we reflect or rotate the tetrahedron the indices get permuted
  and the F-matrices are assumed to remain the same. In this paper we find that
  such tetrahedral symmetry can be dropped thus the string-net model is
  generalized.

  In  Section \ref{qalgsec} we will first drop the
  tetrahedron-reflectional symmetry of the F-matrices but keep the
  tetrahedron-rotational symmetry and reformulate the string-net model. We keep
  the tetrahedron-rotational symmetry because in this case the relation between
  string operators and Q-algebra modules is clear. We give the formula to
  compute quasiparticle statistics, the $T,S$ matrices from Q-algebra modules
  by comparing them to string operators. 

}

Next, in Section \ref{qalgG} we will drop the tetrahedron-rotational symmetry
assumption, and generalize string-net models to arbitrary gauge. In arbitrary
gauge the string operators are not naturally defined, but we can still obtain
the formula of quasiparticle statistics by requiring the formula to be gauge
invariant and reduce to the special case if we choose the
tetrahedron-rotation-symmetric gauge.

\add{
  Finally, in Section \ref{boundary} we briefly discuss the boundary
  theory\cite{KK12} of generalized string-net models which shows the
  holographic bulk-edge relation.  In 2+1D there are many different kinds of
  topological orders, classified by the non-Abelian statistics of the
  quasiparticles plus the chiral central charge of the edge state.
  Mathematically, the non-Abelian statistics, or the fusion and braiding data of
  quasiparticles form a UMTC. On the other hand,
  in 1+1D, there is only trivial topological order.\cite{VCL0501,CGW1107}
  However, if we consider anomalous topological orders that only appear on the
  edge of 2+1D gapped states, we will have nontrivial anomalous 1+1D
  topological orders. In these anomalous 1+1D topological orders, the fusion of
  quasiparticles is also described by a set of F-matrices. Mathematically, the
  F-matrices give rise to a UFC, and anomalous
  1+1D topological orders are classified by UFCs. The F-matrices we use to
  determine a string-net ground state wavefunction turn out to be the same
  F-matrices describing the fusion of quasiparticles on one of the edges of the
  string-net model.\cite{KK12,KW14}  Thus, our algorithm calculating the bulk
  quasiparticle statistics (UMTC) from the F-matrices (UFC) can also be understood
  as calculating the bulk topological order (UMTC) from the anomalous boundary
  topological order (UFC). Since the same bulk topological order may have
  different gapped boundaries, it is a natural consistency question: Do these
  different gapped boundaries lead to the same bulk? The answer is
  ``yes''.\cite{KK12} Mathematically, we give an algorithm to compute the
  Drinfeld center functor $\cZ$ that maps a UFC (that describes a 1+1D
  anomalous topological order) to a UMTC (that describes a 2+1D topological
  order with zero chiral central charge).\cite{M03a} Different gapped
  boundaries of a 2+1D topological phase are described by different UFCs, but
  they share the same Drinfeld center UMTC. In Appendix \ref{znp} we discuss the
  twisted $\bbz_n$ string-net model in detail to illustrate this holographic
  relation.
}

\del{  In 2+1D there are many different kinds of topological orders.  It appears that
  the 2+1D topological orders are labeled and classified by UMTC (that describe
  the non-Abelian statistics of the quasiparticles) plus the chiral central
  charge of the edge state.  On the other hand, in 1+1D, there is only trivial
  topological order.\cite{VCL0501,CGW1107} However, if we consider anomalous
  topological orders that only appear on the edge of 2+1D gapped states, we will
  have nontrivial anomalous 1+1D topological orders. Those  anomalous 1+1D
  topological orders are described and classified by UFC (see section
  \ref{boundary}).\cite{KW14}  Since an anomalous 1+1D topological order is always
  a boundary of a 2+1D topological order, it will be interesting to compute the
  2+1D bulk topological order from the anomalous 1+1D topological order. The
  Drinfeld center functor $\cZ$ that maps a UFC (that describes an 1+1D anomalous
  topological order) to a UMTC (that describes a 2+1D topological order with zero
  chiral central charge) is exactly the mapping from the 1+1D topological order
  on the boundary to the  2+1D topological order in the bulk.\cite{KW14}
  We discussed in detail a twisted $\bbz_n$ string-net model
  to illustrate this holographic relation (see appendix \ref{znp}).
}

\section{Quasiparticle excitations}
\label{qpart}

\subsection{Local quasiparticle excitations and topological quasiparticle
excitations}

Topologically ordered states in 2+1D are characterized by their unusual
particle-like excitations which may carry fractional/non-Abelian statistics.
To understand and to classify particle-like excitations in topologically
ordered states, it is important to understand the notions of local
quasiparticle excitations and topological quasiparticle excitations.  

\begin{figure}[tb]
  \centering
  \includegraphics[scale=0.5]{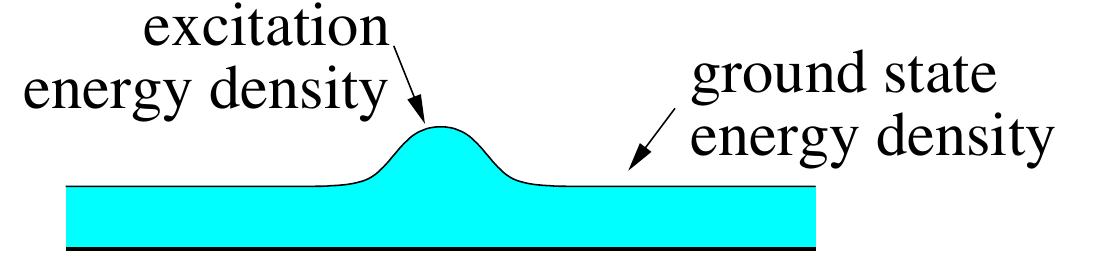}
  \caption{The energy density distribution of a quasiparticle.}
  \label{exceng}
\end{figure}

First we define the notion of ``particle-like'' excitations.  Consider a gapped
system with translation symmetry.  The ground state has a uniform energy
density.  If we have a state with an excitation, we can measure the energy
distribution of the state over the space.  If for some local area, the energy
density is higher than ground state, while for the rest area the energy density
is the same as ground state, one may say there is a ``particle-like''
excitation, or a quasiparticle, in this area (see Figure \ref{exceng}).
\add{
  Among all the quasiparticle excitations, some can be created or annihilated
  by local operators, such as a spin flip.  This kind of particle-like
  excitation is called local quasiparticle. However, in topologically ordered
  systems, there are also quasiparticles that cannot be  created or annihilated
  by any finite number of local operators (in the infinite system size limit).
  In other words, the higher local energy density cannot be created or removed
  by \emph{any} local operators in that area. Such quasiparticles are called
  topological quasiparticles.
}
\del{
  Quasiparticles defined like this can be divided into two types.  The
  first type can be created or annihilated by local operators, such as a spin
  flip.  So, the first type of particle-like excitation is called local
  quasiparticle excitations.  The second type cannot be  created or annihilated
  by any finite number of local operators (in the infinite system size limit).
  In other words, the higher local energy density cannot be created or removed by
  \emph{any} local operators in that area.  The second type of particle-like
  excitation is called topological quasiparticle excitations.
}

From the notions of local quasiparticles and topological quasiparticles, we can
further introduce the notion \emph{topological} quasiparticle type, or simply,
quasiparticle type.  We say that local quasiparticles are of the trivial type,
while topological quasiparticles are of nontrivial types. Two
topological quasiparticles are of the same type if and only if they differ by
local quasiparticles.  In other words, we can turn one topological
quasiparticle into the other one of the same type by applying some local
operators.  

\subsection{Simple type and composite type}
\label{types}

To understand the notion of simple type and composite type, let us discuss
another way to define quasiparticles:\\ 
Consider a gapped local Hamiltonian qubit system defined by a local Hamiltonian
$H_0$ in $d$ dimensional space $M^d$ without boundary.  A collection of
quasiparticle excitations labeled by $i$ and located at $\v x_i$ can be
produced as \emph{gapped} ground states of $H_0+\Delta H$ where $\Delta H$ is
non-zero only near $\v x_i$'s.  By choosing different $\Delta H$ we can create
all kinds of quasiparticles.  We will use $\xi_i$ to label the type of the
quasiparticle at $\v x_i$.

The gapped ground states of $H_0+\Delta H$ may have a degeneracy $D(
M^d;\xi_1,\xi_2,\cdots )$ which depends on the quasiparticle types $\xi_i$ and
the topology of the space $M_d$. The degeneracy is not exact, but becomes exact
in the large space and large particle separation limit.  We will use
$\cV(M^d;\xi_1,\xi_2,\cdots )$ to denote the space of the degenerate ground
states.

If the Hamiltonian $H_0+\Delta H$ is not gapped, we will say $D(
M^d;\xi_1,\xi_2,\cdots )=0$ ({i.e.,}\ $\mathcal V(M^d;\xi_1,\xi_2,\cdots )$
has zero dimension).  If $H_0+\Delta H$ is gapped, but if $\Delta H$ also
creates quasiparticles away from $\v x_i$'s (indicated by the bump in the
energy density away from $\v x_i$'s), we will also say $D(
M^d;\xi_1,\xi_2,\cdots )=0$.  (In this case quasiparticles at $\v x_i$'s do not
fuse to trivial quasiparticles.) So, if $D( M^d;\xi_1,\xi_2,\cdots )>0$,
$\Delta H$ only creates quasiparticles at $\v x_i$'s.

If the degeneracy $D(M^d;\xi_1,\xi_2,\cdots)$ cannot not be lifted by any
small local perturbation near $\v x_1$, then the particle type $\xi_1$ at
$\v x_1$ is said to be simple. Otherwise, the  particle type $\xi_1$ at $\v
x_1$ is said to be composite.  The degeneracy $D(M^d;\xi_1,\xi_2,\cdots)$
for simple particle types $\xi_i$ is a universal property ({i.e.,}\ a
topological invariant) of the topologically ordered state.

\subsection{Fusion of Quasiparticles}

When $\xi_1$ is composite, the space of the degenerate ground states
$\cV(M^d;\xi_1,\xi_2,\xi_3,\cdots)$ has a direct sum
decomposition:
\begin{align}
  &\quad\ \cV(M^d;\xi_1,\xi_2,\xi_3,\cdots)
  \nonumber
  \\
  &=
  \cV(M^d;\zeta_1,\xi_2,\xi_3,\cdots)\oplus
  \cV(M^d;\chi_1,\xi_2,\xi_3,\cdots)
  \nonumber
  \\
  &\oplus
  \cV(M^d;\psi_1,\xi_2,\xi_3,\cdots)\oplus \cdots
\end{align}
where $\zeta_1$, $\chi_1$, $\psi_1$, {\it etc.} are simple types.  To see
the above result, we note that when $\xi_1$ is composite the ground state
degeneracy can be split by adding some small perturbations near $\v x_1$.
After splitting, the original degenerate ground states become groups of
degenerate  states, each group of degenerate  states span the space $\mathcal
V(M^d;\zeta_1,\xi_2,\xi_3,\cdots)$ or $\mathcal
V(M^d;\chi_1,\xi_2,\xi_3,\cdots)$ {\it etc.} which correspond to simple
quasiparticle types at $\v x_1$.  We denote the composite type $\xi_1$ as
\begin{align}
 \xi_1=\zeta_1\oplus \chi_1\oplus \psi_1\oplus \cdots.
\end{align}
When we fuse two simple types of topological particles $\xi$ and $\zeta$
together, it may become a topological particle of a composite type:
\begin{align}
 \xi\otimes \zeta=\eta=\chi_1\oplus \chi_2 \oplus \cdots,
\end{align}
where $\xi,\zeta,\chi_i$ are simple types and $\eta$ is a composite type.
In this paper, we will use an integer tensor $N_{\xi\zeta}^\chi$ to describe
the quasiparticle fusion, where $\xi,\zeta,\chi$ label simple types.
When $N_{\xi\zeta}^\chi=0$, the fusion of $\xi$ and $\zeta$ does not
contain $\chi$.  When $N_{\xi\zeta}^\chi=1$, the fusion of $\xi$ and
$\zeta$ contain one $\chi$: $\xi\otimes b=\chi \oplus \chi_1  \oplus
\chi_2 \oplus \cdots$.  When $N_{\xi\zeta}^\chi=2$, the fusion of $\xi$ and
$\zeta$ contain two $\chi$'s: $\xi\otimes \zeta =\chi \oplus \chi
\oplus \chi_1  \oplus \chi_2 \oplus \cdots$.  This way, we can denote that fusion
of simple types as 
\begin{align} 
\xi\otimes \zeta=\oplus_\chi N_{\xi\zeta}^\chi \chi .  
\end{align} 
In physics, the quasiparticle types always refer to simple types. The fusion
rules $N_{\xi\zeta}^\chi$ is a universal property of the topologically ordered
state.  The degeneracy $D(M^d;\xi_1,\xi_2,\cdots)$ is determined completely by
the fusion rules $N_{\xi\zeta}^\chi$.

\add{
  Let us then consider the fusion of 3 simple quasiparticles $\xi,\zeta,\chi$.
  We may first fuse $\xi,\zeta$, and then with $\chi$,
  $(\xi\ot\zeta)\ot\chi=(\oplus_{\alpha} N_{\xi\zeta}^\alpha \alpha)\ot
  \chi=\oplus_\beta (\sum_\alpha N_{\xi\zeta}^\alpha N_{\alpha\chi}^\beta)
  \beta$. We may also first fuse $\zeta,\chi$ and then with $\xi$,
  $\xi\ot(\zeta\ot\chi)=\xi\ot(\oplus_{\alpha} N_{\zeta\chi}^\alpha
  \alpha)=\oplus_\beta (\sum_\alpha N_{\xi\alpha}^\beta
  N_{\zeta\chi}^\alpha)\beta$. This requires that $\sum_\alpha
  N_{\xi\zeta}^\alpha N_{\alpha\chi}^\beta=\sum_\alpha N_{\xi\alpha}^\beta
  N_{\zeta\chi}^\alpha$. If we further consider the degenerate states $\mathcal
  V (M^d;\xi,\zeta,\chi,\cdots)$, it is not hard to see fusion in different
  orders means splitting the space $\mathcal V (M^d;\xi,\zeta,\chi,\cdots)$ as
  different direct sums of subspaces. Thus, fusion in different orders 
  differ by basis changes of $\mathcal V (M^d;\xi,\zeta,\chi,\cdots)$. The
  \emph{F-matrices} are nothing but the data to describe such basis changes.

  For 1+1D anomalous topological orders (gapped edges of 2+1D topological
  orders), the quasiparticles can only fuse but not braiding. So, the fusion
  rules $N_{\xi\zeta}^\chi$ and the F-matrices are enough to describe 1+1D
  anomalous topological orders. Later, we will see fusion rules and F-matrices
  are also used to determine a string-net wavefunction, which may seem
  confusing. However, as we have mentioned, this is a natural result of the
  holographic bulk-edge relation. Intuitively, one may even view the string-net
  graphs in 2D space as the 1+1D space-time trajectory of the edge
  quasiparticles.

  For 2+1D topological orders, the quasiparticles can also braid. We also need 
  data to describe the braiding of the quasiparticles in addition to the fusion
  rules and the F-matrices, as introduced in the next two subsections.
}
\subsection{Quasiparticle intrinsic spin}

If we twist the quasiparticle at $\v x_1$ by rotating $\Delta H$ at $\v x_1$ by
360$^\circ$ (note that $\Delta H$ at $\v x_1$ has no rotational symmetry), all
the degenerate ground states in $\mathcal V(M^d;\xi_1,\xi_2,\xi_3,\cdots)$ will
acquire the same geometric phase $\ee^{\ri\theta_{\xi_1}}$ provided that the
quasiparticle type $\xi_1$ is a simple type.  We will call
$\ee^{\ri\theta_{\xi}}$ the intrinsic spin (or simply spin) of the simple type
$\xi$, which is a universal property of the topologically ordered state.

\subsection{Quasiparticle mutual statistics}

If we move the quasiparticle $\xi_2$ at $\v x_2$ around the quasiparticle
$\xi_1$ at $\v x_1$, we will generate a non-Abelian geometric phase -- a
unitary transformation acting on the degenerate ground states in $\mathcal
V(M^d;\xi_1,\xi_2,\xi_3,\cdots)$.  Such a unitary transformation not only
depends on the types $\xi_1$ and $\xi_2$, but also depends on the
quasiparticles at other places.  So, here we will consider three quasiparticles
of simple types $\xi$, $\zeta$, $\chi$ on a 2D sphere $S^2$.  The ground state
degenerate space is $\cV(S^2;\xi,\zeta,\chi)$.  For some choices of $\xi$,
$\zeta$, $\chi$,  $D(S^2;\xi,\zeta,\chi) \geq 1$, which is the dimension
of $\cV(S^2;\xi,\zeta,\chi)$.  Now, we move the
quasiparticle $\zeta$ around the quasiparticle $\xi$.  All the degenerate
ground states in $\cV(S^2;\xi,\zeta,\chi)$ will acquire the same geometric
phase $\dfrac{\ee^{\ri \theta_{\chi^*}}}{\ee^{\ri \theta_\xi}\ee^{\ri
\theta_\zeta}}$.  This is because, in $\mathcal V(S^2;\xi,\zeta,\chi)$, the
quasiparticles $\xi$ and $\zeta$ fuse into $\chi^*$, the anti-quasiparticle of
$\chi$.  Moving quasiparticle $\zeta$ around the quasiparticle $\xi$ plus
rotating $\xi$ and $\zeta$ respectively by 360$^\circ$ is like rotating
$\chi^*$ by 360$^\circ$.  So, moving quasiparticle $\zeta$ around the
quasiparticle $\xi$ generates a phase $\dfrac{\ee^{\ri
\theta_{\chi^*}}}{\ee^{\ri \theta_\xi}\ee^{\ri \theta_\zeta}}$.  We see that
the quasiparticle mutual statistics is determined by the  quasiparticle spin
$\ee^{\ri\theta_\xi}$  and the  quasiparticle fusion rules
$N_{\xi\zeta}^\chi$.  For this reason, we call the set of data
$(\ee^{\ri\theta_\xi},N_{\xi\zeta}^\chi)$ quasiparticle statistics.

\add{  It is an equivalent way to describe
  quasiparticle statistics by $T,S$ matrices.
  The $T$ matrix is a diagonal matrix. The diagonal elements are the
  quasiparticle spins
  \begin{align}
    T_{\xi\zeta}=T_{\xi}\delta_{\xi\zeta}=\ee^{\ri \theta_\xi}
    \delta_{\xi\zeta}.
  \end{align}
  The $S$ matrix can be determined from
  the quasiparticle spin $\ee^{\ri \theta_\xi}$ and quasiparticle fusion rules
  $N_{\xi\zeta}^\chi$
  [see Eq. (223) in Ref. \onlinecite{Kit06}] :
  \begin{align}
    S_{\xi\zeta}=\frac{1}{D_{\cZ(\cC)}}\sum_\chi N_{\xi\zeta^*}^\chi
    \frac{\ee^{\ri \theta_\chi}}{ \ee^{\ri \theta_\xi} \ee^{\ri \theta_\zeta}}
    d_\chi ,
  \end{align}
  where $d_\xi>0$ is the largest eigenvalue of the matrix $N_\xi$, whose
  elements are $N_{\xi,\zeta\chi} = N_{\xi\zeta}^\chi$.

  On the other hand $S$ matrix determines the fusion rules $N_{\xi\zeta}^\chi$
  via the Verlinde formula[see \eqref{ver} in Section \ref{TSstatistics}].  So,
  $T_\xi$ and $S_{\xi\zeta}$ fully determine the quasiparticle statistics
  $(\ee^{\ri \theta_\xi}, N_{\xi\zeta}^\chi)$, and the  quasiparticle
  statistics $(\ee^{\ri \theta_\xi}, N_{\xi\zeta}^\chi)$ fully determines
  $T_\xi$ and $S_{\xi\zeta}$.
  \medskip

  We want to emphasize
  that the fusion rules and F-matrices of bulk quasiparticles
  and edge quasiparticles are \emph{different}. In this paper we use only the
  F-matrices of edge quasiparticles, which is also the F-matrices describing
  the bulk string-net wavefunctions. Although our Q-algebra module algorithm
  can be used to compute the F-matrices of bulk quasiparticles, we did not
  explain in detail how to do this, because calculating
  the $T,S$ matrices is enough
  to distinguish and classify 2+1D topological orders with
  gapped boundaries.
}

\section{String-net models with tetrahedron-rotational symmetry}
\label{qalgsec}

The string-net condensation was suggested by Levin and Wen as a mechanism for
topological phases.\cite{LW05} We give a brief review here.

The basic idea of Levin and Wen's construction was to find an ideal fixed-point
ground state wave function for topological phases.  Such an ideal wave function
can be fully determined by a finite amount of data.  The idea is not to
directly describe the wave function, but to describe some local constraints
that the wave function must satisfy.  These local constraints can be viewed as
a scheme of ground state renormalization.

Let us focus on lattice models. We put the lattice on a sphere so that there is
no nontrivial boundary conditions.  Since renormalization will change the
lattice, we will consider a class of ground states on arbitrary lattices on the
sphere.  One way to obtain ``arbitrary lattices'' is to triangulate the sphere in arbitrary
ways.  There may be physical degrees of freedom on the faces, edges, as well as
vertices of the triangles.  Any two triangulations can be related by adding,
removing vertices and flipping edges.  The ideal ground state must renormalize
coherently when re-triangulating.

The string-net picture is dual to the triangulation picture.  As an intuitive
example, one can consider the strings as electric flux lines through the edges
of the triangles.  Like the triangulation picture, there are some basic
local transformations of the string-nets, which we call evaluations.
Physically, evaluations are related to the so-called \emph{local unitary
transformations}\cite{CGW10},
and states related by local unitary transformations
belong to the same phase.  If we evaluate the whole string-net on the sphere,
or in other words, we renormalize the whole string-net so that no degrees of
freedom are left, we should obtain just a number.
We require that this number remains the same no matter how we evaluate the
whole string-net. This gives rise to the desired local constraints
of the ideal ground state wave function.  We now demonstrate in detail the
formulation of the string-net model with the tetrahedron-rotational symmetry.  

\subsection{String-net}

A string-net is a 2-dimensional directed trivalent graph.  The vertices and
edges (strings) are labeled by some physical degrees of freedom. By convention,
we use
$i,j,k,\cdots$ for string labels and $\alpha,\beta,\cdots$ for vertex labels.  We
assume that the string and vertex label sets are finite.

A fully labeled string-net corresponds to a basis vector of the Hilbert space.
If a string-net is not labeled, it stands for the ground state subspace in the
total Hilbert space spanned by the basis string-nets with all possible
labellings.
A partially labeled string-net corresponds to the projection of the ground
state subspace to the subspace of the total Hilbert space where states on the
labeled
edges/vertices are given by the fixed labels.  This way, we have a graph
representation of the  ground state subspace, which will help us to actually
compute the  ground state subspace.

There is an involution of the string label set, $i\mapsto i^*$ satisfying
$i^{**}=i$, corresponding to reversing the string direction
\begin{equation}
  \f{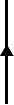}\,i\ =\ \f{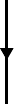}\,i^*\ =\ \f{i}\,i^{**}.
  \label{dual}
\end{equation}
When an edge is vacant, or not occupied by any string, we say it is
a trivial string.  The trivial string is labeled by 0 and $0^*=0$. Trivial strings are
usually omitted or drawn as dashed lines
\begin{equation}
  \text{vacuum}=\ \f{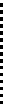}\,0\ =\ \f{0.pdf}\,0^*.
  \label{0}
\end{equation}

In addition we assume that trivial strings are totally invisible, {i.e.,}\ can
be
arbitrarily added, removed and deformed without affecting the ideal ground
state wave function.  To understand this point, suppose we have a unlabeled
string-net on a graph.  It corresponds to a subspace $\cV$ of the
total Hilbert space $\cH$ on the graph.  Now, we add a trivial string to
the string-net which give us a partially labeled string-net on a new graph
(with an extra string carrying the label $0$).  Such a partially labeled
string-net on a new graph corresponds to subspace $\cV_0$ of the total
Hilbert space $\cH_0$ on the new graph. The two subspaces $\mathcal
V$ and $\cV_0$ are very different belonging to different  total Hilbert
spaces.  The statement that trivial strings are totally invisible implies that
the  two subspaces are isomorphic to each other $\cV \cong \cV_0$.
In other words, there exists a local linear map from $\cH_0$ to
$\cH$, such that the map is unitary when restricted on $\cV_0$.
Such a map is called an evaluation, which will be discussed in more detail
below.\subsection{Evaluation and F-move}
\label{evf}

A string-net graph represents a subspace, which corresponds to the ground
state subspace on that graph.  When we do wavefunction renormalization, we
change the graph on which the string-net is defined.  However, the  ground
state subspace represented by the string-net, in some sense, is not changed
since the string-net represents a fixed-point wavefunction under
renormalization.  To understand such a fixed-point property of the string-net
wavefunction, we need to compare ground state subspaces on different graphs.
This leads to the notion of evaluation.

We do not directly specify the ground state subspace
represented by a string-net.  Rather, we specify several evaluations ({\it
i.e.} several local linear maps).  Those evaluations will totally
fix the ground state subspace of the string-net for every graph.

Consider two graphs with total Hilbert space $\cH_1$ and $\cH_2$.
Assume that the two graphs differ only in a local area and dim$\cH_1 \geq$
dim$\cH_2$.  An evaluation is a local linear map from
$\cH_1$ to $\cH_2$.  Here ``local'' means that the map is
identity on the overlapping part of the two graphs.
Note that the evaluation maps a Hilbert space of higher
dimension to a  Hilbert space of lower dimension. It reduces the degrees
of freedom and represents a wave function renormalization.

\add{
  Although evaluation depends on the two graphs with $\cH_1$ and $\cH_2$, since
  the graphs before and after evaluation are normally shown in the equations,
  we will simply use $\ev$ to denote evaluations.  We will point out the two
  graphs only if it is necessary.
}

Let us list the evaluations that totally fix the ground state subspace.
For a single vertex, we have the following evaluation
\begin{equation}
  \ev\f{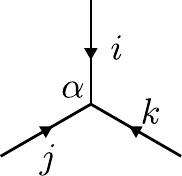}=\delta_{ijk,\alpha}\f{vertex},
  \label{ev-vertex}
\end{equation}
where
\begin{gather}
  \delta_{ijk,\alpha}=0 \text{ or } 1 \label{delta},\\
\delta_{ijk,\alpha}=\delta_{kij,\alpha}=\delta_{k^*j^*i^*,\alpha}
\label{deltas},\\
  \delta_{ij0,\alpha}=\delta_{ij^*}\delta_{0\alpha}\label{delta0},\\
  \sum_m N_{ijm}N_{m^*kl}=\sum_n N_{inl}N_{n^*jk}.
  \label{fusiona}
\end{gather}
We note that the above evaluation does not change the graph and thus $\mathcal
H_1=\cH_2$.  The evaluation is a projection operator in $\cH_1$
whose action on the basis of  $\cH_1$ is given by (\ref{ev-vertex}).

The vertex with $\delta_{ijk,\alpha}=1$ is called a stable vertex.
$N_{ijk}=\sum_{\alpha}\delta_{ijk,\alpha}$ is the dimension of the stable
vertex subspace, called fusion rules.
To determine the order of the $ijk$ labels, one should first use \eqref{dual}
to make the three strings going inwards,
then read the string labels anticlockwise.
If one thinks of strings as electric flux lines, $\delta_{ijk,\alpha}$ enforces
the total flux to be zero for the ground state.

The next few evaluations are for 2-edge plaquettes, $\Theta$-graphs, and
closed loops:
\begin{gather}
\ev\
\f{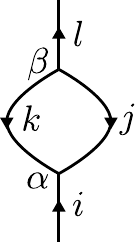}=\frac{\Theta_{ijk}}{O_i}\delta_{ijk,\alpha}
\delta_{\alpha\beta}\delta_{il}\
\f{i}\,i,
  \label{ev-2-plaquette}\\
\ev\ \raise
-23pt\hbox{\includegraphics{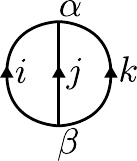}}=
\Theta_{ijk}\delta_{ijk,\alpha}\delta_{\alpha\beta},
\label{thetagraph}\\
  \ev\ \f{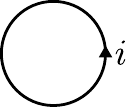}=O_i\label{di},
\end{gather}
where
\begin{gather}
  \Theta_{ijk}=\Theta_{kij}=\Theta_{k^*j^*i^*}\label{theta},\\ 
  \Theta_{ii^*0}=\Theta_{i^*i0}=O_i=O_{i^*}\label{theta0},\\
  O_0=\ev(\text{vacuum})=1 \label{d0}.
\end{gather}
$O_i=\varkappa_i d_i$ where $d_i>0$ is called the quantum dimension of the type
$i$ string.
When $i$ is self-dual $i=i^*$, the phase factor $\varkappa_i$ corresponds to
the Frobenius-Schur indicator. Otherwise $\varkappa_i$ can be adjusted to 1 by
gauge transformations.
$O_i=O_{i^*},\Theta_{ijk}=\Theta_{kij}$ is because for any closed string-net on
the sphere, the
half loop on the right can be moved to the left across the other side of the
sphere.
Those evaluations change the graph. They are described by how every basis
vector of $\cH_1$ is mapped to a vector in $\cH_2$.

The last evaluation is called F-move.  It changes the graph.
In fact, the F-move is the most basic graph changing operation
acting on local areas with two stable vertices. It is given by
\begin{equation}
  \ev\f{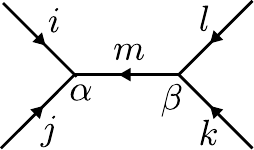}=\sum_{n\lambda\rho}F_{kln,\lambda\rho}^{ijm,\alpha\beta}\f{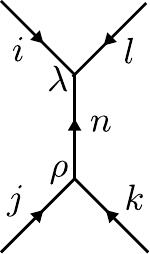}.
  \label{F-move}
\end{equation}
It is equivalent to flipping edges in the triangulation picture.  The rank 10
tensor $F_{kln,\lambda\rho}^{ijm,\alpha\beta}$ are called F-matrices.
$m,\alpha\beta$ are considered as column indices and $n,\lambda\rho$ as row
indices.  $F_{kln,\lambda\rho}^{ijm,\alpha\beta}$ is zero if any of the four
vertices is unstable.
Otherwise, $F_{kl}^{ij}$  is a unitary matrix.

Note that the evaluations can be done recursively.
When two graphs within $\cH_1$ and  $\cH_2$ are connected by
different sequences of evaluations, the induced maps from $\cH_1$ to
$\cH_2$ by different sequences must be the same.
Firstly the F-matrices must satisfy the well known pentagon equations
\begin{gather}
  \sum_{n\tau\lambda\eta}F^{ijq,\alpha\beta}_{kln,\eta\lambda}
  F^{n^*jk,\lambda\gamma}_{rsp,\tau\mu}F^{lin,\eta\tau}_{pst,\rho\nu}
=\sum_{\sigma}F^{lq^*k,\beta\gamma}_{rst,\rho\sigma}
F^{ijq,\alpha\sigma}_{rt^*p,\nu\mu}.
  \label{pent}
\end{gather}

We also assume the tetrahedron-rotational symmetry.
The tetrahedron-rotational symmetry is actually the symmetry of the evaluation,
not of the graphs. For example, if one rotates the graphs in
\eqref{thetagraph}by $180^\circ$, the result of the evaluation should be
$\Theta_{k^*j^*i^*}$ and the
tetrahedron-rotational symmetry requires that
$\Theta_{ijk}=\Theta_{k^*j^*i^*}$. In general, with tetrahedron-rotational
symmetry, doing evaluation is
``rotation-invariant''.
When the evaluation of tetrahedron graphs, and simpler graphs such as
$\Theta$-graphs or
closed loops, is rotation-invariant, the evaluation of all graphs is
rotation-invariant.
Therefore, we call it tetrahedron-rotational symmetry.

The tetrahedron-rotational symmetry puts the following constraints on the
F-matrices.
Firstly, it is necessary that the trivial string is totally invisible.
So, if in \eqref{F-move} we set the label $k$ to 0, the corresponding F-matrix
elements should be 1 when the labels match and 0 otherwise, {i.e.,}\
\begin{gather}
\ev\ \f{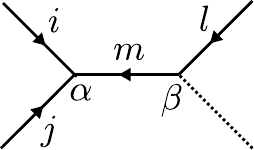}=\sum_{n\lambda\rho}F^{ijm,\alpha\beta}_{0ln,\lambda\rho}\f{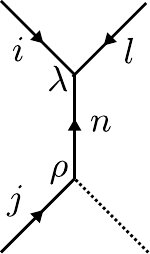},\\
F^{ijm,\alpha\beta}_{0ln,\lambda\rho}
  =\delta_{ijl,\alpha}\delta_{\alpha\lambda}
  \delta_{ml}\delta_{nj}\delta_{\beta 0}\delta_{\rho 0}.
    \label{f0}
\end{gather}
Secondly consider the tetrahedron graphs.
After one step of F-move, the tetrahedron graphs have only 2-edge plaquettes.
Thus, the amplitude can be expressed by $F^{ijm,\alpha\beta}_{kln,\lambda\rho}$,
$\Theta_{ijk}$ and $O_i$, {i.e.,}\
\begin{alignat}{2}
  &\ev\ &\f{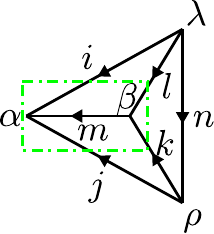}&= F^{ijm,\alpha\beta}_{kln,\lambda\rho}
  \frac{\Theta_{nli}\Theta_{n^*jk}}{O_n},\\
  &\ev\ \ &\f{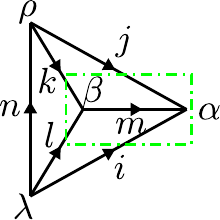}&= F_{ijn^*,\rho\lambda}^{klm^*,\beta\alpha}
  \frac{\Theta_{nli}\Theta_{n^*jk}}{O_n},\\
  &\ev\ &\f{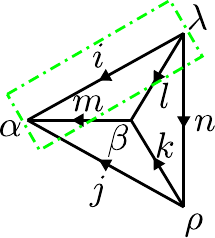} &=F_{l^*n^*k,\rho\beta}^{jmi,\alpha\lambda}
  \frac{\Theta_{m^*kl}\Theta_{n^*jk}}{O_k},\\
  &\ev\ &\f{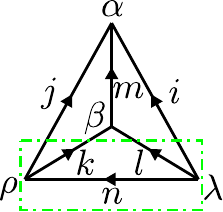} &=F_{i^*l^*m,\beta\alpha}^{k^*j^*n,\rho\lambda}
    \frac{\Theta_{mij}\Theta_{m^*kl}}{O_m},
\end{alignat}
where the F-move is performed in the boxed area.
These four results must be the same.
Thus, we got another constraint on the F-matrices.
\begin{align}
  F^{ijm,\alpha\beta}_{kln,\lambda\rho}
  &=F_{ijn^*,\rho\lambda}^{klm^*,\beta\alpha}
  =F_{l^*n^*k,\rho\beta}^{jmi,\alpha\lambda}
  \frac{O_n \Theta_{m^*kl}}{O_k\Theta_{nli}}\nonumber\\
  &=F_{i^*l^*m,\beta\alpha}^{k^*j^*n,\rho\lambda}
  \frac{O_n \Theta_{mij}\Theta_{m^*kl}}{O_m \Theta_{nli}\Theta_{n^*jk}}
  \label{sph}.
\end{align}
Note that \eqref{sph} is
different from that in Ref. \onlinecite{LW05} because we do not allow
reflection of the tetrahedron.
This is necessary to include cases of fusion rules like $N_{ijk}\neq N_{jik}$,
for example the finite group $G$ model with a non-Abelian group $G$
[see Section \ref{finiteG}].
It turns out that the conditions above are sufficient for evaluation of any
string-net graph to be rotation-invariant.

\add{
  With these consistency conditions, given any two string-net graphs
  with total Hilbert spaces $H_1$ and $H_2$, $\dim H_1\geq \dim H_2$, there is
  a unique evaluation map from $H_1$ to $H_2$, given by the compositions of
  simple evaluations listed above. Thus, evaluation depends on only the graphs
  before and after, or $H_1$ and $H_2$, not on the way we change the graphs.
  As we mentioned before, usually it is not even necessary to explicitly point
  out $H_1$ and $H_2$, since they are automatically shown in the equations and
  graphs.
}

We want to emphasize that the fusion rules {(\ref{ev-vertex}-\ref{fusiona})},
the F-move \eqref{F-move}, and the pentagon
equation \eqref{pent} are the most fundamental ones.  The rest of the equations
(\ref{ev-2-plaquette}-\ref{d0})\eqref{f0}\eqref{sph} are either normalization
conventions, gauge choices, or conditions of the tetrahedron-rotational
symmetry. With the tetrahedron-rotational symmetry, $O_i,\Theta_{ijk}$ are encoded in F-matrices. In \eqref{sph} set some
indices to 0, and we have
\begin{align}
  F^{ii^*0,00}_{ii^*0,00}&=\frac{1}{O_i},\\
  F^{ijk,\alpha\beta}_{j^*i^*0,00}
  &=\frac{\Theta_{ijk}}{O_i O_j}\delta_{ijk,\alpha}\delta_{\alpha\beta},\\
  F^{jj^*0,00}_{i^*ik,\alpha\beta}
  &=\frac{O_k}{\Theta_{ijk}}\delta_{ijk,\alpha}\delta_{\alpha\beta}.
\end{align}
Moreover, in \eqref{pent} set $r$ to 0 and one can get
\begin{equation}
  \sum_{n\lambda\rho}F_{kln,\lambda\rho}^{ijm,\alpha\beta}
F^{jkn^*,\rho\lambda}_{lim',\alpha'\beta'}
=\delta_{mm'}\delta_{\alpha\alpha'}\delta_{\beta\beta'}.
  \label{invF}
\end{equation}
Thus, $O_i$ satisfies
\begin{align}
  \sum_{k}N_{ijk}O_k 
=\sum_{k\alpha\beta}
F^{ijk,\alpha\beta}_{j^*i^*0,00}F^{jj^*0,00}_{i^*ik,\alpha\beta}O_i O_j
  =O_i O_j.
  \label{qd}
\end{align}
This implies that $O_i$ is an eigenvalue of the matrix $N_i$, whose entries are
$N_{i,jk}=N_{ijk}$, and the corresponding eigenvector is
$(O_0,O_1,\dots)^\mathsf{T}$.

\subsection{Fixed-point Hamiltonian}
Does the evaluation defined above really describe the renormalization of some
physical ground states?
What is the corresponding Hamiltonian?
A sufficient condition for the string-nets to be physical ground states is that
the F-move is unitary, or that the F-matrices are unitary
\begin{equation}
  \sum_{n\lambda\rho}F_{kln,\lambda\rho}^{ijm,\alpha\beta}
(F_{kln,\lambda\rho}^{ijm',\alpha'\beta'})^*
=\delta_{mm'}\delta_{\alpha\alpha'}\delta_{\beta\beta'}.
  \label{uni}
\end{equation}
This requires a special choice of $O_i, \Theta_{ijk}$.
From \eqref{uni}\eqref{invF} we know
\begin{equation}
F^{jkn^*,\rho\lambda}_{lim,\alpha\beta}
=(F_{kln,\lambda\rho}^{ijm,\alpha\beta})^*,
  \label{unis}
\end{equation}
which implies that $F^{ii^*0,00}_{ii^*0,00}=(F_{i^*i0,00}^{i^*i0,00})^*$,
$O_i=O_i^*$ are real numbers, or $\varkappa_i=\pm 1$,
and $F^{jj^*0,00}_{i^*ik,\alpha\alpha}=(F^{ijk,\alpha\alpha}_{j^*i^*0,00})^*$,
{i.e.,}\ if $N_{ijk}>0$
\begin{equation}
  \left|\Theta_{ijk}\right|^2=O_iO_jO_k=d_id_jd_k>0.
  \label{thetaijk}
\end{equation}
Moreover, \eqref{qd}\eqref{thetaijk} together imply that
\begin{equation}
  \sum_k N_{ijk}d_k=d_id_j.
  \label{qdabs}
\end{equation}
Hence $d_i$ has to be the largest eigenvalue (Perron-Frobenius eigenvalue) of the matrix $N_i$ and the
corresponding eigenvector is $(d_0,d_1,\dots)^\mathsf{T}$.

To find the corresponding Hamiltonian, note that
\begin{align}
  &\ev\,\ev^\dag \ \f{i}\,i=\ev\sum_{jkl\alpha\beta}
\frac{\Theta^*_{ijk}}{O_i}\delta_{ijk,\alpha}\delta_{\alpha\beta}\delta_{il}\
\f{2-plaquette}
  \nonumber\\&
  =\sum_{jk}N_{ijk}\frac{|\Theta_{ijk}|^2}{O_i^2}\ \f{i}\,i
  =\sum_{k}O_k^2 \ \f{i}\,i
  =D_\cC^2 \ \f{i}\,i,
\end{align}
where $D_\cC=\sqrt{\sum_{k}O_k^2}=\sqrt{\sum_{k}d_k^2}$ is the total quantum
dimension.

\begin{figure}[tb]
  \centering
  \includegraphics{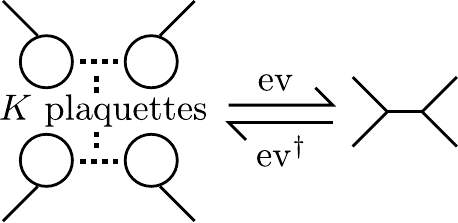}
\caption{A local area with $K$ plaquettes and 4 external legs. The evaluation
removes all the plaquettes.}
  \label{fig:evK}
\end{figure}

For a local area with $K$ plaquettes, consider the evaluation that removes all
the $K$ plaquettes and results in a tree graph, as sketched in Figure
\ref{fig:evK}.
Since F-move does not change the number of plaquettes, we can first use F-move
to deform the local area and make all the plaquettes 2-edge plaquettes.
Thus, we have
\begin{equation}
  \frac{\ev\,\ev^\dag}{D_\cC^{2K}}=\textbf{1}.
  \label{evid}
\end{equation}

Consider
\begin{equation}
  P=\frac{\ev^\dag\,\ev}{D_\cC^{2K}},
\end{equation}
which means that first use $\ev$ to remove all the plaquettes in the local
area, and then use $\ev^\dag$ to recreate the plaquettes and go back to the
original
graph.
It is easy to see that ${P^2=P}$. Thus, $P$ is a Hermitian projection.
Like evaluation, $P$ can also act on any local area of the string-net.
We can take the Hamiltonian as the sum of local projections acting on every
vertex and plaquette
\begin{equation}
  H=\sum_{\substack{\text{vertices}\\\text{plaquettes}}}(\textbf{1}-P),
  \label{hamiltonian}
\end{equation}
which is the fixed-point Hamiltonian.

We see that $P$ is exactly the projection onto the ground state subspace.  $P$
acting on a single vertex projects onto the stable vertex; $P$ acting on a
plaquette is equivalent to the $B_p$ operator.\cite{LW05,KK12,Kon12,Lan12} The
$B_p$ operator is more general because there may be ``nonlocal'' plaquettes,
for example when the string-net is put on a torus, in which case evaluation
cannot be performed.  But in this paper we will not consider such ``nonlocal''
plaquettes.  Evaluation is enough for our purpose.

If we evaluate the whole string-net, the evaluated tree-graph string-net
represents the ground state.
For a fixed lattice on the sphere with $K$ plaquettes, the evaluated tree graph
is just the void graph, or the vacuum.
Therefore, the normalized ground state is
\begin{equation}
  |\psi\rangle_{\text{ground}}=\dfrac{\ev^\dag}{D_\cC^K}|\text{vacuum}\rangle.
  \label{groundstate}
\end{equation}
Generically the ground state subspace is $\cV= \ev^\dag \cV_\text{tree}.$

\subsection{Cylinder ground states, quasiparticle excitations and Q-algebra}

Now we have defined the string-net models with tetrahedron-rotational
symmetry. We continue to study the quasiparticles excitations.

\begin{figure}[tb]
  \centering
  \includegraphics[scale=1]{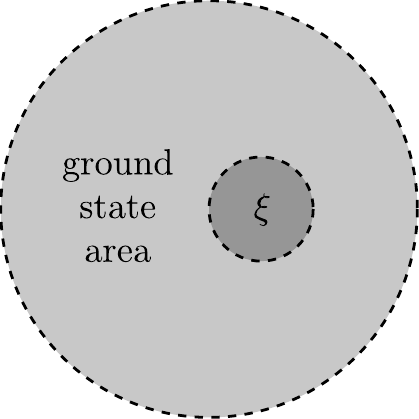}
\caption{Quasiparticle $\xi$: The local energy density is constant in the
ground state area but higher in the $\xi$ area.}
  \label{fig:qpa}
\end{figure}

Let us first discuss the generic properties of quasiparticle excitations from a
different point of view.
By definition, a quasiparticle is a local area with higher energy density,
labeled by $\xi$, surrounded by the ground state area (see Figure
\ref{fig:qpa}).
\add{
  We want to point out that, a topological quasiparticle is scale invariant.
  If we zoom out, put the $\xi$ area and ground state area together, and view
  the larger area as a single quasiparticle area $\xi'$, then $\xi'$ should be
  the same type as $\xi$. Moreover, if we are considering a fixed-point model
  such as the string-net model, the excited states of the quasiparticle won't
  even change no matter how much surrounding ground state area is included.
  Intuitively, we may view this renormalization process as ``gluing'' a cylinder
  ground state to the quasiparticle area. ``Gluing a cylinder ground state'' is
  then an element of the ``renormalization group'' that acts on (renormalizes)
  the quasiparticle states. Thus, quasiparticle states form ``representations''
  of the ``renormalization group''. Of course ``renormalization group'' is not a
  group at all, but the idea to identify quasiparticles as ``representations''
  still works. We develop this idea rigorously in the following. We
  will define the ``gluing'' operation, introduce the algebra induced by
  gluing cylinder ground states and show that quasiparticles are
  representations of, or modules over this algebra. This algebra is nothing but
  the ``renormalization group''.
}

Since any local operators acting inside the $\xi$ area will not change the
quasiparticle type,
we do not quite care about the degrees of freedom inside the $\xi$ area,
Instead, the entanglement between the ground state area and the $\xi$ area is
much more important,
and should capture all the information about the quasiparticle types and
statistics.
Since we are considering systems with local Hamiltonians,
the entanglement should be only in the neighborhood of the boundary between the
ground state area and the $\xi$ area.

To make things clear, we would first forget about the entanglement and study
the properties of ground states on a cylinder with the open boundary condition.
\add{
  Here open boundary condition means that 
 setting all boundary Hamiltonian terms to zero thus strings on the boundary are
 free to be in any state.
 }
Later we will put the entanglement back
\add{by ``gluing'' boundaries and adding back the
Hamiltonian terms near the ``glued'' boundaries.}

On a cylinder with the open boundary condition, the ground states form a
subspace $\vcyl$ of the total Hilbert space.  $\vcyl$ should be scale
invariant, {i.e.,}\ not depend on the size of the cylinder.  We want to show
that, the fixed-point cylinder ground states in $\vcyl$ allows a cut-and-glue
operation.

\begin{figure}[tb]
  \centering
  \includegraphics{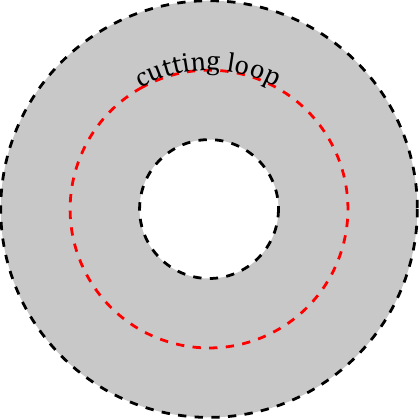}
\caption{(Color online) Cut a cylinder into two cylinders. The entanglement
between the two
cylinders is only in the neighborhood of the cutting loop.}
  \label{fig:cutting}
\end{figure}

Given a cylinder, we can cut it into two cylinders with a loop, as in Figure
\ref{fig:cutting}.  The states in the two cylinders are entangled with each
other; but again, the entanglement is only near the cutting loop.  If we ignore
the entanglement for the moment, in other words, imposing open boundary
conditions for both cylinders, by scale invariance, the ground state subspaces
on the two cylinders should be both $\vcyl$.  Next, we add back the entanglement
(this can be done, {e.g.,} by applying proper local projections in the
neighborhood of the cutting loop), which is like ``gluing'' the two  cylinders
along the cutting loop, and we should obtain the ground states on the bigger
cylinder before cutting, but still states in $\vcyl$.  Therefore, gluing two
cylinders by adding the entanglement back gives a map
\begin{alignat}{2}
  \vcyl&\ot \vcyl &&\overset{\text{glue}}{\longrightarrow} \vcyl
  \nonumber\\ 
  h_1 &\ot h_2 &&\longmapsto h_1h_2
\end{alignat}
It is a natural physical requirement that such gluing is associative, $(h_1
h_2)h_3=h_1(h_2 h_3)$.  Thus, it
can be viewed as a multiplication.  Now, the cylinder ground state
subspace $\vcyl$ is equipped with a multiplication, the gluing map.  Mathematically, $\vcyl$
forms an \emph{algebra} [see Appendix \ref{secalgmod}].

\begin{figure}[tb]
  \centering
  \includegraphics{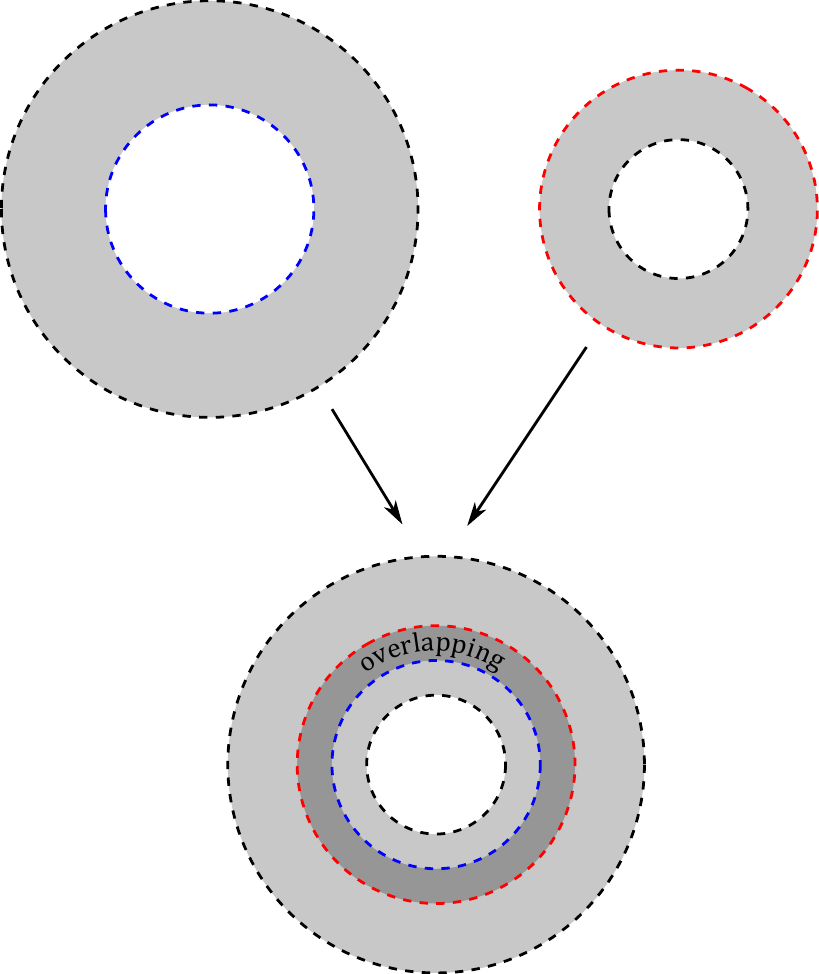}
\caption{(Color online) Gluing two cylinders: make sure there is an overlapping
area between the glued boundaries (red and blue).}
  \label{fig:gluing}
\end{figure}

We can also enlarge a cylinder by gluing another cylinder onto it.
Note that when two cylinders are cut from a larger one as in Figure
\ref{fig:cutting},
there is a natural way to put them back together,
however, when we arbitrarily pick two cylinders,
simply putting them together may not work.
To glue, or enforce entanglements between two cylinders,
we need to first put them in such a way that there is an overlapping area
between their glued boundaries (see Figure \ref{fig:gluing}).
In this overlapping area, we identify degrees of freedom from one cylinder
with those from the other cylinder;
this way we ``connect and match'' the boundaries.
Next, we apply proper local projections in the neighborhood of the overlapping
area,
such that the two cylinders are well glued.
But, the ground state subspace remains the same,
{i.e.,}\ ``multiplying'' $\vcyl$ by $\vcyl$ is still $\vcyl$,
\begin{align}
  \vcyl\vcyl&=\bigg\{\sum_k c^{(k)}h_1^{(k)}h_2^{(k)}\bigg|
  \begin{array}{rl}
    k\in \bbn,&h_1^{(k)}\in \vcyl,\\
    c^{(k)}\in \bbc,& h_2^{(k)}\in\vcyl
  \end{array}
  \bigg\}
  \nonumber\\
&=\vcyl.
\end{align}

Now, we put back the quasiparticle $\xi$.  Since the entanglement between $\xi$
and the ground state area is restricted in the neighborhood of the boundary,
it can be viewed as imposing some nontrivial boundary conditions on the
cylinder.  Equivalently, we may say that the quasiparticle $\xi$ picks a
subspace $M_\xi$ of $\vcyl$.  $M_\xi$ should also be scale invariant.  If we
enlarge the area by gluing a cylinder onto it, in other words, multiply $M_\xi$
by $\vcyl$, $M_\xi$ remains the same, $\vcyl M_\xi=M_\xi.$ Mathematically,
$M_\xi$ is a \emph{module} over the algebra $\vcyl$.  In this way, the
quasiparticle $\xi$ is identified with the module  $M_\xi$ over the algebra
$\vcyl$. 
A reducible module corresponds to a composite type of quasiparticle,
and an irreducible module corresponds to a simple type of quasiparticle (see
section \ref{types}).

\begin{figure}[tb]
  \centering
  \includegraphics{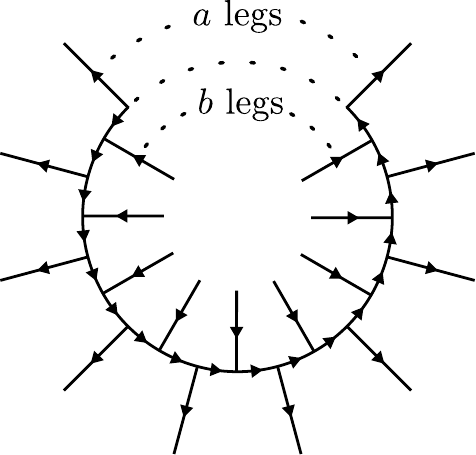}
\caption{A typical tree graph on a cylinder. Here the dashed lines stand for
the omitted part of the graph, but not trivial strings.}
  \label{fig:treegraph}
\end{figure}

As for string-net models, recall that ground state subspaces can be represented
by evaluated tree graphs.
The actual ground state subspace can always be obtained by applying $\ev^\dag$
to the space of evaluated tree graphs.
Thus, we can find out $\vcyl$ by examining the possible tree graphs on a
cylinder.
A typical tree graph on a cylinder is like Figure \ref{fig:treegraph}.
Assuming that there are $a$ legs on the outer boundary and $b$ legs on the
inner boundary,
we denote the space of these graphs by $V^a_b$.
As evaluated graphs, all the vertices in the graphs in $V^a_b$ must be stable.
In principle $a,b$ can take any integer numbers.
But note that if $c<a$, we can add $a-c$ trivial legs on the outer boundary,
and $V^c_b$ can be viewed as a subspace of $V^a_b$.
Similarly $V^a_c\subset V^a_b$ for $c<b$.
Therefore, we know the largest space is $\vcyl=\ev^\dag V^\infty_\infty$.

We find that the gluing of cylinder ground states can be captured by the spaces
$V^a_b$.
The gluing is nothing but adding back the entanglement.
For string-net model the proper local projections are just $\ev^\dag \ev$.
But before doing evaluation we have to ``connect and match'' the boundaries.
{i.e.,}\ make sure the strings are well connected.
Note that $\ev^\dag$ and $\ev$ acting inside each cylinder do not affect the
boundary legs.
$(\ev^\dag V^a_b)$ can be glued onto $(\ev^\dag V^c_d)$ from the outer side
only if $b=c$.
We need to first connect the legs on the inner boundary of $(\ev^\dag V^a_b)$
with those on the outer boundary of $(\ev^\dag V^b_d)$ and make their labels
match each other's;
broken strings are not allowed inside a ground state area.
This defines a map $p: (\ev^\dag V^a_b) \ot (\ev^\dag V^b_d)\to (\ev^\dag
V^a_b)\ot (\ev^\dag V^b_d)|_\text{w.c.}$,
where w.c. means restriction to the subspace in which the strings are well
connected.
Thus, ${\dfrac{\ev^\dag \ev}{D_\cC^{2K}} p:(\ev^\dag V^a_b) \ot (\ev^\dag
V^b_d)\to (\ev^\dag V^a_d)}$
is the desired gluing if there are $K$ plaquettes in $(\ev^\dag V^a_d)$.
Recall that evaluation can be performed in any sequence.
We know the following diagram
\begin{equation}
  \xymatrix{
    (\ev^\dag V^a_b) \ot (\ev^\dag V^b_d)
    \ar[d]^p
    \ar[rr]^-{\ev\ot\ev}
    &&V^a_b\ot V^b_d\ar[d]^p\\
    (\ev^\dag V^a_b) \ot (\ev^\dag V^b_d)|_\text{w.c.}
    \ar[rr]^-{\ev\ot\ev}
    \ar[rrd]^\ev 
    \ar[d]^-{\frac{\ev^\dag \ev}{D_\cC^{2K}}}
    &&V^a_b\ot V^b_d|_\text{w.c.}\ar[d]^\ev\\
    (\ev^\dag V^a_d)
    &&V^a_d
    \ar[ll]^-{\frac{\ev^\dag}{D_\cC^{2K}}}
  }
  \label{vabalg}
\end{equation}
commutes.
Thus, gluing $(\ev^\dag V^a_b)$ with $(\ev^\dag V^c_d)$
to obtain ground states in $(\ev^\dag V^a_d)$ can be done by first
considering the evaluation of the tree graphs, $V^a_b\ot V^c_d\xrightarrow{p}
V^a_b\ot V^c_d|_\text{w.c.}\xrightarrow{\ev}V^a_d$
and then applying $\ev^\dag$ to get the actual ground states.

However, it is impossible to deal with an infinite-dimensional algebra
$\vcyl=\ev^\dag V^\infty_\infty$.
\add{We want to reduce it to an algebra of finite dimension. Again our idea is
to do renormalization. When we glue the cylinder ground states, we renormalize
along the radial direction. Now, we renormalize along the tangential direction,
or reduce the number of boundary legs, to reduce the dimension of the algebra.}

More rigorously, our goal is to study the quasiparticles,
which correspond to modules over $\vcyl$,
rather than the algebra $\vcyl$ itself.
So, if we can find some algebra such that its modules are the ``same'' as
those over $\vcyl$
(here ``same'' means that the categories of modules are equivalent),
this algebra can also be used to study the quasiparticles.
Mathematically, two algebras are called \emph{Morita
equivalent}\cite{Mor58,Kon12}
if they have the ``same'' modules.
Thus, we want to find finite dimensional algebras that are Morita equivalent to
$\vcyl$.

Note that $V^a_a$ with the multiplication $\ev p:V^a_a\ot V^a_a\xrightarrow{p}
V^a_a\ot V^a_a|_\text{w.c.}\xrightarrow{\ev}V^a_a$
forms an algebra.
From \eqref{vabalg} we also know that $(\ev^\dag V^a_a)$ and $V^a_a$ are
isomorphic algebras
(the isomorphisms are just $\ev$ and $\ev^\dag$).
It turns out that all the algebras $V^a_a$ are Morita equivalent for
$a=1,2,\cdots$ [see Section \ref{Morita}].
Thus, we know $V^1_1$ and $\vcyl=\ev^\dag V^\infty_\infty$ have the ``same''
modules.
We choose the algebra $V^1_1$ to study the quasiparticles of string-net models
for $V^1_1$ has the lowest dimension among the algebras $V^a_a$.
\add{Now, we reduced the infinite-dimensional algebra $\vcyl$ to
the finite-dimensional $V^1_1$.}
Since a graph in $V^1_1$ is like a letter Q,
and $V^1_1$ describes the physics of quasiparticles,
we name it the \emph{Q-algebra}, denoted by
\begin{equation}
  Q=V^1_1=\raise -45pt \hbox{\includegraphics{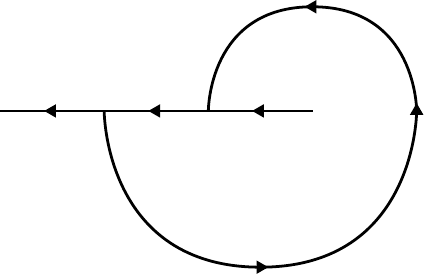}}\ .
\end{equation}
The subtlety of Morita equivalence will be discussed further in Section
\ref{Morita}.

In detail, the natural basis of $Q$ is
\begin{equation}
  Q_{rsj}^{i,\mu\nu}=\f{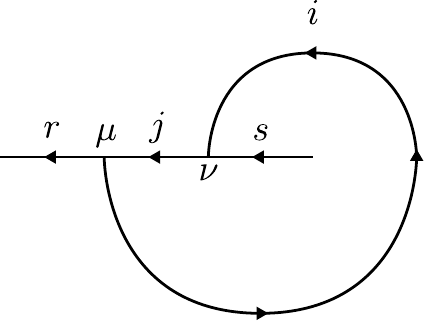}\ .
  \label{q-basis}
\end{equation}
The notation $Q^{i,\mu\nu}_{rsj}$ looks like a tensor.
But $Q^{i,\mu\nu}_{rsj}$ denotes a basis vector rather than a number.
On one hand, $Q^{i,\mu\nu}_{rsj}$ represents a cylinder ground state
$\ev^\dag |Q^{i,\mu\nu}_{rsj}\rangle$;
on the other hand, when glued onto other cylinder ground states,
$Q^{i,\mu\nu}_{rsj}$ can be viewed as a linear operator
$\hat{Q}^{i,\mu\nu}_{rsj}$.
Both of
$|Q^{i,\mu\nu}_{rsj}\rangle$ and $\hat{Q}^{i,\mu\nu}_{rsj}$
are incomplete and misleading.
That is why we choose the simple notation $Q^{i,\mu\nu}_{rsj}$;
just keep in mind that it stands for a vector/operator.
As an evaluated graph, the two vertices are stable,
$\delta_{rj^*i,\mu}=\delta_{sij^*,\nu}=1$. Thus, the dimension of the Q-algebra
is
\begin{equation}
  \dim Q= \sum_{rsij}N_{rj^*i}N_{sij^*}=\sum_{rs}\Tr(N_r N_s).
  \label{dimQ}
\end{equation}
In terms of the natural basis, the multiplication is
\begin{gather}
  Q_{rsj}^{i,\mu\nu}Q_{s'tl}^{k,\sigma\tau}
  =\ev p\ (Q_{rsj}^{i,\mu\nu}\ot Q_{s'tl}^{k,\sigma\tau})
  \label{qalg}
  \\
  =\delta_{ss'}\sum_{mn\lambda\rho}
  Q_{rtn}^{m,\lambda\rho}
  \sum_{\alpha\beta\gamma}
  F_{k^*ln^*,\alpha\beta}^{ij^*s,\nu\sigma}
  F_{k^*nm^*,\lambda\gamma}^{r^*i^*j,\mu\beta}
  F_{in^*m,\rho\gamma}^{tkl^*,\tau\alpha}
  \frac{\Theta_{kim^*}}{O_m} .
  \nonumber 
\end{gather}
We know that the identity is
\begin{equation}
  \textbf{1}=\sum_r Q_{rrr}^{0,00}=\sum_r \f{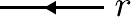}.
\end{equation}

We can study the quasiparticles by decomposing the Q-algebra.
The simple quasiparticle types correspond to simple $Q$-modules.
The number of quasiparticle types is just the number of different simple $Q$-modules.
As of the Morita equivalence of $V_a^a$ algebras, 
we also want to mention that the \emph{centers} [see Appendix \ref{secalgmod}] of
Morita equivalent algebras are isomorphic.
Thus, the center $Z(Q)\cong Z(V_a^a)\cong Z(\vcyl)$ is an \emph{invariant}.
We argue that $Z(Q)$ is exactly the ground state subspace on a torus and
$\dim(Z(Q))$ is the torus ground state degeneracy,
also the number of quasiparticle types.

We give a more detailed discussion on the Q-algebra in Appendix
\ref{appendix-qalg}.

Assume that we have obtained the module $M_\xi$ over the Q-algebra,
or the invariant subspace ${M_\xi\subset Q}$,
that corresponds to the quasiparticle $\xi$.
Since ${M_\xi=\textbf{1}M_\xi=\oplus_r Q_{rrr}^{0,00} M_\xi}$, it is possible
to choose the basis vectors of $M_\xi$ from $Q_{rrr}^{0,00} M_\xi$ respectively.
Such a basis vector can be labeled by $r,\tau$, namely,
\begin{equation}
  e_{r\tau}^{\xi}=\f{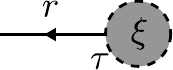}\in Q_{rrr}^{0,00} M_\xi.
  \label{m-basis}
\end{equation}
Then we can calculate the representation matrix of $Q_{rsj}^{i,\mu\nu}$
with respect to this basis
\begin{align}
 Q_{rsj}^{i,\mu\nu}e_{t\sigma}^{\xi} 
  &=\sum_{q\tau} M_{\xi,rsj,q\tau t\sigma}^{i,\mu\nu}e_{q\tau}^\xi
 \nonumber\\
 &= \ev p \left( \f{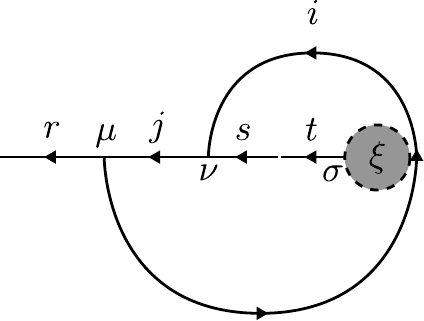} \right)
  \nonumber\\
  &=\delta_{st}\sum_\tau M_{\xi,rsj,\tau\sigma}^{i,\mu\nu}\f{m-basis}
  \nonumber\\
  &=\delta_{st}\sum_\tau M_{\xi,rsj,\tau\sigma}^{i,\mu\nu}e_{r\tau}^\xi ,
  \label{m-rep}
\end{align}
where $p$ is still the map that connects legs and matches labels.
We know that the representation matrix of $Q_{rsj}^{i,\mu\nu}$ is
${M_{\xi,rsj,q\tau
t\sigma}^{i,\mu\nu}=\delta_{rq}\delta_{st}M_{\xi,rsj,\tau\sigma}^{i,\mu\nu}}$,
which is a block matrix.
Since $Q_{rrr}^{0,00}$ is an idempotent,
$M_{\xi,rrr,\tau\sigma}^{0,00}=\delta_{\tau\sigma}$.
Later we will see that the representation matrices
$M_{\xi,rsj,\tau\sigma}^{i,\mu\nu}$ are closely related to the string
operators, and can be used to calculate the quasiparticle statistics.

\subsection{String operators and quasiparticle statistics}
\label{secstat}

\begin{figure}[tb]
  \centering
  \includegraphics[scale=1]{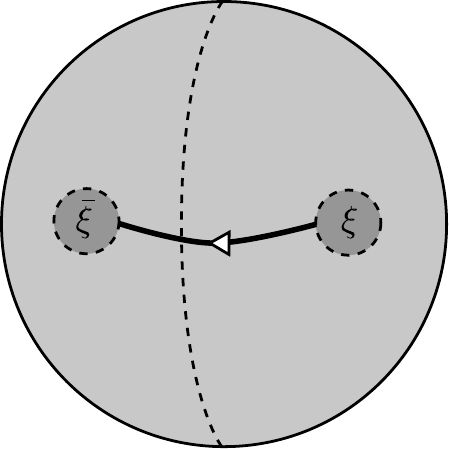}
  \caption{A string operator on the sphere}
  \label{fig:sosph}
\end{figure}

The string operator\cite{LW05} is yet another way to study the quasiparticles.
A string operator creates a pair of quasiparticles at its ends (see Figure
\ref{fig:sosph}).
It is also the hopping operator of the quasiparticles, {i.e.,}\ a
quasiparticle can be moved around with the corresponding string operator.
First recall the matrix representations of string operators.
For consistency we still label the string operator with $\xi$,
\begin{equation}
  \f{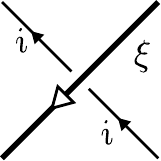} =\sum_{rsj\mu\nu\tau\sigma}
  \Omega_{\xi,rsj,\tau\sigma}^{i,\mu\nu}\f{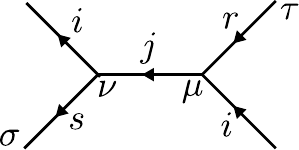},
  \label{str-op}
\end{equation}
where $\Omega_{\xi,rsj,\tau\sigma}^{i,\mu\nu}$ is zero when either vertex is
unstable.

For a longer string operator, one can apply \eqref{str-op} piece by piece,
and contract the $r,\tau$ or $s,\sigma$ labels at the connections.
In particular, $\Omega_{\xi,rrr,\tau\sigma}^{0,00}=\delta_{\tau\sigma}$,
since $\Omega_{\xi,rrr}^{0,00}$ means simply extend the string operator.
We define $N_{\xi,r}=\Tr(\Omega_{\xi,rrr}^{0,00})$,
which means the number of type $r$ strings the string operator $\xi$ decomposes
to.

Consider a closed string operator $\xi$
\begin{equation}
  \ev\left(\f{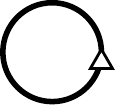}\,\xi\right)
  =\varkappa_\xi d_\xi
  =\sum_r N_{\xi,r} O_r
  =\sum_r N_{\xi,r} \varkappa_r d_r
  \label{qdxi}.
\end{equation}
If $\xi$ is simple and $N_{\xi,r}>0,\ N_{\xi,s}>0$,
there must be some $i,j,\mu,\nu,\tau,\sigma$ such that
$\Omega_{\xi,rsj,\tau\sigma}^{i,\mu\nu}\neq 0$.
Otherwise $\xi$ is reducible, $\xi=\xi_1 \oplus \xi_2$ where $\xi_1$ does not
contain $s$, ${N_{\xi_1,s}=0}$, and $\xi_2$ does not contain $r$,
$N_{\xi_2,r}=0$.
${\Omega_{\xi,rsj,\tau\sigma}^{i,\mu\nu}\neq 0}$ implies that $N_{i^*s^*j}>0,\
N_{irj^*}>0$,
and due to \eqref{thetaijk}, $O_rO_iO_j>0,\ O_sO_iO_j>0$.
Thus, we have $O_rO_s>0$, $\varkappa_r=\varkappa_s$, which is also the same as
$\varkappa_\xi$.
In other words, when $\xi$ is simple, $\varkappa_\xi=\varkappa_r$ for
$N_{\xi,r}>0$.
Therefore, the quantum dimension of simple quasiparticle $\xi$ is
\begin{equation}
  d_\xi=\sum_r N_{\xi,r} d_r.
\end{equation}

The quasiparticle spin and $S$-matrix can be expressed in terms of
string operators.  For simple quasiparticles $\xi,\zeta$
\begin{align}
\overline{T_\xi}&=\ee^{-\ri \theta_\xi}=\frac{1}{d_\xi}\ \ev\left(\
\f{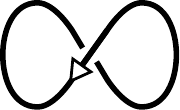}\,\xi\right),\\
  S_{\xi\zeta}&=\frac{1}{D_{\cZ(\cC)}}\ \ev\left(\xi\,\f{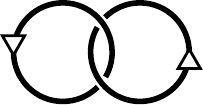}\,\zeta\right),
\end{align}
where $ D_{\cZ(\cC)}=\sqrt{\sum_\xi d_\xi^2}$ is the total quantum dimension of
the quasiparticles.
Applying \eqref{str-op} we have
\begin{gather}
\overline{T_\xi} =\ee^{-\ri \theta_\xi}=\frac{1}{d_\xi}\sum_r O_r^2
\Tr(\Omega_{\xi,rr0}^{r^*,00}),\\
  S_{\xi\zeta} =\frac{1}{D_{\cZ(\cC)}}\sum_{rst\mu\nu}
  \frac{\Theta_{rst}\Theta_{srt}}{O_t}
  \Tr(\Omega_{\xi,rrt^*}^{s,\mu\nu})\Tr(\Omega_{\zeta,s^*s^*t}^{r^*,\mu\nu}).
\end{gather}

One can find that for some $\xi$, $S_{\xi\zeta}=\dfrac{d_\zeta}{D_{\cZ(\cC)}}$.
Such $\xi$ is the trivial quasiparticle, and later will be labeled by 1.
The quasiparticle fusion rules $N_{\xi\zeta}^\chi$
can be  determined from $S_{\xi\zeta}$, which is known as the Verlinde
formula\cite{V8860}:
\begin{align}
N_{\xi\zeta}^\chi= \sum_\psi
\frac{S_{\xi\psi}S_{\zeta\psi}\overline{S_{\chi\psi}}}
{S_{1\psi}} ,
\label{ver}
\end{align}
and then we can then identify the anti-quasiparticle $\xi^*$ of $\xi$,
which satisfies $N_{\zeta\xi^*}^1=N_{\xi^*\zeta}^1=\delta_{\xi\zeta}$.
\label{TSstatistics}

Now look at the graph in \eqref{m-rep}.
We can also use string operator $\xi$ to move the quasiparticle out of the
loop, and then do the evaluation.
The result should be the same as the representation matrix
$M^{i,\mu\nu}_{\xi,rsj,\tau\sigma}$.
\begin{align}
  &\ev p\left( \f{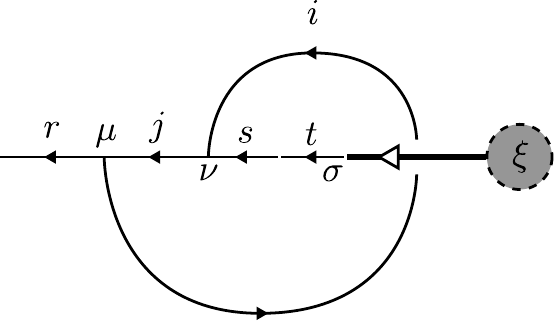} \right) 
  \nonumber\\&
  =\sum_{r's'j'\mu'\nu'\tau'\sigma'}
  \Omega_{\xi,r's'j',\tau'\sigma'}^{i,\mu'\nu'}\xt
  \nonumber\\&
  \ev p\left( \f{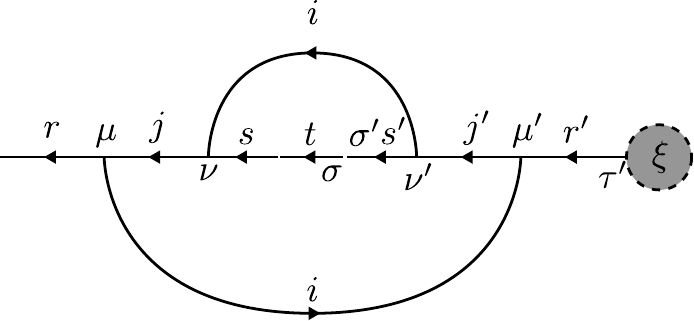} \right)
  \nonumber\\&
=\delta_{st}\sum_\tau \frac{\Theta_{rj^*i}\Theta_{sij^*}}{O_r O_j}
\Omega_{\xi,rsj,\tau\sigma}^{i,\mu\nu}\f{m-basis}\ ,
  \label{so-ev}
\end{align}
Comparing the two results \eqref{m-rep}\eqref{so-ev}, we get the relations
between the module $M_\xi$ and the string operator $\xi$
\begin{gather}
M_{\xi,rsj,\tau\sigma}^{i,\mu\nu}=\frac{\Theta_{rj^*i}\Theta_{sij^*}}{O_r
O_j}\Omega_{\xi,rsj,\tau\sigma}^{i,\mu\nu},
  \label{m-so}\\
  N_{\xi,r}=\Tr(M_{\xi,rrr}^{0,00})=\dim (Q_{rrr}^{0,00} M_\xi).
\end{gather}
It turns out that the matrix representations of Q-algebra modules and
the string operators differ by only some normalizing factors.
The statistics in terms of Q-algebra modules is
\begin{gather}
\overline{T_\xi}=\frac{1}{d_\xi}\sum_r O_r
\Tr(M_{\xi,rr0}^{r^*,00}),\label{Txi}\\
  S_{\xi\zeta}=\frac{1}{D_{\cZ(\cC)}}\sum_{rst\mu\nu}
  \frac{O_rO_sO_t}{\Theta_{rst}\Theta_{srt}}
  \Tr(M_{\xi,rrt^*}^{s,\mu\nu})\Tr(M_{\zeta,s^*s^*t}^{r^*,\mu\nu}).
  \label{Sxi}
\end{gather}

\subsection{Examples}
In the following examples, there are no extra degrees of freedom on the
vertices, $N_{ijk}\leq 1$. Such fusion rules are called multiplicity-free.
We can omit all the vertex labels. We will first list the necessary data
$(N_{ijk},F_{kln}^{ijm})$ to define a specific rotation-invariant string-net
model.
The tensor elements not explicitly given are either 0 or can be calculated from
the constraints given in Section \ref{evf}.
Second we give the corresponding Q-algebra. The multiplication is given as a
table
\[\begin{array}{c|c}
    &e_b\\\hline e_a&e_a e_b
  \end{array}\]
In the end we calculate the simple modules over the Q-algebra and
$N_{\xi,r},d_\xi,T_\xi,S_{\xi\zeta}$.

\subsubsection{Toric code ($\bbz_2$) model}
The toric code model is the most simple string-net model.~\cite{Kit03}
\begin{itemize}
  \item Two types of strings, labeled by 0,1 and $1^*=1$.
  \item $N_{011}=1$, $F_{110}^{110}=1$, $O_1=1$.
\end{itemize}

The Q-algebra is 4-dimensional. The natural basis is
\begin{gather*}
  e_{00}=Q_{000}^0,e_{01}=Q_{001}^1,\\
  e_{10}=Q_{111}^0,e_{11}=Q_{110}^1.
\end{gather*}
The multiplication is 
\[ \begin{array}{c|cccc}
  &e_{00}&e_{01}&e_{10}&e_{11}\\
  \hline
  e_{00}&e_{00}&e_{01}&0&0\\
  e_{01}&e_{01}&e_{00}&0&0\\
  e_{10}&0&0&e_{10}&e_{11}\\
  e_{11}&0&0&e_{11}&e_{10}
\end{array}\]

Easy to see this is the direct sum of two group algebras of $\bbz_2$. There are
4 1-dimensional simple modules.
\[\begin{array}{c||c|c|c|c}
    \xi&1&2&3&4\\
    \hline
\text{basis}&\dfrac{e_{00}+e_{01}}{2}&\dfrac{e_{00}-e_{01}}{2}
&\dfrac{e_{10}+e_{11}}{2}&\dfrac{e_{10}-e_{11}}{2}\\
    \hline
    M_{\xi,000}^0&1&1&0&0\\
    M_{\xi,001}^1&1&-1&0&0\\
    M_{\xi,111}^0&0&0&1&1\\
    M_{\xi,110}^1&0&0&1&-1\\
    \hline
    N_{\xi,0}&1&1&0&0\\
    N_{\xi,1}&0&0&1&1\\
    \hline
    d_\xi&1&1&1&1\\
    \hline
    T_\xi&1&1&1&-1
\end{array}\]
and
\[S=\frac{1}{2}\left(
\begin{array}{rrrr}1&1&1&1\\1&1&-1&-1\\1&-1&1&-1\\1&-1&-1&1\end{array}
\right).\]

\subsubsection{Double-semion model}
\begin{itemize}
  \item Two types of strings, labeled by 0,1 and $1^*=1$.
  \item $N_{011}=1$, $F_{110}^{110}=-1$, $O_1=-1$.
\end{itemize}

The Q-algebra is 4-dimensional. The natural basis is
\begin{gather*}
  e_{00}=Q_{000}^0,e_{01}=Q_{001}^1,\\
  e_{10}=Q_{111}^0,e_{11}=Q_{110}^1.
\end{gather*}
The multiplication is 
\[ \begin{array}{c|cccc}
  &e_{00}&e_{01}&e_{10}&e_{11}\\
  \hline
  e_{00}&e_{00}&e_{01}&0&0\\
  e_{01}&e_{01}&e_{00}&0&0\\
  e_{10}&0&0&e_{10}&e_{11}\\
  e_{11}&0&0&e_{11}&-e_{10}
\end{array}\]

If we change the basis $e_{11}\mapsto -\mathrm{i}e_{11}$,
this is still the direct sum of two group algebras of $\bbz_2$.
There are 4 1-dimensional simple modules.
\[\begin{array}{c||c|c|c|c}
    \xi&1&2&3&4\\
    \hline
\text{basis}&\dfrac{e_{00}+e_{01}}{2}&\dfrac{e_{00}-e_{01}}{2}
&\dfrac{e_{10}-\mathrm{i}e_{11}}{2}&\dfrac{e_{10}+\mathrm{i}e_{11}}{2}\\
    \hline
    M_{\xi,000}^0&1&1&0&0\\
    M_{\xi,001}^1&1&-1&0&0\\
    M_{\xi,111}^0&0&0&1&1\\
    M_{\xi,110}^1&0&0&\mathrm{i}&-\mathrm{i}\\
    \hline
    N_{\xi,0}&1&1&0&0\\
    N_{\xi,1}&0&0&1&1\\
    \hline
    d_\xi&1&1&1&1\\
    \hline
    T_\xi&1&1&\mathrm{i}&-\mathrm{i}
\end{array}\]
and
\[S=\frac{1}{2}\left(
\begin{array}{rrrr}1&1&1&1\\1&1&-1&-1\\1&-1&-1&1\\1&-1&1&-1\end{array}
\right).\]

\subsubsection{$\bbz_N$ model}\label{zn}
\begin{itemize}
  \item $N$ types of strings, labeled by $0,1,\dots,N-1$ and $i^*=N-i$.
  \item We use $\res{\cdots}_N$ to denote the residual modulo $N$.
  \item $N_{ijk}=1$ iff $\res{i+j+k}_N=0 $.
\item $F_{kln}^{ijm}=1$ iff $m=\res{k+l}_N $, $n=\res{j+k}_N $,
$\res{i+j+k+l}_N=0 $.
\end{itemize}

The Q-algebra is $N^2$-dimensional.
The natural basis is \[e_{ri}=Q_{rr\res{ r-i}_N}^{\res{ -i}_N}.\]
The multiplication is \[e_{ri}e_{sj}=\delta_{rs}e_{r\res{ i+j}_N}.\]

This is the direct sum of $N$ group algebras of $\bbz_N$. There are $N^2$
1-dimensional simple modules.
We use $\cl{\cdots}$ to denote a composite label.
The simple modules can be labeled with two numbers $\cl{ ri}$.
The basis is \[M_{\cl{ ri}}^r=\frac{1}{N}\sum_{k=0}^{N-1}
\mathrm{e}^{-\frac{2\pi \mathrm{i}}{N}i k}e_{rk}.\]
The matrix representations are \[M_{\cl{
ri},ss\res{s+j}_N}^j=\delta_{rs}\mathrm{e}^{-\frac{2\pi \mathrm{i}}{N}ij}.\]
Then, we get
\begin{align*}
  N_{\cl{ ri},s}&=\delta_{rs},\\
  d_{\cl{ ri}}&=1,\\
  T_{\cl{ ri}}&=\mathrm{e}^{-\frac{2\pi \mathrm{i}}{N}ri},\\
S_{\cl{ ri}\cl{ sj}}&=\frac{1}{N}\mathrm{e}^{\frac{2\pi \mathrm{i}}{N}\left(
rj+si \right)}.
\end{align*}
\subsubsection{Finite group \textit{G} model}\label{finiteG}

Similar to the $\bbz_N$ case we can define the rotation-invariant string-net
model for a finite group $G$:
\begin{itemize}
\item $|G|$ types of strings, labeled by the group elements $g\in G$ and
$g^*=g^{-1}$.
  \item The trivial string is now labeled by $\one$, the identity element of $G$.
  \item $N_{g_1 g_2 g_3}=1$ iff $g_1 g_2 g_3=\one$.
\item $F_{g_3 g_4 g_6}^{g_1 g_2 g_5}=1$ iff $g_5=g_3 g_4$, $g_6=g_2 g_3$, $g_1
g_2 g_3 g_4=\one$.
\end{itemize}

The Q-algebra is $|G|^2$-dimensional and the natural basis is 
\[e_{gh}=Q_{g\cl{ h^{-1}gh}\cl{ h^{-1}g}}^{\cl{ h^{-1}}}.\]
The multiplication is \[e_{gh}e_{g'h'}=\delta_{g\cl{ hg'h^{-1}}}e_{g\cl{
hh'}}.\]

It turns out that the Q-algebra is the Drinfeld double $D(G)$ of the finite
group $G$.
The modules over $D(G)$ have been well studied. Some examples of the $T,S$
matrices of $D(G)$ can be found in Refs. \onlinecite{CGR00,Coq12}.
In particular if $G$ is Abelian, $D(G)$ is the direct sum of $|G|$ group
algebras of $G$, and there are $|G|^2$ 1-dimensional simple modules.

\subsubsection{Doubled Fibonacci phase}\label{eg}
\begin{itemize}
  \item Two types of strings, labeled by 0,1 and $1^*=1$.
  \item $N_{011}=N_{111}=1$, $O_1=\gamma=\dfrac{1+\sqrt{5}}{2}$.
\item
$F_{110}^{110}=\gamma^{-1},F_{110}^{111}=F_{111}^{110}=\gamma^{-1/2},
F_{111}^{111}=-\gamma^{-1}$.
\end{itemize}

The Q-algebra is 7-dimensional. The natural basis is
\begin{gather*}
  e_1=Q_{000}^0,e_2=Q_{001}^1,\\
  e_3=Q_{111}^0,e_4=Q_{110}^1,e_5=Q_{111}^1,\\
  e_6=Q_{011}^1,\\
  e_7=Q_{101}^1.
\end{gather*}
\begin{widetext}
  The multiplication is
  \[\begin{array}{c|ccccccc}
      &e_1&e_2&e_3&e_4&e_5&e_6&e_7\\
      \hline
      e_1&e_1&e_2&0&0&0&e_6&0\\
      e_2&e_2&e_1+e_2&0&0&0&-\frac{1}{\gamma}e_6&0\\
      e_3&0&0&e_3&e_4&e_5&0&e_7\\
e_4&0&0&e_4&\frac{1}{\gamma}e_3+\frac{1}{\sqrt{\gamma}}e_5&
\frac{1}{\sqrt{\gamma}}e_3-\frac{1}{\gamma}e_5&0&e_7\\
      e_5&0&0&e_5&\frac{1}{\sqrt{\gamma}}e_3-\frac{1}{\gamma}e_5
&-\frac{1}{\gamma}e_3+e_4-\frac{1}{\gamma^2\sqrt{\gamma}}e_5&0&\frac{1}{\gamma\sqrt{\gamma}}e_7\\
e_6&0&0&e_6&e_6&\frac{1}{\gamma\sqrt{\gamma}}e_6&0&\sqrt{\gamma}e_1-\frac{1}{\sqrt{\gamma}}e_2\\
e_7&e_7&-\frac{1}{\gamma}e_7&0&0&0&\frac{1}{\sqrt{\gamma}}e_3+\frac{1}{\sqrt{\gamma}}e_4+\frac{1}{\gamma^2}e_5&0
  \end{array}\]
\end{widetext}
To decompose this algebra we can do idempotent decomposition. Firstly
$\textbf{1}=e_1+e_3$.
Secondly as stated in Appendix \ref{appendix-qalg}, since $Q_{01}=\langle
e_6\rangle,Q_{10}=\langle e_7\rangle$, we immediately obtain two primitive
orthogonal idempotents
\begin{align*}
h_1&=\sqrt{\frac{\gamma}{5}}e_6e_7=\frac{1}{\sqrt{5}\gamma}(\gamma^2 e_1-\gamma
e_2),\\
h_2&=\sqrt{\frac{\gamma}{5}}e_7e_6=\frac{1}{\sqrt{5}\gamma}(\gamma e_3+\gamma
e_4+\frac{1}{\sqrt{\gamma}}e_5 ).
\end{align*}
Thirdly since $\dim \left[(e_1-h_1)Q(e_1-h_1)\right]=1$,
\[h_3=e_1-h_1=\frac{1}{\sqrt{5}\gamma}(e_1+\gamma e_2)\]
is another primitive orthogonal idempotent.
Fourthly since $\dim \left[(e_3-h_2)Q(e_3-h_2)\right]=2$ we can solve for the two primitive
orthogonal idempotents in ${(e_3-h_2)Q(e_3-h_2)}$
\begin{gather*}
  h_4+h_5=e_3-h_2,\\
h_4=\frac{1}{\sqrt{5}\gamma}\left(
e_3+\mathrm{e}^{-\frac{4\pi\mathrm{i}}{5}}e_4+
  {\sqrt{\gamma}}\mathrm{e}^{\frac{3\pi\mathrm{i}}{5}}e_5\right),\\
h_5=\frac{1}{\sqrt{5}\gamma}\left( e_3+\mathrm{e}^{\frac{4\pi\mathrm{i}}{5}}e_4
+
  {\sqrt{\gamma}}\mathrm{e}^{-\frac{3\pi\mathrm{i}}{5}}e_5\right).
\end{gather*}
The final primitive orthogonal idempotent decomposition is
\[\textbf{1}=h_1+h_2+h_3+h_4+h_5.\]

The Q-algebra can now be decomposed as it own module
\[Q=Q h_1\oplus Q h_2 \oplus Q h_3 \oplus Q h_4 \oplus Q h_5\]
and the two 2-dimensional modules are isomorphic
\[ Q h_1=\langle h_1,e_7\rangle \cong Q h_2 =\langle e_6, h_2\rangle.\]
We have made a special choice of the basis so that the representation matrices
look nice.
However this is not necessary. The statistics depends on only the traces.
\[\begin{array}{c||c|c|c|c}
    \xi&1&2&3&4\\
    \hline
    \text{basis}&h_3&h_4&h_5&
    \begin{array}{rr}
      &h_1,\,\sqrt[4]{\gamma/5}e_7\\
      \text{or}&\sqrt[4]{\gamma/5}e_6,\,h_2
    \end{array}\\
    \hline
    M_{\xi,000}^0&1&0&0&\mz{1}{0}{0}{0}\\
    M_{\xi,001}^1&\gamma&0&0&\mz{-\gamma^{-1}}{0}{0}{0}\\
    M_{\xi,111}^0&0&1&1&\mz{0}{0}{0}{1}\\
M_{\xi,110}^1&0&\mathrm{e}^{\frac{4\pi\mathrm{i}}{5}}&
\mathrm{e}^{-\frac{4\pi\mathrm{i}}{5}}&\mz{0}{0}{0}{1}\\
M_{\xi,111}^1&0&\sqrt{\gamma}\mathrm{e}^{-\frac{3\pi\mathrm{i}}{5}}&\sqrt{\gamma}\mathrm{e}^{\frac{3\pi\mathrm{i}}{5}}
    &\mz{0}{0}{0}{\gamma^{-3/2}}\\
    M_{\xi,011}^1&0&0&0 &\mz{0}{\sqrt[4]{5/\gamma}}{0}{0}\\
    M_{\xi,101}^1&0&0&0 &\mz{0}{0}{\sqrt[4]{5/\gamma}}{0}\\
    \hline
    N_{\xi,0}&1&0&0&1\\
    N_{\xi,1}&0&1&1&1\\
    \hline
    d_\xi&1&\gamma&\gamma&\gamma^2\\
    \hline
T_\xi&1&\mathrm{e}^{-\frac{4\pi\mathrm{i}}{5}}&\mathrm{e}
^{\frac{4\pi\mathrm{i}}{5}}&1
\end{array}\]
and
\[S=\frac{1}{\sqrt{5}\gamma}\left(
\begin{array}{cccc}1&\gamma&\gamma&\gamma^2\\\gamma &-1&
\gamma^2&-\gamma\\
\gamma&\gamma^2&-1&-\gamma\\
\gamma^2&-\gamma&-\gamma&1\end{array} \right).\]

\section{Generalized string-net models}
\label{qalgG}

From now on we will drop the assumption that evaluation of the string-nets is
rotation-invariant.
We are going to choose a preferred orientation of the string-nets, from bottom
to top,
and we can then safely drop the arrows in the graphs.
We will also change our notations of fusion rules and F-matrices to a less
symmetric version $N_{ij}^k,F^{ijk}_{l;nm}$.
The trivial string type is not assumed to be totally invisible.

The generalized string-net model in arbitrary gauge is defined as follows.

\subsection{String types and fusion rules}

The string types are given by a label set $L$.
Strings can fuse and split. For simplicity we consider multiplicity-free fusion
rules $N_{ij}^k=\delta_{ij}^k\in\{0,1\}$ in this section, so there are no
vertex labels.
But it is quite straightforward to generalize to fusion rules with
multiplicity, as in the previous section. The fusion rules satisfy
\begin{gather}
  \sum_m N_{ij}^m N_{mk}^l=\sum_n N_{in}^l N_{jk}^n.
  \label{fusion}
\end{gather}
For each splitting or fusion vertex, there is a nonzero number $Y^{ij}_k$
\begin{gather}
  \ev\ \f{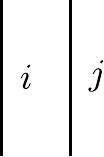}=\sum_k \frac{\delta_{ij}^k}{Y^{ij}_k}\f{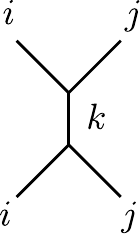},\\
  \ev\ \f{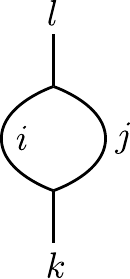}=\delta_{ij}^k Y^{ij}_k\ \f{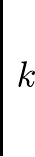}.
  \label{ag-sf}
\end{gather}

There is a trivial string type, labeled by 0, and
${N_{0i}^k=N_{i0}^k=\delta_{ik}}$.
There is an involution of the label set $L$, $i\mapsto i^*$.
$i^*$ is called the dual type of $i$, and
${N_{ij}^0=N_{ji}^0=\delta_{ij^*}}$.

\subsection{F-move and pentagon equations}
We only need to assume one kind of F-move
\begin{gather}
 \ev \f{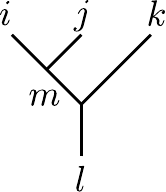}=\sum_n F^{ijk}_{l;nm}\f{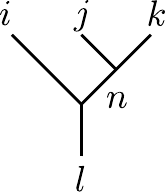},
\end{gather}
$F^{ijk}_{l;nm}=0$ if $N_{ij}^m N_{mk}^l N_{in}^l N_{jk}^n=0$.
$F^{ijk}_{l}$ are invertible matrices and satisfy the pentagon equations
\begin{gather}
\sum_n
F^{jkl}_{q;pn}F^{inl}_{s;qr}F^{ijk}_{r;nm}=F^{ijp}_{s;qm}F^{mkl}_{s;pr}.
\label{penta}
\end{gather}

We see that
$F^{ijk}_{0;nm}=\omega^{ijk}\delta_{in^*}\delta_{km^*}\delta_{ij}^{k^*}$ is
just a number.
We can express the invertible matrix of $F^{ijk}_l$ in terms of
$F^{ijk}_l,\omega^{ijk}$.
Consider the following pentagon equations
\begin{align}
\sum_n
F^{jkl^*}_{i^*;pn}F^{inl^*}_{0;i^*l}F^{ijk}_{l;nm}
&=F^{ijp}_{0;i^*m}F^{mkl^*}_{0;pl},\\
\sum_m
F^{ijk}_{l;nm}F^{l^*mk}_{0;lk^*}F^{l^*ij}_{k^*;mp}
&=F^{l^*in}_{0;lp}F^{pjk}_{0;nk^*},
\end{align}
we have
\begin{equation}
(F^{ijk}_l)^{-1}_{mn}=\frac{\omega^{inl^*}}{\omega^{ijm^*}
\omega^{mkl^*}}F^{jkl^*}_{i^*;m^*n}
  =\frac{\omega^{l^*mk}}{\omega^{l^*in}\omega^{n^*jk}}F^{l^*ij}_{k^*;mn^*}.
  \label{invFt}
\end{equation}
This is like ``rotating'' the F-matrix by $90^\circ$.
We see that evaluation is no longer rotation-invariant,
and the difference after rotations is controlled by $F_{0}^{ijk}$.
This explains why we have to assume that the trivial strings are not totally
invisible.
Trivial strings can still be added, removed, or deformed,
which will introduce isomorphisms between different ground state subspaces.
But unlike the rotation-invariant case, these isomorphisms can be highly
non-trivial.

If we ``rotate'' once more, we find
\begin{equation}
F^{ijk}_{l;nm}=\frac{\omega^{jm^*i}\omega^{ijm^*}\omega^{mkl^*}}{\omega^{jkn^*}\omega^{nl^*i}\omega^{inl^*}}F^{kl^*i}_{j^*;n^*m^*},
  \label{rotF}
\end{equation}
thus $360^\circ$ rotation implies
\begin{equation}
\frac{\omega^{jm^*i}\omega^{ijm^*}\omega^{mkl^*}\omega^{l^*mk}
\omega^{kl^*m}\omega^{m^*ij}}
{\omega^{jkn^*}\omega^{nl^*i}\omega^{inl^*}
\omega^{l^*in}\omega^{n^*jk}\omega^{kn^*j}}=1.
  \label{pivotal}
\end{equation}

Since we choose a special orientation for the generalized string-net model,
there are also other kinds of F-moves. If we stack \[\f{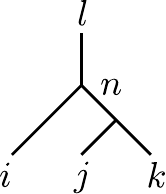},\quad\f{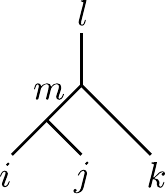}\] onto \[\f{fdl},\quad\f{fdr},\]
we can evaluate the amplitude using $F^{ijk}_l$ and $Y^{ij}_{k}$.
This way we can find out what should the F-moves between
\[\f{fur},\quad\f{ful}\]
look like. Now, we have 4 kinds of F-move:
\begin{gather}
  \ev\f{fdl}=\sum_n F^{ijk}_{l;nm}\f{fdr},\\
  \ev\f{fdr}=\sum_m (F^{ijk}_l)^{-1}_{mn} \f{fdl},\\
\ev\f{fur}=\sum_m \frac{Y^{jk}_nY^{in}_l}{Y^{ij}_mY^{mk}_l}F^{ijk}_{l;nm}
\f{ful},\\
\ev\f{ful}=\sum_n
\frac{Y^{ij}_mY^{mk}_l}{Y^{jk}_nY^{in}_l}(F^{ijk}_l)^{-1}_{mn}\f{fur}.
\end{gather}

\subsection{Gauge transformation and quantum dimension}

A gauge of the string-net model is a choice of fusion or splitting vertices.
Thus, a gauge transformation is nothing but a change of basis.
For the case of multiplicity-free fusion rules, it can be given by a set of
nonzero numbers $f^{ij}_k,f_{ij}^k$
\begin{gather}
  \f{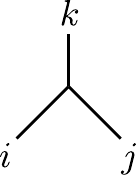}\mapsto f_{ij}^k \f{ijk},\quad \f{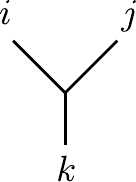}\mapsto f^{ij}_k\f{kij},\\
  Y^{ij}_k\mapsto \tilde{Y}^{ij}_k=f^{ij}_k f_{ij}^k Y^{ij}_k,\\
F^{ijk}_{l;nm}\mapsto
\tilde{F}^{ijk}_{l;nm}=\frac{f^{ij}_mf^{mk}_l}{f^{jk}_nf^{in}_l}F^{ijk}_{l;nm}.
\end{gather}
Gauge transformations should not affect the physics of the system.
Physical quantities, such as the $T,S$ matrices, should be gauge invariant.

In addition we assume that $F^{ijk}_l=\one$ if any of $i,j,k$ is the trivial
type 0 and $Y^{i0}_i=Y^{0i}_i=1$.
But essentially these correspond to a convenient gauge choice. [See Appendix
\ref{normgauge}.]
With this assumption the gauge transformation is slightly restricted
\begin{equation}
f^{i0}_i=f^{0i}_i=f_0^{00},\
f_{i0}^i=f_{0i}^i=f_{00}^0=(f_0^{00})^{-1}.\label{ngauge}
\end{equation}

We want to point out that, by choosing a special direction,
the string-net model with tetrahedron-rotational symmetry can be mapped to the
generalized string-net model with a rotation-invariant gauge.
One can see that the resulting rotation-invariant gauge must satisfy
$N^{k}_{ij}=N_{ijk^*},Y^{i^*i}_0=O_i,\ Y^{ij}_k=\dfrac{\Theta_{ijk^*}}{O_k},\
F^{ijk}_{l;nm}=F^{j^*i^*m}_{lk^*n}$
and other conditions of the tetrahedron-rotational symmetry.
A generalized string-net model may not always allow a rotation-invariant gauge.

In the rotation-invariant case, we assumed that F-matrices are unitary, which is
a physical requirement.
But now we allow arbitrary gauge transformations, which may break the unitary
condition of F-matrices.
Thus, we slightly weaken the condition: There exists a \emph{unitary gauge} such
that $F^{ijk}_l$ are unitary matrices.
For generalized string-net models, we prefer to work in the unitary gauge
where all $Y^{ij}_k=1$.
Note that in a unitary gauge
${F^{i^*ii^*}_{i^*;00}=(F^{ii^*i}_{i})^{-1}_{00}=\overline{F^{ii^*i}_{i;00}}}$.
We can define a gauge invariant quantity
\begin{equation}
  {d_i=\dfrac{1}{\sqrt{F^{ii^*i}_{i;00}F^{i^*ii^*}_{i^*;00}}}},
\end{equation}
which is the quantum dimension of the type $i$ string.
Thus, it is also required that for any string type $i$,
${F^{ii^*i}_{i;00}\neq 0}$,
which is necessary for defining quantum dimensions.

\subsection{Q-algebra and quasiparticle statistics}
The Q-algebra in arbitrary gauge is
\begin{gather}
  Q_{rsj}^i=\f{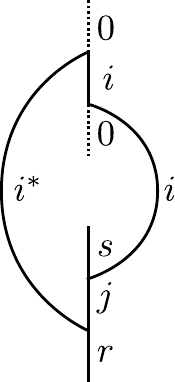},
\end{gather}
\begin{align}
  Q_{rsj}^i Q_{s'tl}^k&=\delta_{ss'}\sum_{mn}F^{k^*li}_{j;ns}F^{tki}_{n;ml}
  (F^{i^*k^*n}_r)^{-1}_{m^*j}
  \nonumber\\&\xt(F^{k^*ki}_i)^{-1}_{0m}F^{i^*k^*m}_{0;im^*}
  \frac{Y^{k^*k}_0Y^{i^*i}_0}{Y^{m^*m}_0}
  Q_{rtn}^m.
  \label{qalgt}
\end{align}

In the rotation-invariant gauge the string operators are well defined and
can be obtained from the matrix representations of the Q-algebra.
But in arbitrary gauge, since there is a preferred direction,
it is not quite obvious how to construct a closed string operator.
However, note that different gauges just mean that we choose different bases of
the Q-algebra,
we know that the difference between string operators and matrix representations
of the Q-algebra is
at most some factors depending on the choice of gauge.

Therefore, similarly to the rotation-invariant case, if we have found the
irreducible matrix representations of the Q-algebra,
we can calculate the quasiparticle statistics.
The number of quasiparticle types is just the number of different irreducible
representations up to similarity transformation.
We can also calculate the $T,S$ matrices.
Use $\xi$ to label irreducible representations,
assuming the representation matrix of $Q_{rsj}^i$ is $M_{\xi,rsj}^i$, and we
have
\begin{align}
\overline{T_\xi} &=\frac{1}{d_\xi}\sum_r d_r^2 C(T,r)\Tr(M_{\xi,rr0}^{r^*}),\\
S_{\xi\zeta}& =\frac{1}{D_{\cZ(\cC)}}\sum_{rst} d_rd_sC(S,r,s,t)
  \nonumber\\
  &\xt\Tr(M_{\xi,rrt^*}^{s})\Tr(M_{\zeta,s^*s^*t}^{r^*}),
\end{align}
where $d_\xi=\sum_r \Tr(M_{\xi,rrr}^0) d_r$, $D_{\cZ(\cC)}=\sqrt{\sum_\xi
d_\xi^2}$, and
$C(T,r),C(S,r,s,t)$ are undetermined factors that make the expressions gauge
invariant.
To determine $C(T,r),C(S,r,s,t)$, the basic idea is to use the vertices in
$Q_{rsj}^i$ to rebuild a ground state graph,
whose gauge transformation will cancel that of the trace term.
The result should agree with the special case \eqref{Txi}\eqref{Sxi} in the
rotation-invariant gauge.

The graph to cancel the gauge transformation of $\Tr(M_{\xi,rr0}^{r^*})$ is
easy to find,
simply a closed $r$ loop. Thus, we have $C(T,r)=1/Y^{rr^*}_{0}$.
However, there are two graphs for $C(S,r,s,t)$. One is
\begin{equation}
  \frac{\ev\left( \f{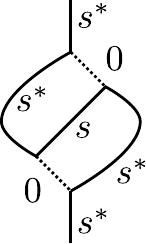} \right)}{\ev\left( \f{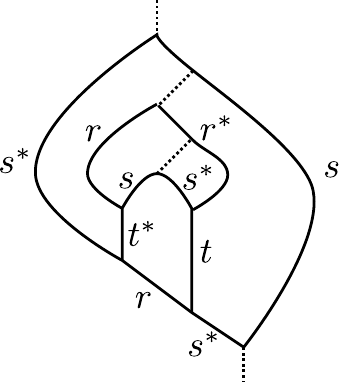} \right)}=
  \frac{F^{s^*ss^*}_{s^*;00}F^{t^*s^*r^*}_{0;tr}}
  {F^{s^*t^*t}_{s^*;0r}F^{rss^*}_{r;0t^*}Y^{rr^*}_0Y^{s^*s}_0}.
  \label{sgauge}
\end{equation}
The other can be obtained by permuting the labels ${r\to s^*,s\to r^*,t\to
t^*}$.
The two graphs should give the same amplitude, {i.e.,}\
\begin{equation}
  \frac{F^{s^*ss^*}_{s^*;00}F^{t^*s^*r^*}_{0;tr}}
  {F^{s^*t^*t}_{s^*;0r}F^{rss^*}_{r;0t^*}}
  =
  \frac{F^{rr^*r}_{r;00}F^{trs}_{0;t^*s^*}}
  {F^{rtt^*}_{r;0s^*}F^{s^*r^*r}_{s^*;0t}}.
  \label{ssame}
\end{equation}
Amazingly this is true due to the pentagon equations.
One can prove this using \eqref{rotF}\eqref{pivotal} and the following pentagon
equation
\begin{equation}
F^{srt}_{0;s^*t^*}F^{s^*t^*t}_{s^*;0r}F^{s^*sr}_{r;t^*0}=
F^{s^*ss^*}_{s^*;00}.\end{equation}

Finally, we obtain the gauge invariant formulas of $T,S$ matrices
\begin{align}
\overline{T_\xi} &=\frac{1}{d_\xi}\sum_r d_r^2\frac{1}{Y^{rr^*}_0}
\Tr(M_{\xi,rr0}^{r^*}),\label{aT}\\
  S_{\xi\zeta} &=\frac{1}{D_{\cZ(\cC)}}\sum_{rst} d_rd_s
  \frac{F^{s^*ss^*}_{s^*;00}F^{t^*s^*r^*}_{0;tr}}
  {F^{s^*t^*t}_{s^*;0r}F^{rss^*}_{r;0t^*}Y^{rr^*}_0Y^{s^*s}_0}
  \nonumber\\&
  \xt\Tr(M_{\xi,rrt^*}^{s})\Tr(M_{\zeta,s^*s^*t}^{r^*}).\label{aS}
\end{align}

We want to mention that the mathematical structure underlying generalized
string-net models is \emph{category theory}.
After generalizing to arbitrary gauge, the data $(N_{ij}^k,F^{ijk}_{l;nm})$ of a generalized string-net model
correspond to a fusion category $\cC$. Moreover with the unitary assumption
$\cC$ is UFC.
The Q-algebra modules correspond to the Drinfeld center $\cZ(\cC)$,
which is the unitary modular tensor category that describes the fusion and
braiding of the quasiparticles.
\label{gsmQ}

\subsection{Example: twisted quantum double}\label{tqd}
Now we give a simple example built on a finite group $G$ and its 3-cocycles
$H^3(G,U(1))$
\begin{itemize}
  \item Label set $L=G$, $N_{ab}^{c}=\delta_{\cl{ab}c}$, $Y^{ab}_c=1$.
  \item $F^{abc}_{\cl{abc};\cl{bc}\cl{ab}}=
    \alpha_{abc}$. $\alpha_{abc}\in H^3(G,U(1))$ is the 3-cocycle.
    $\alpha_{abc}=1$ if any of $a,b,c$ is identity.
    $\alpha_{abc}$ satisfies the cocycle condition
    \begin{equation}
      \alpha_{abc}\alpha_{a\cl{bc}d}\alpha_{bcd}
      =\alpha_{ab\cl{cd}}\alpha_{\cl{ab}cd}.
      \label{3-cocycle}
    \end{equation}
\end{itemize}

A basis of the Q-algebra $Q$ is
\begin{gather}
  Q_g^x=\f{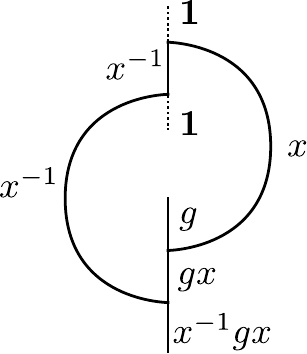}
\end{gather}
and
\begin{align}
  Q_h^yQ_g^x
  &=
  \frac{\alpha_{\cl{y^{-1}}\cl{x^{-1}}x}\alpha_{\cl{x^{-1}}\cl{gx}y} 
  \alpha_{gxy}} {\alpha_{\cl{y^{-1}}\cl{x^{-1}}\cl{gxy}}
  \alpha_{\cl{y^{-1}x^{-1}}xy}}\nonumber\\
  &\xt
  \delta_{\cl{x^{-1}gx}h}Q_{g}^{\cl{xy}}.
\end{align}
It turns out that $Q^\mathrm{op}$ (the same algebra $Q$ with the multiplication
performed in the reverse order)
is isomorphic to the twisted quantum double $D^\alpha(G)$. See Appendix
\ref{prooftqd} for the proof.

It is well known that 2D symmetry protected topological (SPT) phases are classified by the 3-cocycles
$H^3(G,U(1))$.\cite{CGLW13}
While in this example, when the fusion rules are give by the group $G$, the
generalized string-net models,
up to gauge transformations, are also in one-to-one correspondence with
3-cocycles in $H^3(G,U(1))$.
This example indicates that there may be deeper relations between generalized
string-net models and SPT phases.\cite{LG12,HW12,HWW13}

\section{Boundary theory of string-net models}
\label{boundary}

We have used tensors $(N_{ij}^k, F^{ijk}_{l;mn})$ to label different
string-net models processing different topological orders.  Here we like to
follow a similar scheme as in the bulk to construct the (gapped) boundary
theory of string-net models.\cite{KK12} In particular, we want to find the
tensors that
label different types of boundaries for a given bulk string-net model
labeled by the UFC $\cC$, or $(N_{ij}^k, F^{ijk}_{l;mn})_\cC$.

Firstly we still assume the degrees of freedom at the boundary have the form of
string-nets.
We need a label set $B$ to label the boundary string types.
To distinguish from the bulk string types, we add a underline to the boundary
string type labels:
$\underline x,\underline y,\dots\in B$.
Again the bulk strings can fuse with the boundary strings. There are fusion
rules
\begin{equation}
  N_{i\underline x}^{\underline y}=\dim\left( \f{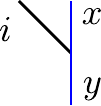} \right),
\end{equation} satisfying
\begin{equation}
\sum_y N_{i\underline x}^{\underline y}N_{j\underline y}^{\underline z}=\sum_k
N_{ij}^k N_{k\underline x}^{\underline z},\
  N_{0\underline x}^{\underline y}=\delta_{xy},
  \label{b-fuse}
\end{equation}
or in matrix form
\begin{equation}
  N_i N_j =\sum_k N_{ij}^k N_k,\ N_0=\one,
  \label{b-fuse-matrix}
\end{equation}
where the entries of matrix $N_i$ are $N_{i,xy}=N_{i\underline x}^{\underline
y}$.
There are similar F-moves on the boundary
\begin{equation}
\f{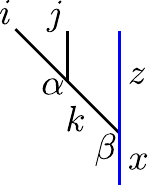}=\sum_{y\lambda\rho} F^{ij\underline z}_{\underline x;\underline
y\lambda\rho, k\alpha\beta}\f{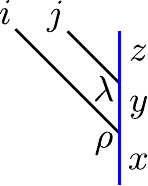}
  \label{b-F-move}
\end{equation}
which also satisfy the pentagon equations
\begin{gather}
  \sum_{n\tau\lambda\eta}F^{ijk}_{l;n\eta\lambda,m\alpha\beta}
F^{in\underline z}_{\underline w;\underline
x\tau\mu,l\lambda\gamma}F^{jk\underline z}_{\underline x;\underline
y\rho\nu,n\eta\tau}
  \nonumber\\
=\sum_{\sigma}F^{mk\underline z}_{\underline w;\underline
y\rho\sigma,l\beta\gamma}F^{ij\underline y}_{\underline w;\underline
x\nu\mu,m\alpha\sigma}.
  \label{b-pent}
\end{gather}

\add{
With the boundary fusion rules $ N_{i\underline x}^{\underline y}$ and
the boundary F-matrices $F^{ij\underline z}_{\underline x;\underline
y\lambda\rho, k\alpha\beta}$ we can similarly define evaluation maps and then
the Hamiltonians on the boundary as what we did in the bulk. This way we have a
gapped boundary theory of the string-net model, labeled by $(N_{i\underline
x}^{\underline y},F^{ij\underline z}_{\underline x;\underline
y\lambda\rho, k\alpha\beta})$.
}

The boundary quasiparticles can also be classified by modules over the boundary
Q-algebra\cite{KK12,Kon12} shown in the following sketch graph
\begin{equation}
  \f{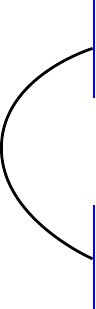}\ \bullet\ \f{b-algl}=\ev\  \f{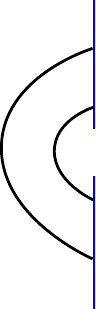}\quad .
  \label{b-alg}
\end{equation}

The modules over the boundary Q-algebra form a fusion category $\cB$, with
another set of data $(N_{ij}^k, F^{ijk}_{l;mn})_\cB$.
$\cB$ describes the fusion of the boundary quasiparticles, and
$\cB$ can also be used to construct a string-net model.
It is interesting that no matter which boundary we choose, such string-net
model constructed from $\cB$ always
describes the same bulk phase constructed from $\cC$,
or $\cZ(\cB)\cong \cZ(\cC)$.\cite{KK12}

We provide an example of this.
Consider the bulk phase described by $\bbz_N$ string-net model as in Section
\ref{zn}.
The gapped boundaries and boundary quasiparticles of $\bbz_N$ model are easy to
find.
[In Ref. \onlinecite{Lan12} this has been done using the language of module
category theory.]
The boundaries are classified by the integer factors of $N$. For each
integer factor $M$ of
$N$, there is a gapped boundary
\begin{itemize}
  \item The boundary string type label set is ${B=\{0,1,\cdots,M-1\}}$.
\item The boundary fusion rules are $N_{i\underline x}^{\underline y}=1$ iff $y=i+x \mod
M$, otherwise $N_{i\underline x}^{\underline y}=0$.
\item The boundary F-matrices are $F^{ij\underline z}_{\underline
x;\underline yk}=1$ for all stable vertices.
\end{itemize}

There are $N$ types of boundary quasiparticles on this $M$-boundary.
The string-net model given by the fusion category of these boundary quasiparticles
is
\begin{itemize}
  \item The string type label set is ${L=\bbz_M\xt \bbz_{{\frac{N}{M}}}}$.
More precisely the labels are $(x,y),$ ${x=0,1,\cdots,M-1,}$
${y=0,1,\cdots,{{\frac{N}{M}}}-1}$.
\item The fusion rules are given by the group
$\bbz_M\xt\bbz_{{\frac{N}{M}}}$, or $N_{(a_1,b_1)(a_2,b_2)}^{(a_3,b_3)}=
\delta_{a_3\res{a_1+a_2}_M}\delta_{b_3\res{b_1+b_2}_{{\frac{N}{M}}}}$.
\item The F-matrices, as in Section \ref{tqd}, are given by the nontrivial
3-cocycle
$$\alpha_{(a_1,b_1)(a_2,b_2)(a_3,b_3)}=\ee^{-2\pi
\ri\frac{a_1}{N}(b_2+b_3-\res{b_2+b_3}_{{\frac{N}{M}}})}.$$
\end{itemize}
By straightforward calculation, one can show that the modular data $T,S$ of the
above string-net model is the same as that of $\bbz_N$ model.
This relation is independent of the choice of boundary type $M$.

We can even extend our method to study the boundary changing operators.
They should be classified by modules over the boundary changing Q-algebras at
the junction of two different boundaries, as the following sketch (color
online)\begin{equation}
  \f{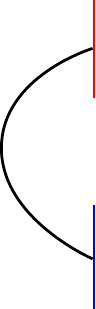}\ \bullet\ \f{bc-algl}=\ev\ \f{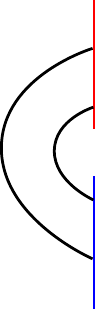}\quad ,
  \label{bc-alg}
\end{equation}
where the upper red lines and the lower blue lines represent different
boundaries.

The formulation of the Q-algebras at the boundaries is very much similar to
that in the bulk.  We will not elaborate on general formulas of the Q-algebras
in this section.  Instead, we will give a rather detailed discussion about the
twisted $(\bbz_N,p)$ string-net model and its boundary theory in Appendix
\ref{znp}, which we expect to be helpful for the readers to understand this
subject.

\section{Morita equivalence and fusion of excitations}
\label{Morita}

In Section \ref{gsmQ} we discussed the Q-algebra
\begin{equation}
  Q=\f{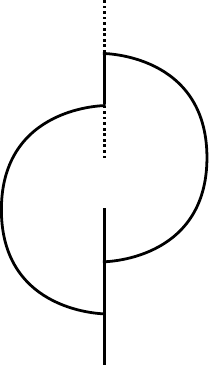}\ ,
\end{equation}
but as we have mentioned, the Q-algebra is not the only one that is related to
quasiparticle excitations.
We should also consider, for example, the $\nphia$-algebra
\begin{equation}
  \nphia=\f{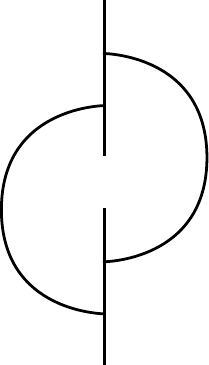}\ .
\end{equation}
$Q$-modules are in one-to-one correspondence with $\nphia$-modules.
To see this, consider the following subspaces of $\nphia$
\begin{gather}
  B_{Q\phia}=\f{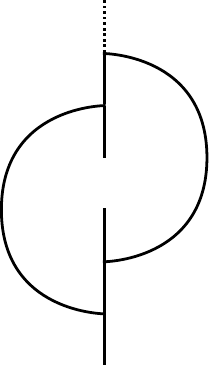}\ ,\ B_{\phia Q}=\f{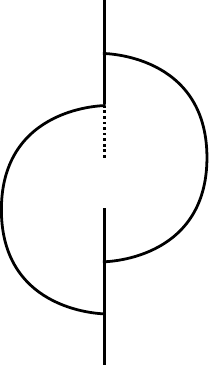}
\end{gather}
It not difficult to check that $B_{Q\phia}B_{\phia Q}=Q$ and $B_{\phia
Q}B_{Q\phia}=\nphia$.
Therefore, for a $\nphia$-module $M_\phia$, $B_{Q\phia}M_\phia$ is a
$Q$-module, and for a $Q$-module $M_Q$, $B_{\phia Q}M_Q$ is a $\nphia$-module.
Such maps of modules are invertible, $B_{Q\phia}B_{\phia Q}M_Q=M_Q$ and
$B_{\phia Q}B_{Q\phia}M_\phia=M_\phia$.

Moreover, there are more complicated algebras, such as
\begin{equation*}
  \f{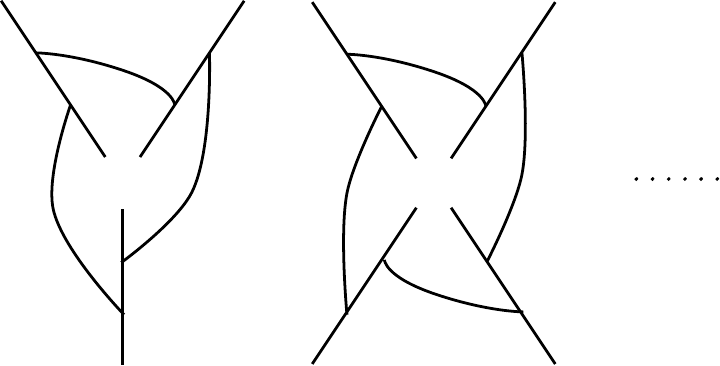}
\end{equation*}
One can similarly show that the modules over these algebras are in one-to-one
correspondence; these algebras are Morita equivalent.
Therefore, one can take any of these algebras to study the quasiparticles.
The physical properties of the quasiparticles do not depend on the choice of
algebras.

There are similar Morita equivalent relations for the local operator algebras
on boundaries.
The most general case is the following graph
\begin{equation}
  A_{\cM\cN}^{(m,n)}=\ \f{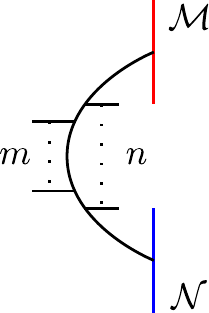}\ ,
\end{equation}
where $m,n$ are the number of legs (not string labels).
$A_{\cM\cM}^{(0,0)}$ and $A_{\cM\cN}^{(0,0)}$ are the boundary Q-algebra and
boundary changing Q-algebra discussed before.
According to Ref. \onlinecite{Kon12} Lemma 2,
$A_{\cM\cN}^{(m,m)}$ and $A_{\cM\cN}^{(n,n)}$ are Morita equivalent algebras;
the $A_{\cM\cN}^{(m,m)}$-$A_{\cM\cN}^{(n,n)}$-bimodule $A_{\cM\cN}^{(m,n)}$ is
invertible and defines the Morita equivalence, {i.e.,}\
${A_{\cM\cN}^{(m,n)}\ot_{A_{\cM\cN}^{(n,n)}}A_{\cM\cN}^{(n,m)}\cong
A_{\cM\cN}^{(m,m)}}$ as $A_{\cM\cN}^{(m,m)}$-$A_{\cM\cN}^{(m,m)}$-bimodules.

Moreover, it was pointed out by Kong that there are co-multiplication-like
maps\begin{equation}
  \Delta_{m,n,\cM,\cN,\cR}: A_{\cM\cR}^{(m+n,m+n)}\to A_{\cM\cN}^{(m,m)}\ot
  A_{\cN\cR}^{(n,n)},
\end{equation}
which control the fusion of boundary quasiparticles or boundary changing operators
\begin{gather}
  \ot_{m,n,\cM,\cN,\cR}: \cC_{\cM\cN}^{(m)}\xt 
  \cC_{\cN\cR}^{(n)}\to\cC_{\cM\cR}^{(m+n)},
\end{gather}
where $\cC_{\cM\cN}^{(m)}$ is the category of modules over
$A_{\cM\cN}^{(m,m)}$.

Graphically (color online)
\begin{equation}
  \f{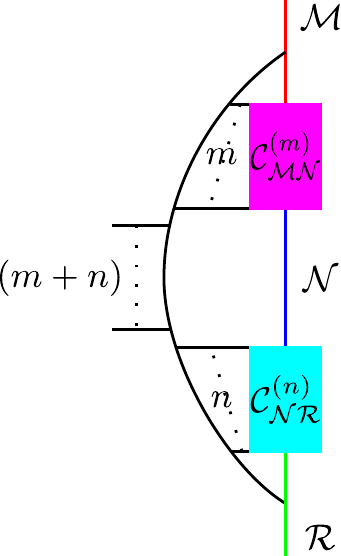}\ \longrightarrow\ \f{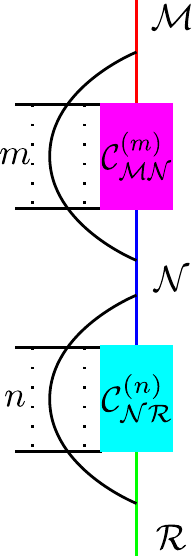}\ .
\end{equation}
This picture can be used to compute the F-matrices of the quasiparticles on the
boundary.

By the folding trick\cite{KK12}, a $\cC$-$\cD$-domain-wall $\cM$ can be viewed
as a $\cC\boxtimes\cD^\mathrm{op}$-boundary $\cM$,
\begin{equation}
  \f{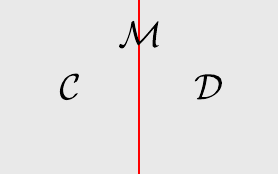}\xrightarrow{\text{folding}}\f{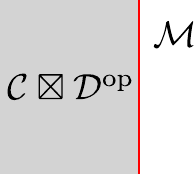}\ .
\end{equation}
As a special case the \nphia-algebra in the $\cC$-bulk can be viewed as the
boundary Q-algebra on the $\cC\boxtimes\cC^\mathrm{op}$-boundary $\cC$.
Therefore, the bulk quasiparticle excitations can also be studied via boundary
quasiparticles, as in Ref. \onlinecite{KK12}.
But, for bulk quasiparticles we already know how to compute the $T,S$ matrices,
using the simpler Q-algebra,
which fully determines the quasiparticle statistics.
This approach is only useful if we also want to compute, {e.g.,} the
F-matrices and braiding R-matrices of the UMTC $\cZ(\cC)$ that describes the
bulk quasiparticles.
\section{The mathematical structure of our construction}

We start with a \emph{unitary fusion category} (UFC).
In this paper a UFC $\cC$ is given by the \emph{fusion rules} and
\emph{F-matrices},
which satisfy a series of self-consistent conditions.
We then use the UFC $\cC$ to construct the fixed-point ground state
wavefunction, and the corresponding Levin-Wen Hamiltonian, {i.e.,}\ a
string-net model.

The quasiparticle excitations of such a model are given by the Drinfeld center
$\cZ(\cC)$ of the UFC $\cC$, which is a UMTC.
One can take the definition of $\cZ(\cC)$ and solve the corresponding
conditions to search for the quasiparticles. However, this is not a finite
algorithm.
Instead we introduce a finite algorithm, the Q-algebra approach, to calculate
$\cZ(\cC)$.
We use the data of $\cC$ to construct the Q-algebra, and the quasiparticles
correspond to the modules over the Q-algebra.
In other words, the UMTC $\cZ(\cC)$ is equivalent to the category of modules
over Q-algebra.
\add{Our Q-algebra approach to compute the Drinfeld center functor may be a special
  case of \emph{annularization}\cite{Wal06}.}

Then we consider the ``natural'' boundary of a string-net model given by UFC
$\cC$.
The ground state wavefunction of the ``natural'' boundary, similarly,
is given by \emph{boundary fusion rules} and \emph{boundary F-matrices},
which are compatible with those in the bulk.
Mathematically, such a boundary corresponds to a \emph{module category} $\cM$
over $\cC$. (Note that a module category over a tensor category is a different
notion from a category of modules.)
One can use a similar Q-algebra approach to study the quasiparticle
excitations on the boundary,
{i.e.,}\ the boundary quasiparticles are modules over the boundary Q-algebra.
It turns out that, the category of excitations on the $\cM$-boundary is again a
UFC $\cC_\cM$,
and string-net models given by $\cC$ and $\cC_\cM$ describes the same phase. 
In other words $\cZ(\cC)\cong \cZ(\cC_\cM)$, and $\cM$ is an invertible
$\cC$-$\cC_\cM$-bimodule (or a transparent domain wall between $\cC$ and
$\cC_\cM$).
Moreover, $\cC$ is naturally an $\cC$-module, and we know that $\cC_\cC\cong
\cC$.
That is to say, the UFC $\cC$ which we start with, can be viewed as a boundary
theory of the $\cZ(\cC)$ bulk.
The data of excitations on 1D boundaries can be used to construct the 2D bulk
string-net ground states.
This is the boundary-bulk duality of string-net models.
We conclude the discussion above with Table \ref{tab:mathstruc}.
\begin{table}[tb]
  \centering
  \begin{tabular}{|c|c|c|}
    \hline
    &2+1D bulk& 1+1D boundary\\
    \hline
    Ground states&UFC $\cC$& $\cC$-module $\cM$\\
    \hline
    Excitations
    &UMTC $\cZ(\cC)\cong \cZ(\cC_\cM)$& UFC $\cC_\cM$\\
    \hline
  \end{tabular}
  \caption{Mathematical structure of string-net models:\\
  The excitations are obtained by taking modules over Q-algebras.}
  \label{tab:mathstruc}
\end{table}

\begin{figure}[tb]
  \centering
  \includegraphics{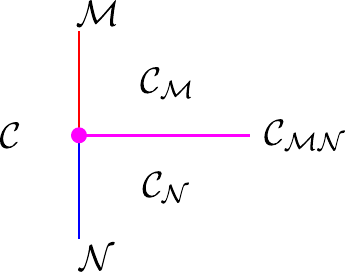}
  \caption{(Color online) Holographic Relation}
  \label{fig:holo}
\end{figure}

We also want to point out that the boundary changing operators can also be
calculated using the Q-algebra approach.
The boundary changing operators between boundary $\cM$ and boundary $\cN$ are
the modules over the Q-algebra at the junction of $\cM,\cN$,
and form a category which is the invertible $\cC_\cM$-$\cC_\cN$-bimodule
$\cC_{\cM\cN}$.
This provide us another holographic picture:
The 0D boundary changing operators can be used to construct the 1D transparent
domain walls.
We conclude the holographic relation in the Figure \ref{fig:holo}.
The $\cC_\cM$, $\cC_\cN$ and $\cC_{\cM\cN}$ on the right side can be viewed
either as boundary quasiparticles on $\cM$, $\cN$ and boundary changing operators
between them, or as bulk string-net models and the transparent domain wall
between them.
In particular, if we take $\cM=\cN$, the boundary changing operators reduce to
the boundary quasiparticles on $\cM$, {i.e.,}\ $\cC_{\cM\cM}\cong\cC_\cM$ is a
UFC.
Also recall that $\cC_\cC\cong \cC$, we have $\cM\cong \cC_{\cC\cM}$.
Since string-net models given by $\cC$ and $\cC_\cM$ are equivalent,
we may start with $\cC'=\cC_\cM$ instead of $\cC$, but in the end we should
arrive at exactly the same structure, in particular,
$\cC_{\cM\cN}=\cC'_{\cM\cN}$.
All in all, we conclude all the information of string-net models
are included in the point-like objects, either boundary changing operators or
excitations, in the categories $\cC_{\cM\cN}$.

\section{Conclusion}

Given a many-body ground state wave function and its Hamiltonian, how to
compute the topological excitations and their properties?  This is one of the
fundamental problems in the theory of topologically ordered states.  In this
paper, we address this issue in a simple situation: We compute the topological
excitations and their properties from an ideal  many-body ground state wave
function (and its ideal Hamiltonian).

The ideal ground state wave function and its ideal Hamiltonian ({i.e.,}\ the
string-net model) is constructed on the data of a UFC, {i.e.,} fusion rules and F-matrices.  They satisfy a series of
consistent conditions.  Using the data of the UFC, we can construct the
Q-algebra.  We showed that the topological
excitations in a string-net model can be classified by the modules over the
corresponding Q-algebra. The dimensions of Q-algebras are finite. Like the
groups, the canonical representation of the Q-algebra contains all types of
irreducible representations. In other words the Q-algebra contains all types of
simple modules as its subspaces.
So, it is an efficient approach to study the properties of the
quasiparticles by studying the Q-algebra and its modules.  Using this approach
we calculated the modular data $T,S$ of the quasiparticles.  Since the
topological excitations are described by a UMTC which is the Drinfeld center of
the UFC describing the ground state, our Q-algebra approach can also be viewed
as an efficient method to compute the Drinfeld center of a UFC.

The whole scheme to construct string-net models is very general, systematic and
can be naturally generalized to construct the boundary theory. The
boundary quasiparticles and boundary changing operators can also be studied via
Q-algebras at the boundaries.

It is interesting to note that the particle-like excitations at the boundary of
a string-net model are also described by a UFC.  The boundary UFC fully determines
the bulk, including the UMTC that describe the bulk topological
quasiparticles.\cite{KW14} The bulk UMTC is again given by the Drinfeld center of
the boundary UFC.  Thus, our Q-algebra approach is an efficient method to
compute the bulk properties from the edge properties.  It is also a concrete
example of the holographic relation between topological orders in different
dimensions.\cite{KW14}

We thank Liang Kong for many very helpful discussions.  This research is
supported by NSF Grant No. DMR-1005541, NSFC 11074140, and NSFC 11274192. It
is also supported by the John Templeton Foundation. Research at Perimeter
Institute is supported by the Government of Canada through Industry Canada and
by the Province of Ontario through the Ministry of Research.

\appendix

\section{A brief introduction to algebras and modules}
\label{secalgmod}
An \emph{algebra} $A$ is a vector space equipped with a multiplication.
\begin{alignat}{2}
  A&\ot A &&\to A\nonumber\\
  a&\ot b &&\mapsto ab
\end{alignat}
The multiplication must be bilinear and associative.
The identity of the multiplication must exist,
{i.e.,}\ there exists $\one \in A$ such that
$\forall a\in A, \one a=a \one=a$.

Given an algebra $A$,
we can define the multiplication of the subspaces of $A$.
Let $A_1$ and $A_2$ be subspaces of $A$.
\begin{align}
  A_1A_2& := \bigg\{\sum_k c^{(k)}a_1^{(k)}a_2^{(k)}\bigg|
  \begin{array}{rl}
    k\in \bbn,&a_1^{(k)}\in A_1,\\
    c^{(k)}\in \bbc,& a_2^{(k)}\in A_2
  \end{array}
  \bigg\}
\end{align}
is still a subspace of $A$.
This is analog to the multiplication of subgroups,
but note that here we need to take linear combinations.

Another important notion is the idempotent,
which is analog to projection operators.
An \emph{idempotent} $h$ in an algebra $A$ is a vector such that $hh=h$.
Two idempotents $h_1,h_2$ are \emph{orthogonal} iff $h_1h_2=h_2h_1=0$.
Note that the sum of orthogonal idempotents $h=h_1+h_2$ is still an idempotent.
An idempotent $h$ is \emph{primitive} iff it can not be written as sum of
nontrivial ({i.e.,}\ not 0 or $h$ itself) orthogonal idempotents.

We also like to consider \emph{central} elements.
A vector $a$ in $A$ is central if it commutes with all other vectors,
$ab=ba,\forall b\in A$.
The \emph{center} of $A$ is the subspace formed by all central elements,
denoted by $Z(A)$.

The most simple example is the matrix algebra.
Consider the $n\times n$ square matrices $\bbm_n$.
Under usual matrix multiplication $\bbm_n$ forms an algebra. The identity
matrix $I_n$ is the identity of the algebra.
A canonical basis of $\bbm_n$ is $E_{ab}$. $E_{ab}$ is the matrix with only the
$(a,b)$ entry $1$ and other entries $0$.
Then the matrix multiplication can be written as
\begin{equation}
  E_{ab}E_{b'c}=\delta_{bb'} E_{ac}
\end{equation}
$\{E_{aa}\}$ is a set of primitive orthogonal idempotents
\begin{equation}
  I_n=\sum_{a=1}^n E_{aa}
\end{equation}

A slightly more complicated case is the direct sum of matrix algebras.
Assume that
$A=\bbm_{n_1}\oplus\bbm_{n_2}\oplus\cdots\oplus\bbm_{n_\xi}
\oplus\cdots\oplus\bbm_{n_K}$.
We know the dimension of $A$ satisfies 
\begin{equation}
  \dim A=\sum_{\xi=1}^K n_\xi^2
\end{equation}
The elements of $A$ can be written as $(A_1,A_2,\dots,A_\xi,\dots,A_K)$,
$A_\xi\in\bbm_{n_\xi}$.
The multiplication is component-wise,
$(\dots,A_\xi,\dots)(\dots,B_\xi,\dots)=(\dots,A_\xi B_\xi,\dots)$.
Equivalently one may think the elements of $A$ as block-diagonal matrices,
with $K$ blocks and the $\xi$th block is $n_\xi\times n_\xi$.
Similarly we have a canonical basis
$E^{\xi}_{ab}=(0,\dots,0,E_{ab},0,\dots,0)$, where $E_{ab}$ is the $\xi$th
component
\begin{equation}
  E^{\xi}_{ab}E^{\zeta}_{b'c}=\delta_{\xi\zeta}\delta_{bb'}E^{\xi}_{ac}
\end{equation}
$\{E^\xi_{aa}\}$ is a set of primitive orthogonal idempotents
\begin{equation}
\textbf{1}=(I_{n_1},\dots,I_{n_K})=\sum_{\xi=1}^K \sum_{a=1}^{n_\xi}
E^{\xi}_{aa}
\end{equation}
Note that $(0,\dots,0,I_{n_\xi},0,\dots,0)$ are \emph{central} primitive
orthogonal idempotents,
and $Z(A)=\bbc(I_{n_1})\oplus\cdots\oplus\bbc(I_{n_K})$.

If an algebra $A$ is isomorphic to the direct sum of matrix algebras, we say
$A$ is \emph{semisimple}.
In other words if $A$ is semisimple, there exists a basis $e^{\xi}_{ab}$ of $A$,
satisfying
\begin{equation}
  e^{\xi}_{ab}e^{\zeta}_{b'c}=\delta_{\xi\zeta}\delta_{bb'}e^{\xi}_{ac}
\end{equation}
We call such basis $e^{\xi}_{ab}$ \emph{canonical}.
Finding a canonical basis means that we fully decomposed the algebra, which is
usually a nontrivial task.
But we can do idempotent decomposition,
{i.e.,}\ decomposing the identity as the sum of primitive orthogonal
idempotents.
Each set of primitive orthogonal idempotents correspond to the ``diagonal
elements'' of a canonical basis, $e^{\xi}_{aa}$.
The ``off-diagonal elements'', $e^{\xi}_{ab}$, can be picked out from
$e^{\xi}_{aa}A e^{\xi}_{bb}$.

A \emph{module} over an algebra $A$ is a vector space $M$ equipped with an
$A$-action.
$A$-action means that the elements of $A$ can act on $M$ as linear
transformations of $M$.
We also require that the $A$-action is linear and associative, and that the
identity of $A$ acts on $M$ as the identity transformation.
$M$ is invariant under the $A$-action.
It is obvious that $A$ can be considered as the module over itself.

Equivalently we can say there is an algebra homomorphism from $A$ to the linear
transformations of $M$.
After choosing a basis of $M$, one can represent the elements of $A$ by
matrices.
The matrix representations are equivalent up to basis changes of $M$, or up to
similarity transformations.
If we take $A$ as the module over itself, the
corresponding matrix representation is called the \emph{canonical
representation}.

It is possible that $M$ has some subspace $V$ that is invariant under the
$A$-action. Such $V$ is a submodule of $M$.
If $M$ has no submodules other than $0$ and itself, we say $M$ is a
\emph{simple} module over $A$.

It is easy to check that, up to isomorphism, the matrix algebra $\bbm_n$ has
only one simple module,
the $n$-dimensional vector space, or the column vector space $\bbm_{n\times
1}$. If we choose the canonical basis of $\bbm_{n\times 1}$, the matrix
representation is just $\bbm_n$ itself.
We can also think $\bbm_n$ as it own module. As $\bbm_n$-module, $\bbm_n$ is
the direct sum of $n$ column vector spaces $\bbm_{n\times 1}$.
The corresponding matrix representation has dimension $n^2 \times n^2$, and is
block-diagonal with $n$ blocks of dimension $n\times n$,
if we choose the canonical basis $E_{ab}$ of $\bbm_n$.

Now we can easily get the properties of modules over semisimple algebras.
Assuming that $A$ is an semisimple algebra,
$A\cong\bbm_{n_1}\oplus\cdots\oplus\bbm_{n_K}$.
We know that up to isomorphism, $A$ has $K$ different simple modules of
dimension $n_1,\dots,n_K$.
$A$ as its own module is the direct sum of these simple modules, in which
the simple module of dimension $n_\xi$ appears $n_\xi$ times.
Thus, we have the ``sum of squares'' law, $\dim A=\sum_\xi n_\xi^2$.
One can also easily check that $\dim(Z(A))$ = number of central primitive
orthogonal idempotents = number of different simple modules.

\section{Q-algebras in string-net models with tetrahedron-rotational
symmetry}\label{appendix-qalg}
We discuss the Q-algebra in a string-net model with the tetrahedron-rotational
symmetry in detail in this section.

We know that the Q-algebra is semisimple.\cite{M03a}
Immediately we get the powerful ``sum of squares'' law.
Let $\xi$ be the label of simple quasiparticles, and $M_\xi$ be the
corresponding simple module
\begin{equation}
  \dim Q=\sum_\xi \left( \dim M_\xi \right)^2.
  \label{sos}
\end{equation}
This puts a strict constraint on the number of simple quasiparticle types.
For example, in doubled Fibonacci phase there are two types of strings and
the fusion rules are $N_{000}=N_{011}=N_{111}=1$. The Q-algebra has dimension
7. Since $7=7\times 1=3\times 1+1\times 2^2$,
we know the number of simple quasiparticle types in doubled Fibonacci phase can
be only either 7 or 4.
Moreover, since the Q-algebra of doubled Fibonacci phase is not a commutative
algebra,
we must have $\dim(Z(Q))<7$, therefore double Fibonacci phase must have 4 types
of quasiparticles.

How do we decompose the Q-algebra?
A straightforward approach is trying to simultaneously block diagonalize the
representation matrices.
But this is tedious and impractical. A better way is to do idempotent
decomposition.
Decomposing the algebra is equivalent to decomposing its identity as the sum of
primitive orthogonal idempotents
\begin{equation}
  \textbf{1}=\sum_a h_a,\quad h_a h_b=\delta_{ab} h_a
  \label{idem}
\end{equation}
and $h_a$ cannot be further decomposed. With such idempotent decomposition, $Q
h_a$ are simple modules and ${Q=\oplus_a Q h_a}$.

Still it is not recommended to search for all the idempotents and then try to
decompose the identity.
As long as the algebra has simple modules of dimension 2 or more, there are
infinite many idempotents.
It is more practical to decompose the idempotents recursively.
Given an idempotent $h$, if by any means we find an idempotent $h'\in hQh,
h'\neq h$,
we can decompose $h$ as $h=h'+(h-h')$; otherwise if such $h'$ does not exist,
$h$ is primitive.
This way we only need to find one idempotent in $hQh$, so it is much more
efficient.
We can always do this recursive decomposition numerically.

Also note that the identity of subalgebras are essentially idempotents.
We can as well search for subalgebras of $hQh$. For the Q-algebra case this is
very useful.
To see this we first define the following subspace of $Q$
\begin{equation}
  Q_{rs}=\raise -45pt \hbox{\includegraphics{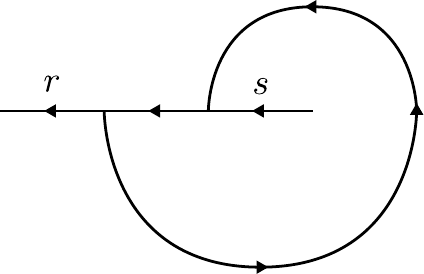}}
  \label{q-subspace}
\end{equation}
and
\begin{gather}
  Q_{rs}Q_{s't}\subseteq \delta_{ss'}Q_{rt},\\
  \dim Q_{rs}=\sum_{ij}N_{rji}N_{sij}=\Tr(N_r N_s).
  \label{dimQrs}
\end{gather}
$Q_{rr}$ are subalgebras of $Q$. The identity in $Q_{rr}$ is
\begin{equation}
  Q_{rrr}^{0,00}=\f{qr-id}.
  \label{qr-id}
\end{equation}
We know that the identity of $Q$ can be decomposed as
\begin{equation}
  \textbf{1}=\sum_r Q_{rrr}^{0,00}=\sum_r \f{qr-id}.
  \label{id-trd}
\end{equation}
Such decomposition is almost trivial. But imagine if we continue to decompose
$Q_{rrr}^{0,00}$,
eventually we will arrive at a ``canonical'' basis $e_{rs,ab}^{\xi}$ of $Q$
such that
\begin{gather}
  e_{rs,ab}^{\xi}\in Q_{rs},\label{qacb1}\\
e_{rs,ab}^{\xi}e_{s't,b'c}^\zeta=
\delta_{\xi\zeta}\delta_{ss'}\delta_{bb'}e_{rt,ac}^\xi.
\label{qacb2}
\end{gather}
$\{e_{rr,aa}^\xi\}$ is a set of primitive orthogonal idempotents
\begin{equation}
  \textbf{1}=\sum_{\xi ra}e_{rr,aa}^\xi.
\end{equation}
If we fix the labels $\xi,s,b$ in $e^\xi_{rs,ab}$ and let $r,a$ vary,
they span a subspace $Qe^\xi_{ss,bb}$ which is a simple module corresponding to
the simple quasiparticle type $\xi$.

Although for now we cannot explicitly calculate the canonical basis, we know
they exist.
The existence of such nice basis is significantly helpful for understanding the
structure of the Q-algebra.
For example, if for some string labels $r\neq s$, $\dim Q_{rs}=1$,
we know that $Q_{sr}Q_{rs}$ and $Q_{rs}Q_{sr}$ are subalgebras of dimension 1.
Thus, we obtain two primitive orthogonal idempotents ${h_1\in Q_{sr}Q_{rs}}$,
${h_2\in Q_{rs}Q_{sr}}$, which are identities of $Q_{sr}Q_{rs}$ and
$Q_{rs}Q_{sr}$.
We can immediately construct two isomorphic simple modules
$Qh_1\cong Q h_2$.
The dimension of $Qh_1$ or $Qh_2$ is at least 2.

For doubled Fibonacci phase it is exactly this case.
The dimensions of $Q_{rs}$ subspaces are $\dim Q_{00}=2$, $\dim Q_{11}=3$,
$\dim Q_{01}=\dim Q_{10}=1$.
Therefore, the Q-algebra of doubled Fibonacci phase has simple modules of
dimension at least 2,
and due to the ``sum of squares'' law \eqref{sos} the number of simple
quasiparticle types must be 4, as we claimed.
With the help of the two primitive orthogonal idempotents obtained from
$Q_{01}Q_{10}$ and $Q_{10}Q_{01}$,
it becomes very easy to do further idempotent decomposition and find out the
rest 3 simple modules of dimension 1.
More details about the doubled Fibonacci phase can be found in Section
\ref{eg}. We see the power of the Q-algebra \nocite{Kit03}approach. By simply
examining
the dimensions of the Q-algebra and its subspaces,
which depend on only the fusion rules $N_{ijk}$,
we obtain the number of simple quasiparticle types of doubled Fibonacci phase.
For complicated phases at least we can restrict the number of simple
quasiparticle types to several possible values.
To get full information of the quasiparticles, such as string operators and the
statistics,
we still need to fully decompose the algebra and explicitly calculate the
simple modules.

\section{The gauge transformation that fixes $F^{ijk}_l=\one$ for $i,j$ or $k$
trivial and $Y^{0i}_i=Y^{i0}_i=1$}
\label{normgauge}

Recall the pentagon equation \eqref{penta}. Set indices $j,k$ or $i,j$ or $k,l$
to 0, and we have
\begin{align}
  F^{00l}_{l;l0}F^{i0l}_{s;li}F^{i00}_{i;0i}&=F^{i0l}_{s;li}F^{i0l}_{s;li},\\
  F^{0kl}_{q;qk}F^{0kl}_{q;qk}F^{00k}_{k;k0}&=F^{00q}_{q;q0}F^{0kl}_{q;qk},\\
  F^{j00}_{j;0j}F^{ij0}_{r;jr}F^{ij0}_{r;jr}&=F^{ij0}_{r;jr}F^{r00}_{r;0r}.
\end{align}
Thus, we know
\begin{align}
  F^{i0j}_{k;ji}&=F^{00j}_{j;j0}F^{i00}_{i;0i},\\
  F^{0ij}_{k;kj}&=\frac{F^{00k}_{k;k0}}{F^{00i}_{i;i0}},\\
  F^{ij0}_{k;jk}&=\frac{F^{k00}_{k;0k}}{F^{j00}_{j;0j}}.
\end{align}
Therefore, we just to need transform $F^{00i}_{i;0i}$ and $F^{i00}_{i;i0}$ to 1
for all $i$, then all $F^{ijk}_l$ with $i,j$ or $k$ trivial will be transformed
to $\one$ automatically. Since
\begin{alignat}{2}
  Y^{i0}_i&\mapsto &\tilde{Y}^{i0}_i&=f^{i0}_i f_{i0}^i Y^{i0}_i,\\
  Y^{0i}_i&\mapsto &\tilde{Y}^{0i}_i&=f^{0i}_i f_{0i}^i Y^{0i}_i,\\
  F^{i00}_{i;0i}&\mapsto &\tilde{F}^{i00}_{i;0i}
  &=\frac{f^{i0}_i}{f^{00}_0}F^{i00}_{i;0i},\\
  F^{00i}_{i;i0}&\mapsto &\ \tilde{F}^{00i}_{i;i0}
  &=\frac{f^{00}_0}{f^{0i}_i}F^{00i}_{i;i0},
\end{alignat}
choosing
\begin{align}
  f^{i0}_i&=f^{00}_0 (F^{i00}_{i;0i})^{-1},\\
  f^{0i}_i&=f^{00}_0 F^{00i}_{i;i0},\\
  f^i_{i0}&=(f^{i0}_i Y^{i0}_i)^{-1},\\
  f^{i}_{0i}&=(f^{0i}_i Y^{0i}_i)^{-1},
\end{align}
we see that
${\tilde{Y}^{i0}_i=\tilde{Y}^{0i}_i=\tilde{F}^{i00}_{i;0i}=
\tilde{F}^{00i}_{i;i0}=1}$.
But this does not totally fix $f^{i0}_i,f^{0i}_i,f_{i0}^i,f^{0i}_i$; we can
still choose arbitrary $f^{00}_0$.
This degree of freedom can be covered by further gauge transformations
satisfying \eqref{ngauge}.

\begin{widetext}

\section{Isomorphism between Q-algebra in Section \ref{tqd} and twisted quantum
double $D^\alpha(G)$}\label{prooftqd}
  Recall the definition of $D^\alpha(G)$.
  The underlining vector space is $(\mathbb{C}G)^*\ot\mathbb{C}G$
  and the multiplication is given by
  \begin{equation}
    (g^*\ot x )(h^*\ot y)=\delta_{g\cl{xhx^{-1}}}
    \frac{\alpha_{gxy}\alpha_{xy\cl{(xy)^{-1}gxy}}}{\alpha_{x\cl{x^{-1}gx}y}}
    g^*\ot xy.
  \end{equation}
(Note that here the ${}^*$ symbol denotes dual vectors, but not dual string
types.)

  By multiplying the following cocycle conditions
  \begin{align}
    \alpha_{xy\cl{y^{-1}}}\alpha_{y\cl{y^{-1}}\cl{x^{-1}gxy}}
   & =\alpha_{\cl{xy}\cl{y^{-1}}\cl{x^{-1}gxy}}\alpha_{xy\cl{(xy)^{-1}gxy}},\\
    \alpha_{x\cl{x^{-1}}\cl{gx}}\alpha_{x\cl{x^{-1}gx}y}
    \alpha_{\cl{x^{-1}}\cl{gx}y}
    &=\alpha_{x\cl{x^{-1}}\cl{gxy}},\\
    \alpha_{x\cl{x^{-1}}\cl{gxy}}\alpha_{\cl{xy}\cl{y^{-1}}\cl{x^{-1}gxy}}
   & =\alpha_{\cl{xy}\cl{y^{-1}}\cl{x^{-1}}}
    \alpha_{\cl{xy}\cl{(xy)^{-1}}\cl{gxy}}
    \alpha_{\cl{y^{-1}}\cl{x^{-1}}\cl{gxy}},
  \end{align}
  one can get
  \begin{align}
    &\alpha_{xy\cl{y^{-1}}}\alpha_{x\cl{x^{-1}}\cl{gx}}
    \alpha_{\cl{x^{-1}}\cl{gx}y} 
    \alpha_{x\cl{x^{-1}gx}y}\alpha_{y\cl{y^{-1}}\cl{x^{-1}gxy}}
    \nonumber\\
    =\ &\alpha_{\cl{xy}\cl{y^{-1}}\cl{x^{-1}}}
    \alpha_{\cl{xy}\cl{(xy)^{-1}}\cl{gxy}}
    \alpha_{\cl{y^{-1}}\cl{x^{-1}}\cl{gxy}}\alpha_{xy\cl{(xy)^{-1}gxy}}
  \end{align}
  and thus
  \begin{align}
    \frac{\alpha_{\cl{x^{-1}}\cl{gx}y}}
    {\alpha_{\cl{y^{-1}}\cl{x^{-1}}\cl{gxy}}}&=
    \frac{\alpha_{xy\cl{(xy)^{-1}gxy}}}{\alpha_{x\cl{x^{-1}gx}y}}
    \frac{\alpha_{\cl{xy}\cl{(xy)^{-1}}\cl{gxy}}}
    {\alpha_{x\cl{x^{-1}}\cl{gx}}\alpha_{y\cl{y^{-1}}\cl{x^{-1}gxy}}}
    \frac{\alpha_{\cl{xy}\cl{y^{-1}}\cl{x^{-1}}}} {\alpha_{xy\cl{y^{-1}}}}.
  \end{align}
  Similarly, the following cocycle condition
  \begin{equation}
    \alpha_{\cl{y^{-1}}\cl{x^{-1}}x}\alpha_{\cl{x^{-1}}xy}
    =\alpha_{\cl{y^{-1}x^{-1}}xy} \alpha_{\cl{y^{-1}}\cl{x^{-1}}\cl{xy}},
  \end{equation}
  implies that
  \begin{align}
    \frac{\alpha_{\cl{y^{-1}}\cl{x^{-1}}x}} {\alpha_{\cl{y^{-1}x^{-1}}xy}}
    &=\left( \frac{\alpha_{\cl{x^{-1}}\cl{x}y}}
    {\alpha_{\cl{y^{-1}}\cl{x^{-1}}\cl{xy}}}\right)^{-1}
    =\left( \left.\frac{\alpha_{\cl{x^{-1}}\cl{gx}y}}
    {\alpha_{\cl{y^{-1}}\cl{x^{-1}}\cl{gxy}}}\right|_{g=\textbf{1}}\right)^{-1}
    \nonumber\\
    &=\left(\frac{\alpha_{\cl{xy}\cl{(xy)^{-1}}\cl{xy}}}
    {\alpha_{x\cl{x^{-1}}\cl{x}}\alpha_{y\cl{y^{-1}}\cl{x^{-1}xy}}}
\frac{\alpha_{\cl{xy}\cl{y^{-1}}\cl{x^{-1}}}}
{\alpha_{xy\cl{y^{-1}}}}\right)^{-1}.
  \end{align}
  Therefore, we know that
  \begin{equation}
    \left(
    \frac{\alpha_{y\cl{y^{-1}}\cl{hy}}}{\alpha_{y\cl{y^{-1}}y}}Q_h^y
    \right)
    \left(
    \frac{\alpha_{x\cl{x^{-1}}\cl{gx}}}{\alpha_{x\cl{x^{-1}}x}}Q_g^x
    \right)
    =\delta_{g\cl{xhx^{-1}}}
    \frac{\alpha_{gxy}\alpha_{xy\cl{(xy)^{-1}gxy}}}{\alpha_{x\cl{x^{-1}gx}y}}
    \left(
    \frac{\alpha_{\cl{xy}\cl{(xy)^{-1}}\cl{gxy}}}
    {\alpha_{\cl{xy}\cl{(xy)^{-1}}\cl{xy}}}Q_{g}^{\cl{xy}}
    \right)
  \end{equation}
which means that $Q^\mathrm{op}\cong D^\alpha(G)$ as algebras.
Actually both $Q$ and $D^\alpha(G)$ are quasi-Hopf algebras.
One may further check that they are isomorphic as quasi-Hopf algebras.
\end{widetext}

\section{Twisted ($\bbz_N,p$) string-net model}\label{znp}

In this section, we discuss the twisted $(\bbz_N,p)$ string-net model
and its boundary theory in detail.
We know that the generator in $H^3(\bbz_N,U(1))$ is 
\begin{equation}
  \alpha_{ijk}=\ee^{2\pi \ri \frac{1}{N^2}i(j+k-\res{j+k}_{N})}.
  \label{H3gen}
\end{equation}
This model is given by $\bbz_N$ fusion rule with the $p$-th 3-cocycle
$\alpha^p_{ijk}$, {i.e.,}\
\begin{itemize}
  \item String label set $L=\bbz_N$.
  \item Fusion rule $N_{ij}^k=\delta_{k\res{i+j}_{N}}$.
  \item F-matrices ${F^{ijk}_{\res{i+j+k}_N;\res{j+k}_{N}\res{i+j}_{N}}=
    \alpha^p_{ijk}}.$
\end{itemize}

\subsection{Q-algebra and bulk quasiparticle excitations}
As discussed in Section \ref{tqd}, the Q-algebra of twisted $(\bbz_N,p)$ model
is given by
\begin{align}
  Q_s^jQ_r^i
  &=\delta_{rs}
  \ee^{2 \pi \ri \frac{p}{N^2}i(2r+i-\res{r+i}_N)}
\ee^{2 \pi \ri \frac{p}{N^2}j(2r+j-\res{r+j}_N)}
    \nonumber\\
&\xt \ee^{ -2 \pi \ri \frac{p}{N^2}\res{i+j}_N(2r+\res{i+j}_N-\res{r+i+j}_N) }
  Q_{r}^{\res{i+j}_N}.
\end{align}
If we choose the basis
\begin{equation}
  \tilde{Q}_r^i=\ee^{ -2 \pi \ri \frac{p}{N^2}i(2r+i-\res{r+i}_N)} Q_r^i,
\end{equation}
we see that $\tilde{Q}_s^j\tilde{Q}_r^i=\delta_{rs}\tilde{Q}_r^{\res{i+j}_N}$.
Therefore, we find the irreducible representations (labeled by $\cl{ri}$)
\begin{equation}
  M_{\cl{ri},s}^j=\delta_{rs}\ee^{\left({-2\pi \ri\frac{ij}{N}}\right)}
\ee^{\left[ 2 \pi \ri \frac{p}{N^2}j(2r+j-\res{r+j}_N)\right]}.
\end{equation}
Applying \eqref{aT}\eqref{aS} we get
\begin{align}
  T_{\cl{ri}}&=\ee^{-2\pi\ri\left( \frac{ri}{N}-\frac{pr^2}{N^2} \right)},\\
S_{\cl{ri}\cl{sj}}&=\frac{1}{N}\ee^{ 2\pi\ri \left(
\frac{rj+si}{N}-\frac{2prs}{N^2} \right) }.
\end{align}
When $p\neq 0$, the fusion rule of the quasiparticles is not simply
$\bbz_N\xt\bbz_N$. Using the Verlinde formula,
\begin{align}
  N_{\cl{ri}\cl{sj}}^{\cl{tk}}&=
  \sum_{ql}\frac{S_{\cl{ri}\cl{ql}}S_{\cl{sj}\cl{ql}}
  \overline{S_{\cl{tk}\cl{ql}}}}{S_{\cl{00}\cl{ql}}}\nonumber\\
  &=\delta_{0\res{r+s-t}_N}\delta_{0\res{i+j-k-2p\frac{r+s-t}{N}}_N}.
\end{align}
We also see the equivalent relations of the quasiparticles are
\begin{gather}
  \cl{ri}\sim\cl{r'i'}\Longleftrightarrow r'=r+k_1 N,i'=i+2k_1 p+k_2 N
\end{gather}
where $k_1,k_2$ are integers.

\subsection{Boundary types}
Firstly we search for possible boundary fusion rules.
Note that \eqref{b-fuse-matrix} now becomes
\begin{equation}
  N_i N_j = N_{\res{i+j}_{N}},\ N_0=\one,
  \label{b-fuse-zn-m}
\end{equation}
thus it suffices to work out $N_1$, which is a matrix with non-negative integer
entries and $(N_1)^N=N_0=\one$.

We may write down the conditions explicitly
\begin{equation}
\sum_{x_1,\dots,x_{N-1}}N_{1\underline {x_0}}^{\underline {x_1}}
N_{1\underline{x_1}}^{\underline {x_2}}\cdots 
N_{1\underline {x_{N-1}}}^{\underline
{x_N}}=\delta_{x_0 x_N}.
  \label{b-fuse-zn}
\end{equation}
This is like a ``path integral''.
Since all the entries $N_{1\underline {x_i}}^{\underline {x_{i+1}}}$ are
non-negative integers,
we know that, starting from a fixed boundary string label $x_0=X_0$, there is
only one path $(X_0,X_1,\dots,X_{N-1},X_N=X_0)$ with
$N_{1\underline{X_i}}^{\underline{X_{i+1}}}=1$,
and for all other paths $(X_0,x_1,\dots,x_{N-1},x_N)$, there is at least one
segment $N_{1\underline {x_i}}^{\underline {x_{i+1}}}=0$.

Similarly, we may start from $Y_0$ and find a path
$(Y_0,Y_1,\dots,Y_{N-1},Y_N=Y_0)$,
$N_{1\underline{Y_i}}^{\underline{Y_{i+1}}}=1$.
Consider the path $(X_0,Y_0,Y_1,\dots,Y_{N-1})$.
It is a different path from $(X_0,X_1,\dots,X_{N-1},X_0)$ as long as $Y_0\neq
X_1$, and we have $N_{1\underline{X_0}}^{\underline{Y_0}}=0$.
Considering the path $(Y_0,Y_1,\dots,Y_{N-1},X_0)$ we know that if
$Y_{N-1}\neq X_{N-1}$, $N_{1\underline{Y_{N-1}}}^{\underline{X_0}}=0$.
Therefore, there is only one $x$ satisfying $N_{1\underline{X_0}}^{\underline
x}=1$ and only one $y$ satisfying $N_{1\underline y}^{\underline{X_0}}=1$.
We may say two labels $x,y$ are \emph{1-step-connected} if $N_{1\underline
x}^{\underline y}=1$.
Then $X_0$ is only 1-step-connected to $X_1$ forwards and only 1-step-connected
to $X_{N-1}$ backwards.
Such analysis applies to any label $X_0$.

The connection of labels forms an equivalent relation.
The discussions above then imply that connected labels form closed paths.
If $M$ is the number of different labels in $(X_0,X_1,\dots,X_{N-1},X_0)$,
we know that $X_M=X_0, X_{M+1}=X_1,\dots$, in general, $ X_i=X_{\res{i}_M}$,
and since $X_N=X_0$, $M$ must be a factor of $N$, {i.e.,}\ $M|N$. Since
different closed paths have no intersections,
for an \emph{indecomposable} boundary, it suffices to consider the
boundary fusion rules
\begin{itemize}
  \item Boundary string label set $B=\{0,1,2,\dots,M-1\}$, where $M|N$.
  \item Boundary fusion rules
    $N_{i\underline x}^{\underline y}=\delta_{y\res{i+x}_M}$.
\end{itemize}

However, this not the end of story. We need to find the solutions to the boundary
pentagon equations \eqref{b-pent}.
With the boundary fusion rules above we may simplify our notation of the
boundary F-matrices
$F^{ij\underline
x}_{\underline{\res{i+j+x}_M};\underline{\res{j+x}_M}\res{i+j}_{N}}=
\beta_{ij\underline
x}$.
The pentagon equations \eqref{b-pent} become
\begin{equation}
\alpha^p_{ijk}\beta_{i\res{j+k}_{N}\underline x}\beta_{jk\underline
x}=\beta_{\res{i+j}_{N}k\underline x}\beta_{ij\underline{\res{k+x}_M}}.
  \label{b-cocycle-cond}
\end{equation}
There are not always solutions to \eqref{b-cocycle-cond}. 
To see this, we multiply the following $M$ equations
\begin{widetext}
  \begin{align}
\alpha^p_{ijk}\beta_{i\res{j+k}_{N}\underline x}\beta_{jk\underline
x}&=\beta_{\res{i+j}_{N}k\underline
x}\beta_{ij\underline{\res{k+x}_M}},\nonumber\\
\alpha^p_{ijk}\beta_{i\res{j+k}_{N}\underline{\res{x+1}_M}}
\beta_{jk\underline{\res{x+1}_M}}&=\beta_{\res{i+j}_{N}k
\underline{\res{x+1}_M}}\beta_{ij\underline{\res{k+x+1}_M}},\nonumber\\
\alpha^p_{ijk}\beta_{i\res{j+k}_{N}\underline{\res{x+2}_M}}
\beta_{jk\underline{\res{x+2}_M}}&=
\beta_{\res{i+j}_{N}k\underline{\res{x+2}_M}}
\beta_{ij\underline{\res{k+x+2}_M}},\nonumber\\
    &\vdots\nonumber\\
\alpha^p_{ijk}\beta_{i\res{j+k}_{N}\underline{\res{x+M-1}_M}}
\beta_{jk\underline{\res{x+M-1}_M}}&=
\beta_{\res{i+j}_{N}k\underline{\res{x+M-1}_M}}
\beta_{ij\underline{\res{k+x+M-1}_M}}.
  \end{align}
\end{widetext}
and obtain
\begin{equation}
  \alpha^{pM}_{ijk}f_{i\res{j+k}_N}f_{jk}=f_{\res{i+j}_N k}f_{ij},
  \label{b-pmtrivial}
\end{equation}
where $f_{ij}=\prod_{x=0}^{M-1}\beta_{ij\underline x}$. This implies that
$\alpha^{pM}_{ijk}$ is equivalent to the trivial cocycle.
Therefore, we know $M$ must also satisfy $N|pM$.

On the other hand, for any integer $M$ satisfying $N| pM$ and $M|N$,
\eqref{b-cocycle-cond} does have solutions.
But as in the bulk, there are gauge transformations between equivalent
solutions.
It is not hard to check that, for each $M$ there is only one equivalent class
of solutions.
We pick a canonical form of the solutions
\begin{equation}
  \beta_{ij\underline x}=\ee^{2\pi \ri \frac{p}{N^2}i(j+x-\res{j+x}_{M})}.
  \label{b-cocycle}
\end{equation}

To conclude, the boundary types of twisted $(\bbz_N,p)$ model are classified by
integers $M$ satisfying $N|pM,M|N$.
The $M$-boundary is given by
\begin{itemize}
  \item Boundary string label set $B=\{0,1,2,\dots,M-1\}$.
\item Boundary fusion rules $N_{i\underline x}^{\underline
y}=\delta_{y\res{i+x}_M}$.
\item Boundary F-matrices $${F^{ij\underline
x}_{\underline{\res{i+j+x}_M};\underline{\res{j+x}_M}\res{i+j}_{N}}
=\beta_{ij\underline x}}.$$
\end{itemize}

\subsection{Boundary quasiparticles} 
For the $M$-boundary of $(\bbz_N,p)$ model, we classify the boundary
quasiparticles by studying the modules over the boundary Q-algebra
\begin{align}
  Q^i_{xy}=\ \f{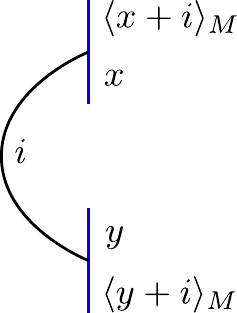},
  \begin{aligned}
    x&=0,1,\dots,{M}-1,
    \\
    y&=0,1,\dots,{M}-1,
    \\
    i&=0,1,\dots,N-1,
  \end{aligned}
\end{align}
\begin{align}
  Q^i_{x'y'}Q^j_{xy}&=\delta_{x'\res{j+x}_M}\delta_{y'\res{j+y}_M}
  \frac{\beta_{ij\underline x}}{\beta_{ij\underline y}}Q^{\res{i+j}_N}_{xy}
  \nonumber\\
  &=\ee^{ 2\pi\ri\frac{p}{N^2}i(x-y+\res{j+y}_M-\res{j+x}_M)}
  \nonumber\\
  &\xt\delta_{x'\res{j+x}_M}\delta_{y'\res{j+y}_M}Q^{\res{i+j}_N}_{xy}.
\end{align}
The dimension of this Q-algebra is $NM^2$.
It is easy to get $N$ different $M$-dimensional simple modules via a bit of
observation, guess, and calculation. We know that these are all the simple
modules.

In the multiplication rule, $\res{x'-y'}_M=\res{x-y}_M$.
Thus, we guess that, a simple module can be labeled by $(a,b)$, where $a$
corresponds to the difference between $x,y$, and $b$ corresponds to the choice
the phase factors. The basis of the $(a,b)$ module is
\begin{gather}
  e^{(a,b)}_x=\ \f{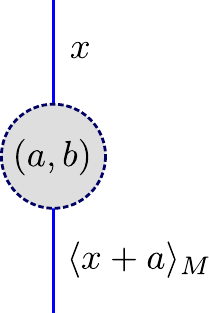},x=0,1,\dots,M-1,
\end{gather}
and the dimension of the $(a,b)$ module is $M$. The algebra action on the
module is
\begin{align}
  Q^i_{x'y'}e^{(a,b)}_{x}&=
  \ee^{2\pi\ri\frac{bi}{N}}
  \ee^{ 2\pi\ri\frac{p}{N^2}i(y'-x'-a)}
  \nonumber\\
  &\xt\delta_{x'x}\delta_{y'\res{x+a}_M}
  e^{(a,b)}_{\res{x+i}_M}.
\end{align}

It is not hard to check that two modules $(a,b)$ and $(a',b')$
are isomorphic iff 
\begin{equation}
  a'=a+k_1 M,\ b'=b+k_1\frac{pM}{N}+k_2\frac{N}{M}.
\end{equation}
Thus, we have $N$ different modules and also we got all the possible simple
modules. In other words, we got all the boundary quasiparticle types.

We can consider the fusion of the boundary quasiparticles, given by the tensor
product of the modules
\begin{alignat}{2}
  (a_1,b_1)&\ot (a_2,b_2)&&\to (a_3,b_3)
  \nonumber\\
  e^{(a_1,b_1)}_x&\ot e^{(a_2,b_2)}_{x'}&&\mapsto
\ee^{-2\pi\ri\frac{x}{N}\left[ b_1+b_2-b_3-\frac{p}{N}\left( a_1+a_2-a_3)
\right) \right] }
  \nonumber\\
  &&&\times\delta_{x'\res{x+a_1}_M} e^{(a_3,b_3)}_x
\end{alignat}
where
\begin{align}
  a_3&=\res{a_1+a_2}_M,
  \nonumber\\
  b_3&=\res{b_1+b_2-\frac{p}{N}\left( a_1+a_2-a_3 \right)}_{{\frac{N}{M}}}.
\end{align}
Thus, the fusion category $\cB_M$ of the excitations on the $M$-boundary is
\begin{itemize}
  \item Fusion rule $$N_{(a_1,b_1)(a_2,b_2)}^{(a_3,b_3)}=
    \delta_{a_3\res{a_1+a_2}_M}
\delta_{b_3\res{b_1+b_2-\frac{p}{N}\left( a_1+a_2-a_3
\right)}_{{\frac{N}{M}}}}.$$
  \item For stable vertices, F-matrices 
$$F^{(a_1,b_1)(a_2,b_2)(a_3,b_3)}_{(a_4,b_4);(a_6,b_6)(a_5,b_5)}=
\displaystyle\ee^{
-2\pi\ri\frac{a_1}{N}\left[ b_2+b_3-b_6-\frac{p}{N}\left( a_2+a_3-a_6)
\right) \right]}.$$
\end{itemize}
One can calculate the modular data $T,S$ of $\cB_M$ string-net model,
which are always the same as those of $(\bbz_N,p)$ model,
no matter which $M$-boundary we choose.
Therefore, $(\bbz_N,p)$ and $\cB_M$ string-net models describe the same physical
phase.
Moreover, $M$-boundary is actually the \emph{transparent domain wall}
(mathematically, the invertible bimodule category)
between $(\bbz_N,p)$ and $\cB_M$.

\subsection{Boundary changing operators}

Similarly we can find the boundary changing operators via the Q-algebra
approach. We now focus at the junction of $M_1$-boundary (red line) and
$M_2$-boundary (blue line).
The corresponding Q-algebra is
\begin{align}
  Q_{xy}^i=\f{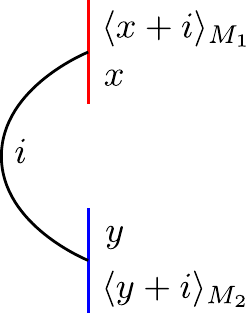},
  \begin{aligned}
    x&=0,1,\dots,{M_1}-1,
    \\
    y&=0,1,\dots,{M_2}-1,
    \\
    i&=0,1,\dots,N-1,
  \end{aligned}
\end{align}
\begin{align}
  Q^i_{x'y'}Q^j_{xy}&=\delta_{x'\res{j+x}_{M_1}}\delta_{y'\res{j+y}_{M_2}}
\frac{\beta^{(M_1)}_{ij\underline x}}{\beta^{(M_2)}_{ij\underline
y}}Q^{\res{i+j}_N}_{xy}
  \nonumber\\
&  =\ee^{ 2\pi\ri\frac{p}{N^2}i(x-y+\res{j+y}_{M_2}-\res{j+x}_{M_1})}
  \nonumber\\
&\xt
  \delta_{x'\res{j+x}_{M_1}}\delta_{y'\res{j+y}_{M_2}}
Q^{\res{i+j}_N}_{xy}.
\end{align}
The dimension of the Q-algebra is $NM_1M_2$.

Let $R$ be the greatest common divisor of $M_1,M_2$, denoted by
$R=\gcd(M_1,M_2)$;
we can write the basis of a simple module $(a,b)_{12}$
\begin{gather}
  e^{(a,b)_{12}}_{w_1w_2 z}=\f{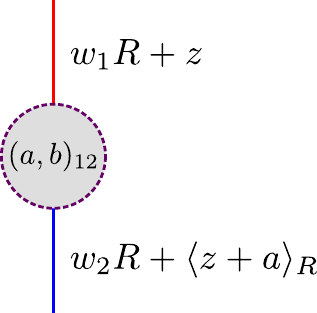},
  \begin{aligned}
    w_1&=0,1,\dots,\tfrac{M_1}{R}-1,
    \\
    w_2&=0,1,\dots,\tfrac{M_2}{R}-1,
    \\
    z&=0,1,\dots,R-1,
  \end{aligned}
\end{gather}
\begin{align}
  &Q^i_{xy}e^{(a,b)}_{w_1w_2 z}=
  \ee^{ 2\pi\ri\frac{bi}{N} }
  \ee^{ 2\pi\ri\frac{p}{N^2}i(y-x-a) }
  \nonumber\\
  &\xt\delta_{x\res{w_1R+z}_{M_1}}
  \delta_{y\res{w_2R+\res{z+a}_R}_{M_2}}
  \nonumber\\
  & \xt e^{(a,b)}_{\res{w_1+\frac{z+i-\res{z+i}_R}{R}}_{\frac{M_1}{R}}
  \res{w_2+\frac{\res{z+a}_R+i-\res{z+a+i}_R}{R}}_{\frac{M_2}{R}}\res{z+i}_R},
\end{align}
We see the dimension of the module $(a,b)_{12}$ is $\dfrac{M_1M_2}{R}$.

Two simple modules $(a,b)_{12}$ and $(a',b')_{12}$
are isomorphic iff
\begin{equation}
  a'=a+k_1 R,\ b'=b+k_1\frac{pR}{N}+k_2\frac{NR}{M_1M_2}.
\end{equation}
Therefore, there are $\dfrac{NR^2}{M_1M_2}$ different simple modules,
which satisfies the sum of squares law: $NM_1M_2=\dfrac{NR^2}{M_1M_2}\left(
\dfrac{M_1M_2}{R} \right)^2$.
We know the $(a,b)_{12}$ modules are all the possible simple modules.

Again we can say the modules $(a,b)_{12}$ form a category $\cD_{12}$.
One can always fuse the boundary quasiparticles
$(a_1,b_1)_1$ on $M_1$-boundary and 
$(a_2,b_2)_2$ on $M_2$-boundary 
with the boundary changing operator $(a,b)_{12}$
to get new composite boundary changing operators.
Mathematically, this means the tensor products $(a_1,b_1)_1\ot (a,b)_{12}$ and
$(a,b)_{12}\ot (a_2,b_2)_2$ are still modules in $\cD_{12}$.
Therefore, $\cD_{12}$ is a $\cB_{M_1}$-$\cB_{M_2}$-bimodule category.

\subsection{Quasiparticles condensing to the boundary: relation to Lagrangian
subgroup}

A given topologically ordered state can have many-different types of
boundaries.\cite{BSH0903,KS11,KK12,WW1263,Lev13,FSV13} A  boundary can be
understood in the following way.  We can always move a bulk quasiparticle
excitation to the boundary, and obtain a boundary quasiparticle.  If a
quasiparticle moves to the boundary and becomes a trivial boundary quasiparticle,
we say the quasiparticle \emph{condenses}\cite{BS0916,Kon14,ERB1301} to the
boundary. For Abelian topological phases, it is believed that quasiparticles that can condense to a boundary
form a \emph{Lagrangian subgroup}, and Lagrangian subgroups are in one-to-one
correspondence to boundary types.\cite{KS11,WW1263,Lev13,FSV13,BJQ13,BJQ13a,Kon14} We
will show this correspondence explicitly for the $(\bbz_N,p)$ string-net
models.

A Lagrangian subgroup $\cK$ is a subset of
quasiparticle types, such that
\begin{align}
  &\forall \xi,\zeta\in \cK, T_\xi=1, D_{\cZ(\cC)} S_{\xi\zeta}=1,
  \nonumber\\
  &\forall \xi'\notin \cK, \exists \xi\in \cK, D_{\cZ(\cC)} S_{\xi\xi'}\neq 1.
\end{align}

For the $(\bbz_N,p)$ model case, moving a quasiparticle $\cl{ri}$ to the
$M$-boundary, we should get a boundary quasiparticle $(a,b)$, as shown in the
following graphs
\begin{widetext}
\begin{gather}
\ev\quad\f{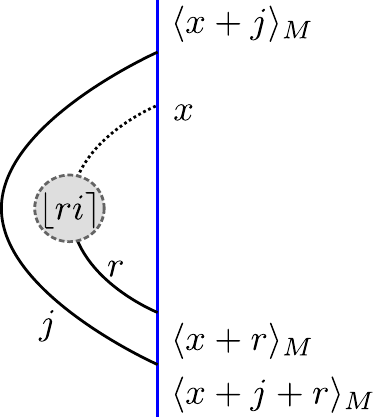}=\frac{\alpha^p_{j\res{-j}_N
j}\alpha^p_{jr\res{-j}_N}}{\alpha^p_{\res{j+r}_N\res{-j}_N j}}
\frac{\beta_{rj\underline x}}{\beta_{jr\underline x}}M^{\res{-j}_N}_{\cl{ri},r}
\ \f{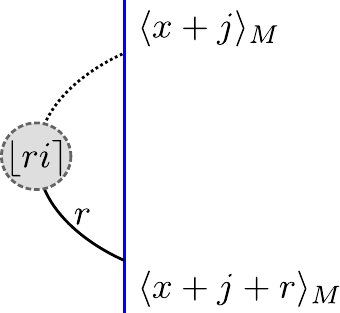},\\
  \f{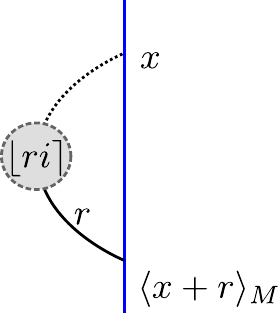}\longmapsto\qquad \ee^{ 2\pi\ri\frac{p}{N^2}rx }\ \f{b-mod-znp},
\end{gather}
\end{widetext}
where
\begin{gather}
  a=\res{r}_M,\ 
  b=\res{i-\frac{p(r-\res{r}_M)}{N}}_{{\frac{N}{M}}}.
\end{gather}
Let $\cK_M$ be the set of quasiparticle types $\cl{ri}$ that maps to the
trivial boundary quasiparticle $(0,0)$, and we see that
\begin{gather}
  \cK_M=\left\{\cl{ri}\left|r=k_1 M,\ i=k_2 \frac{N}{M}+ \frac{p
  r}{N}\right.\right\},\label{znplagsub}
\end{gather}
where $k_1,k_2$ are integers.
One can easily check that $\cK_M$ is indeed a Lagrangian subgroup.

The next question is: Do all the Lagrangian subgroups of $(\bbz_N,p)$ model
have the form of \eqref{znplagsub}? The answer is ``Yes''.

Firstly, note that $T_{\cl{ri}}=1$ requires $\dfrac{ri}{N}-\dfrac{pr^2}{N^2}$
to be some integer number $k$, {i.e.,}\
\begin{gather}
  Nri-pr^2=kN^2.
\end{gather}
Let $m=\gcd(r,N)$,
and $N=um, r=vm, \gcd(u,v)=1$, we have
\begin{equation}
  uvi-pv^2=ku^2,
\end{equation}
which implies that $u|pv^2,v|ku^2$. Since $\gcd(u,v)=1$ we know that $u|p,v|k$
and $N=um|pm$. Thus,
\begin{gather}
  r=vm,\ i=\frac{k}{v}\frac{N}{m}+\frac{pr}{N},\label{znplagnes}
\end{gather}
or equivalently
\begin{equation}
  i=t+\frac{pr}{N},\ N|rt,\ N|pr.
\end{equation}

Then we can show that any Lagrangian subgroup $\cK$ must be equal to some
$\cK_M$.
For convenience, say
${\cK=\left\{\cl{s_1j_1},\cl{s_2j_2},\dots,\cl{s_{|\cK|}j_{|\cK|}}\right\}}$, where
$|\cK|$ is the number of different quasiparticle types in $\cK$.
As discussed above, $T_{\cl{s_nj_n}}=1$ requires that
\begin{gather}
  j_n=t_n+\frac{ps_n}{N},\ N|s_nt_n,\ N|ps_n.\label{sj}
\end{gather}
Let 
\begin{align}
  M&=\gcd (N,s_1,s_2,\dots,s_{|\cK|}),\\
  P&=\gcd(N,t_1,t_2,\dots,t_{|\cK|}).
\end{align}
We have
\begin{gather}
  s_n=k_n M,\ t_n=l_n P,
  \nonumber\\
\gcd(\frac{N}{M},k_1,k_2,\dots,k_{|\cK|})=\gcd(\frac{N}{P},l_1,l_2,\dots,l_{|\cK|})=1.
\end{gather}
We have $N|PMk_nl_n$. 
$D_{\cZ(\cC)} S_{\cl{s_nj_n}\cl{s_{m}j_m}}=1$ requires that
$N|PM(k_nl_m+k_ml_n)$.
With these constraints we can show that $N|PM$:
\begin{align*}
  &N|PM(k_nl_m+k_ml_n)
  \Rightarrow
  N|k_nPM(k_nl_m+k_ml_n)
  \\&\Rightarrow
  N|PM k_n^2 l_m
  \Rightarrow
  N|PM k_n^2 \gcd(\frac{N}{P},l_1,l_2,\dots,l_{|\cK|})
  \\&\Rightarrow
  N|PM k_n^2
  \Rightarrow
  N|PM \gcd(\frac{N}{M},k_1^2,k_2^2,\dots,k_{|\cK|}^2)
  \\&\Rightarrow
  N|PM.
\end{align*}
We then have $PM=u N$ for some integer $u$. We see that
\begin{gather}
  s_n=k_n M,\ j_n=l_n u \frac{N}{M}+\frac{ps_n}{N},
\end{gather}
and we know that $\cl{s_nj_n}\in \cK_M$, in other words $\cK\subseteq \cK_M$.
Due to the properties of Lagrangian subgroups, this is the same as $\cK=\cK_M$.
This can be proved by contradiction: Suppose there is a quasiparticle $\xi \in
\cK_M$ but $\xi\notin \cK$.
$\xi \notin \cK$ means that there should exist a quasiparticle $\zeta\in \cK$,
such that $D_{\cZ(\cC)} S_{\xi\zeta}\neq 1$.
But, for $\cK\subseteq \cK_M$, both $\xi,\zeta$ are in $\cK_M$ and
$\xi\neq\zeta$, we should also have $D_{\cZ(\cC)} S_{\xi\zeta}=1$.
Contradiction.

Now we have shown that for the $(\bbz_N,p)$ model, each $M$-boundary will give
a Lagrangian subgroup $\cK_M$ and these $\cK_M$ are all the possible Lagrangian
subgroups. The Lagrangian subgroups are indeed in one-to-one correspondence to
boundary types.

There is also correspondence between boundary quasiparticles,
boundary changing operators and
Lagrangian subgroups.
Roughly speaking,
if we use $\cZ(\cC)$ to denote the set of all bulk quasiparticle types and
$\cK$ a Lagrangian subgroup,
then $\cZ(\cC)/\cK$ are the quasiparticles on the $\cK$-boundary that survive the
condensation.
Similarly, the boundary changing operators between $\cK_1$-boundary and
$\cK_2$-boundary
should be given by $\cZ(\cC)/\cK_1\boxtimes\cK_2$,
where $\cK_1\boxtimes\cK_2$ are the quasiparticles fused by quasiparticles in
$\cK_1$ and $\cK_2$.

For the $(\bbz_N,p)$ case, suppose $\cK_{M_1},\cK_{M_2}$ are two Lagrangian
subgroups.
Quasiparticles $\cl{ri}$ in $\cK_{M_1}\boxtimes\cK_{M_2}$ are
\begin{equation}
  r=k_1 M_1+ k_2 M_2,\ i=l_1 \frac{N}{M_1}+l_2 \frac{N}{M_2} +\frac{pr}{N}.
\end{equation}
It is easy to see
\begin{gather}
|\cK_{M_1}\boxtimes\cK_{M_2}|=\frac{N}{\gcd(M_1,M_2)}
\frac{N}{\gcd(\dfrac{N}{M_1},\dfrac{N}{M_2})}.
\end{gather}
Let $R=\gcd(M_1,M_2)$,
$  \left|\cK_{M_1}\boxtimes\cK_{M_2}\right|=\dfrac{NM_1M_2}{R^2}$.
Thus,
\begin{gather}
  \left|\cZ(\cC)/\cK_{M_1}\boxtimes\cK_{M_2}\right|=\frac{NR^2}{M_1M_2}
\end{gather}
is the number of different boundary changing operators between $M_1$-boundary
and $M_2$-boundary,
which agrees with our previous results obtained via the Q-algebra approach.

\bibliography{refnew,wencross,all,publst}

\begin{thebibliography}{39}%
\makeatletter
\providecommand \@ifxundefined [1]{%
 \@ifx{#1\undefined}
}%
\providecommand \@ifnum [1]{%
 \ifnum #1\expandafter \@firstoftwo
 \else \expandafter \@secondoftwo
 \fi
}%
\providecommand \@ifx [1]{%
 \ifx #1\expandafter \@firstoftwo
 \else \expandafter \@secondoftwo
 \fi
}%
\providecommand \natexlab [1]{#1}%
\providecommand \enquote  [1]{``#1''}%
\providecommand \bibnamefont  [1]{#1}%
\providecommand \bibfnamefont [1]{#1}%
\providecommand \citenamefont [1]{#1}%
\providecommand \href@noop [0]{\@secondoftwo}%
\providecommand \href [0]{\begingroup \@sanitize@url \@href}%
\providecommand \@href[1]{\@@startlink{#1}\@@href}%
\providecommand \@@href[1]{\endgroup#1\@@endlink}%
\providecommand \@sanitize@url [0]{\catcode `\\12\catcode `\$12\catcode
  `\&12\catcode `\#12\catcode `\^12\catcode `\_12\catcode `\%12\relax}%
\providecommand \@@startlink[1]{}%
\providecommand \@@endlink[0]{}%
\providecommand \url  [0]{\begingroup\@sanitize@url \@url }%
\providecommand \@url [1]{\endgroup\@href {#1}{\urlprefix }}%
\providecommand \urlprefix  [0]{URL }%
\providecommand \Eprint [0]{\href }%
\providecommand \doibase [0]{http://dx.doi.org/}%
\providecommand \selectlanguage [0]{\@gobble}%
\providecommand \bibinfo  [0]{\@secondoftwo}%
\providecommand \bibfield  [0]{\@secondoftwo}%
\providecommand \translation [1]{[#1]}%
\providecommand \BibitemOpen [0]{}%
\providecommand \bibitemStop [0]{}%
\providecommand \bibitemNoStop [0]{.\EOS\space}%
\providecommand \EOS [0]{\spacefactor3000\relax}%
\providecommand \BibitemShut  [1]{\csname bibitem#1\endcsname}%
\let\auto@bib@innerbib\@empty
\bibitem [{\citenamefont {Landau}(1937)}]{Lan37}%
  \BibitemOpen
  \bibfield  {author} {\bibinfo {author} {\bibfnamefont {L.~D.}\ \bibnamefont
  {Landau}},\ }\href@noop {} {\bibfield  {journal} {\bibinfo  {journal} {Phys.
  Z. Sowjetunion}\ }\textbf {\bibinfo {volume} {11}},\ \bibinfo {pages} {26}
  (\bibinfo {year} {1937})}\BibitemShut {NoStop}%
\bibitem [{\citenamefont {Tsui}\ \emph {et~al.}(1982)\citenamefont {Tsui},
  \citenamefont {Stormer},\ and\ \citenamefont {Gossard}}]{TSG8259}%
  \BibitemOpen
  \bibfield  {author} {\bibinfo {author} {\bibfnamefont {D.~C.}\ \bibnamefont
  {Tsui}}, \bibinfo {author} {\bibfnamefont {H.~L.}\ \bibnamefont {Stormer}}, \
  and\ \bibinfo {author} {\bibfnamefont {A.~C.}\ \bibnamefont {Gossard}},\
  }\href@noop {} {\bibfield  {journal} {\bibinfo  {journal} {Phys. Rev. Lett.}\
  }\textbf {\bibinfo {volume} {48}},\ \bibinfo {pages} {1559} (\bibinfo {year}
  {1982})}\BibitemShut {NoStop}%
\bibitem [{\citenamefont {Wen}(1989)}]{Wtop}%
  \BibitemOpen
  \bibfield  {author} {\bibinfo {author} {\bibfnamefont {X.-G.}\ \bibnamefont
  {Wen}},\ }\href@noop {} {\bibfield  {journal} {\bibinfo  {journal} {Phys.
  Rev. B}\ }\textbf {\bibinfo {volume} {40}},\ \bibinfo {pages} {7387}
  (\bibinfo {year} {1989})}\BibitemShut {NoStop}%
\bibitem [{\citenamefont {Wen}\ and\ \citenamefont {Niu}(1990)}]{WNtop}%
  \BibitemOpen
  \bibfield  {author} {\bibinfo {author} {\bibfnamefont {X.-G.}\ \bibnamefont
  {Wen}}\ and\ \bibinfo {author} {\bibfnamefont {Q.}~\bibnamefont {Niu}},\
  }\href@noop {} {\bibfield  {journal} {\bibinfo  {journal} {Phys. Rev. B}\
  }\textbf {\bibinfo {volume} {41}},\ \bibinfo {pages} {9377} (\bibinfo {year}
  {1990})}\BibitemShut {NoStop}%
\bibitem [{\citenamefont {Wen}(1990)}]{Wrig}%
  \BibitemOpen
  \bibfield  {author} {\bibinfo {author} {\bibfnamefont {X.-G.}\ \bibnamefont
  {Wen}},\ }\href@noop {} {\bibfield  {journal} {\bibinfo  {journal} {Int. J.
  Mod. Phys. B}\ }\textbf {\bibinfo {volume} {4}},\ \bibinfo {pages} {239}
  (\bibinfo {year} {1990})}\BibitemShut {NoStop}%
\bibitem [{\citenamefont {Keski-Vakkuri}\ and\ \citenamefont
  {Wen}(1993)}]{KW9327}%
  \BibitemOpen
  \bibfield  {author} {\bibinfo {author} {\bibfnamefont {E.}~\bibnamefont
  {Keski-Vakkuri}}\ and\ \bibinfo {author} {\bibfnamefont {X.-G.}\ \bibnamefont
  {Wen}},\ }\href@noop {} {\bibfield  {journal} {\bibinfo  {journal} {Int. J.
  Mod. Phys. B}\ }\textbf {\bibinfo {volume} {7}},\ \bibinfo {pages} {4227}
  (\bibinfo {year} {1993})}\BibitemShut {NoStop}%
\bibitem [{\citenamefont {Wen}(2012)}]{W1221}%
  \BibitemOpen
  \bibfield  {author} {\bibinfo {author} {\bibfnamefont {X.-G.}\ \bibnamefont
  {Wen}},\ }\href@noop {} {\  (\bibinfo {year} {2012})},\ \Eprint
  {http://arxiv.org/abs/arXiv:1212.5121} {arXiv:1212.5121} \BibitemShut
  {NoStop}%
\bibitem [{\citenamefont {Zhang}\ and\ \citenamefont
  {Vishwanath}(2013)}]{ZV1313}%
  \BibitemOpen
  \bibfield  {author} {\bibinfo {author} {\bibfnamefont {Y.}~\bibnamefont
  {Zhang}}\ and\ \bibinfo {author} {\bibfnamefont {A.}~\bibnamefont
  {Vishwanath}},\ }\href@noop {} {\bibfield  {journal} {\bibinfo  {journal}
  {Phys. Rev. B}\ }\textbf {\bibinfo {volume} {87}},\ \bibinfo {pages} {161113}
  (\bibinfo {year} {2013})},\ \Eprint {http://arxiv.org/abs/arXiv:1209.2424}
  {arXiv:1209.2424} \BibitemShut {NoStop}%
\bibitem [{\citenamefont {Cincio}\ and\ \citenamefont {Vidal}(2013)}]{CV1308}%
  \BibitemOpen
  \bibfield  {author} {\bibinfo {author} {\bibfnamefont {L.}~\bibnamefont
  {Cincio}}\ and\ \bibinfo {author} {\bibfnamefont {G.}~\bibnamefont {Vidal}},\
  }\href@noop {} {\bibfield  {journal} {\bibinfo  {journal} {Phys. Rev. Lett.}\
  }\textbf {\bibinfo {volume} {110}},\ \bibinfo {pages} {067208} (\bibinfo
  {year} {2013})},\ \Eprint {http://arxiv.org/abs/arXiv:1208.2623}
  {arXiv:1208.2623} \BibitemShut {NoStop}%
\bibitem [{\citenamefont {{Levin}}\ and\ \citenamefont {{Wen}}(2005)}]{LW05}%
  \BibitemOpen
  \bibfield  {author} {\bibinfo {author} {\bibfnamefont {M.~A.}\ \bibnamefont
  {{Levin}}}\ and\ \bibinfo {author} {\bibfnamefont {X.-G.}\ \bibnamefont
  {{Wen}}},\ }\href {\doibase 10.1103/PhysRevB.71.045110} {\bibfield  {journal}
  {\bibinfo  {journal} {\prb}\ }\textbf {\bibinfo {volume} {71}},\ \bibinfo
  {eid} {045110} (\bibinfo {year} {2005})},\ \Eprint
  {http://arxiv.org/abs/cond-mat/0404617} {cond-mat/0404617} \BibitemShut
  {NoStop}%
\bibitem [{\citenamefont {{Kitaev}}\ and\ \citenamefont {{Kong}}(2012)}]{KK12}%
  \BibitemOpen
  \bibfield  {author} {\bibinfo {author} {\bibfnamefont {A.}~\bibnamefont
  {{Kitaev}}}\ and\ \bibinfo {author} {\bibfnamefont {L.}~\bibnamefont
  {{Kong}}},\ }\href {\doibase 10.1007/s00220-012-1500-5} {\bibfield  {journal}
  {\bibinfo  {journal} {Communications in Mathematical Physics}\ }\textbf
  {\bibinfo {volume} {313}},\ \bibinfo {pages} {351} (\bibinfo {year}
  {2012})},\ \Eprint {http://arxiv.org/abs/1104.5047} {arXiv:1104.5047
  [cond-mat.str-el]} \BibitemShut {NoStop}%
\bibitem [{\citenamefont {{Kong}}(2012)}]{Kon12}%
  \BibitemOpen
  \bibfield  {author} {\bibinfo {author} {\bibfnamefont {L.}~\bibnamefont
  {{Kong}}},\ }\href@noop {} {\bibfield  {journal} {\bibinfo  {journal}
  {Proceedings of XVII{\tiny TH} International Congress of Mathematical
  Physics}\ ,\ \bibinfo {pages} {444}} (\bibinfo {year} {2012})},\ \Eprint
  {http://arxiv.org/abs/1211.4644} {arXiv:1211.4644 [cond-mat.str-el]}
  \BibitemShut {NoStop}%
\bibitem [{\citenamefont {Verstraete}\ \emph {et~al.}(2005)\citenamefont
  {Verstraete}, \citenamefont {Cirac}, \citenamefont {Latorre}, \citenamefont
  {Rico},\ and\ \citenamefont {Wolf}}]{VCL0501}%
  \BibitemOpen
  \bibfield  {author} {\bibinfo {author} {\bibfnamefont {F.}~\bibnamefont
  {Verstraete}}, \bibinfo {author} {\bibfnamefont {J.~I.}\ \bibnamefont
  {Cirac}}, \bibinfo {author} {\bibfnamefont {J.~I.}\ \bibnamefont {Latorre}},
  \bibinfo {author} {\bibfnamefont {E.}~\bibnamefont {Rico}}, \ and\ \bibinfo
  {author} {\bibfnamefont {M.~M.}\ \bibnamefont {Wolf}},\ }\href@noop {}
  {\bibfield  {journal} {\bibinfo  {journal} {Phys. Rev. Lett.}\ }\textbf
  {\bibinfo {volume} {94}},\ \bibinfo {pages} {140601} (\bibinfo {year}
  {2005})}\BibitemShut {NoStop}%
\bibitem [{\citenamefont {Chen}\ \emph {et~al.}(2011)\citenamefont {Chen},
  \citenamefont {Gu},\ and\ \citenamefont {Wen}}]{CGW1107}%
  \BibitemOpen
  \bibfield  {author} {\bibinfo {author} {\bibfnamefont {X.}~\bibnamefont
  {Chen}}, \bibinfo {author} {\bibfnamefont {Z.-C.}\ \bibnamefont {Gu}}, \ and\
  \bibinfo {author} {\bibfnamefont {X.-G.}\ \bibnamefont {Wen}},\ }\href@noop
  {} {\bibfield  {journal} {\bibinfo  {journal} {Phys. Rev. B}\ }\textbf
  {\bibinfo {volume} {83}},\ \bibinfo {pages} {035107} (\bibinfo {year}
  {2011})},\ \Eprint {http://arxiv.org/abs/arXiv:1008.3745} {arXiv:1008.3745}
  \BibitemShut {NoStop}%
\bibitem [{\citenamefont {{Kong}}\ and\ \citenamefont {{Wen}}(2014)}]{KW14}%
  \BibitemOpen
  \bibfield  {author} {\bibinfo {author} {\bibfnamefont {L.}~\bibnamefont
  {{Kong}}}\ and\ \bibinfo {author} {\bibfnamefont {X.-G.}\ \bibnamefont
  {{Wen}}},\ }\href@noop {} {\bibfield  {journal} {\bibinfo  {journal} {ArXiv
  e-prints}\ } (\bibinfo {year} {2014})},\ \Eprint
  {http://arxiv.org/abs/1405.5858} {arXiv:1405.5858 [cond-mat.str-el]}
  \BibitemShut {NoStop}%
\bibitem [{\citenamefont {M{\"u}ger}(2003)}]{M03a}%
  \BibitemOpen
  \bibfield  {author} {\bibinfo {author} {\bibfnamefont {M.}~\bibnamefont
  {M{\"u}ger}},\ }\href@noop {} {\bibfield  {journal} {\bibinfo  {journal} {J.
  Pure Appl. Alg.}\ }\textbf {\bibinfo {volume} {180,}},\ \bibinfo {pages}
  {159} (\bibinfo {year} {2003})},\ \Eprint {http://arxiv.org/abs/math/0111205}
  {math/0111205} \BibitemShut {NoStop}%
\bibitem [{\citenamefont {{Kitaev}}(2006)}]{Kit06}%
  \BibitemOpen
  \bibfield  {author} {\bibinfo {author} {\bibfnamefont {A.}~\bibnamefont
  {{Kitaev}}},\ }\href {\doibase 10.1016/j.aop.2005.10.005} {\bibfield
  {journal} {\bibinfo  {journal} {Annals of Physics}\ }\textbf {\bibinfo
  {volume} {321}},\ \bibinfo {pages} {2} (\bibinfo {year} {2006})},\ \Eprint
  {http://arxiv.org/abs/cond-mat/0506438} {cond-mat/0506438} \BibitemShut
  {NoStop}%
\bibitem [{\citenamefont {Chen}\ \emph {et~al.}(2010)\citenamefont {Chen},
  \citenamefont {Gu},\ and\ \citenamefont {Wen}}]{CGW10}%
  \BibitemOpen
  \bibfield  {author} {\bibinfo {author} {\bibfnamefont {X.}~\bibnamefont
  {Chen}}, \bibinfo {author} {\bibfnamefont {Z.~C.}\ \bibnamefont {Gu}}, \ and\
  \bibinfo {author} {\bibfnamefont {X.~G.}\ \bibnamefont {Wen}},\ }\href
  {\doibase 10.1103/PhysRevB.82.155138} {\bibfield  {journal} {\bibinfo
  {journal} {\prb}\ }\textbf {\bibinfo {volume} {82}},\ \bibinfo {eid} {155138}
  (\bibinfo {year} {2010})},\ \Eprint {http://arxiv.org/abs/1004.3835}
  {arXiv:1004.3835 [cond-mat.str-el]} \BibitemShut {NoStop}%
\bibitem [{\citenamefont {Lan}(2012)}]{Lan12}%
  \BibitemOpen
  \bibfield  {author} {\bibinfo {author} {\bibfnamefont {T.}~\bibnamefont
  {Lan}},\ }\emph {\bibinfo {title} {Boundaries and Excitations in String-net
  Models}},\ \href@noop {} {\bibinfo {type} {{B}achelor's thesis}},\ \bibinfo
  {school} {Tsinghua University} (\bibinfo {year} {2012})\BibitemShut {NoStop}%
\bibitem [{\citenamefont {Morita}(1958)}]{Mor58}%
  \BibitemOpen
  \bibfield  {author} {\bibinfo {author} {\bibfnamefont {K.}~\bibnamefont
  {Morita}},\ }\href@noop {} {\bibfield  {journal} {\bibinfo  {journal}
  {Science reports of the Tokyo Kyoiku Daigaku. Section A}\ }\textbf {\bibinfo
  {volume} {6}},\ \bibinfo {pages} {83} (\bibinfo {year} {1958})}\BibitemShut
  {NoStop}%
\bibitem [{\citenamefont {Verlinde}(1988)}]{V8860}%
  \BibitemOpen
  \bibfield  {author} {\bibinfo {author} {\bibfnamefont {E.}~\bibnamefont
  {Verlinde}},\ }\href@noop {} {\bibfield  {journal} {\bibinfo  {journal}
  {Nuclear Physics B}\ }\textbf {\bibinfo {volume} {300}},\ \bibinfo {pages}
  {360} (\bibinfo {year} {1988})}\BibitemShut {NoStop}%
\bibitem [{\citenamefont {{Kitaev}}(2003)}]{Kit03}%
  \BibitemOpen
  \bibfield  {author} {\bibinfo {author} {\bibfnamefont {A.~Y.}\ \bibnamefont
  {{Kitaev}}},\ }\href {\doibase 10.1016/S0003-4916(02)00018-0} {\bibfield
  {journal} {\bibinfo  {journal} {Annals of Physics}\ }\textbf {\bibinfo
  {volume} {303}},\ \bibinfo {pages} {2} (\bibinfo {year} {2003})},\ \Eprint
  {http://arxiv.org/abs/quant-ph/9707021} {quant-ph/9707021} \BibitemShut
  {NoStop}%
\bibitem [{\citenamefont {{Coste}}\ \emph {et~al.}(2000)\citenamefont
  {{Coste}}, \citenamefont {{Gannon}},\ and\ \citenamefont {{Ruelle}}}]{CGR00}%
  \BibitemOpen
  \bibfield  {author} {\bibinfo {author} {\bibfnamefont {A.}~\bibnamefont
  {{Coste}}}, \bibinfo {author} {\bibfnamefont {T.}~\bibnamefont {{Gannon}}}, \
  and\ \bibinfo {author} {\bibfnamefont {P.}~\bibnamefont {{Ruelle}}},\ }\href
  {\doibase 10.1016/S0550-3213(00)00285-6} {\bibfield  {journal} {\bibinfo
  {journal} {Nuclear Physics B}\ }\textbf {\bibinfo {volume} {581}},\ \bibinfo
  {pages} {679} (\bibinfo {year} {2000})},\ \Eprint
  {http://arxiv.org/abs/hep-th/0001158} {hep-th/0001158} \BibitemShut {NoStop}%
\bibitem [{\citenamefont {Coquereaux}(2012)}]{Coq12}%
  \BibitemOpen
  \bibfield  {author} {\bibinfo {author} {\bibfnamefont {R.}~\bibnamefont
  {Coquereaux}},\ }\href
  {http://pos.sissa.it/archive/conferences/175/024/ICMP%202012_024.pdf}
  {\bibfield  {journal} {\bibinfo  {journal} {PoS}\ }\textbf {\bibinfo {volume}
  {ICMP2012}},\ \bibinfo {eid} {24} (\bibinfo {year} {2012})},\ \Eprint
  {http://arxiv.org/abs/1212.4010} {arXiv:1212.4010 [math.QA]} \BibitemShut
  {NoStop}%
\bibitem [{\citenamefont {{Chen}}\ \emph {et~al.}(2013)\citenamefont {{Chen}},
  \citenamefont {{Gu}}, \citenamefont {{Liu}},\ and\ \citenamefont
  {{Wen}}}]{CGLW13}%
  \BibitemOpen
  \bibfield  {author} {\bibinfo {author} {\bibfnamefont {X.}~\bibnamefont
  {{Chen}}}, \bibinfo {author} {\bibfnamefont {Z.-C.}\ \bibnamefont {{Gu}}},
  \bibinfo {author} {\bibfnamefont {Z.-X.}\ \bibnamefont {{Liu}}}, \ and\
  \bibinfo {author} {\bibfnamefont {X.-G.}\ \bibnamefont {{Wen}}},\ }\href
  {\doibase 10.1103/PhysRevB.87.155114} {\bibfield  {journal} {\bibinfo
  {journal} {\prb}\ }\textbf {\bibinfo {volume} {87}},\ \bibinfo {eid} {155114}
  (\bibinfo {year} {2013})},\ \Eprint {http://arxiv.org/abs/1106.4772}
  {arXiv:1106.4772 [cond-mat.str-el]} \BibitemShut {NoStop}%
\bibitem [{\citenamefont {{Levin}}\ and\ \citenamefont {{Gu}}(2012)}]{LG12}%
  \BibitemOpen
  \bibfield  {author} {\bibinfo {author} {\bibfnamefont {M.}~\bibnamefont
  {{Levin}}}\ and\ \bibinfo {author} {\bibfnamefont {Z.-C.}\ \bibnamefont
  {{Gu}}},\ }\href {\doibase 10.1103/PhysRevB.86.115109} {\bibfield  {journal}
  {\bibinfo  {journal} {\prb}\ }\textbf {\bibinfo {volume} {86}},\ \bibinfo
  {eid} {115109} (\bibinfo {year} {2012})},\ \Eprint
  {http://arxiv.org/abs/1202.3120} {arXiv:1202.3120 [cond-mat.str-el]}
  \BibitemShut {NoStop}%
\bibitem [{\citenamefont {{Hung}}\ and\ \citenamefont {{Wan}}(2012)}]{HW12}%
  \BibitemOpen
  \bibfield  {author} {\bibinfo {author} {\bibfnamefont {L.-Y.}\ \bibnamefont
  {{Hung}}}\ and\ \bibinfo {author} {\bibfnamefont {Y.}~\bibnamefont {{Wan}}},\
  }\href {\doibase 10.1103/PhysRevB.86.235132} {\bibfield  {journal} {\bibinfo
  {journal} {\prb}\ }\textbf {\bibinfo {volume} {86}},\ \bibinfo {eid} {235132}
  (\bibinfo {year} {2012})},\ \Eprint {http://arxiv.org/abs/1207.6169}
  {arXiv:1207.6169 [cond-mat.other]} \BibitemShut {NoStop}%
\bibitem [{\citenamefont {{Hu}}\ \emph {et~al.}(2013)\citenamefont {{Hu}},
  \citenamefont {{Wan}},\ and\ \citenamefont {{Wu}}}]{HWW13}%
  \BibitemOpen
  \bibfield  {author} {\bibinfo {author} {\bibfnamefont {Y.}~\bibnamefont
  {{Hu}}}, \bibinfo {author} {\bibfnamefont {Y.}~\bibnamefont {{Wan}}}, \ and\
  \bibinfo {author} {\bibfnamefont {Y.-S.}\ \bibnamefont {{Wu}}},\ }\href
  {\doibase 10.1103/PhysRevB.87.125114} {\bibfield  {journal} {\bibinfo
  {journal} {\prb}\ }\textbf {\bibinfo {volume} {87}},\ \bibinfo {eid} {125114}
  (\bibinfo {year} {2013})},\ \Eprint {http://arxiv.org/abs/1211.3695}
  {arXiv:1211.3695 [cond-mat.str-el]} \BibitemShut {NoStop}%
\bibitem [{\citenamefont {Walker}(2006)}]{Wal06}%
  \BibitemOpen
  \bibfield  {author} {\bibinfo {author} {\bibfnamefont {K.}~\bibnamefont
  {Walker}},\ }\href {http://canyon23.net/math/tc.pdf} {\emph {\bibinfo {title}
  {TQFTs}}}\ (\bibinfo {year} {2006})\ pp.\ \bibinfo {pages}
  {52--53}\BibitemShut {NoStop}%
\bibitem [{\citenamefont {Bais}\ \emph {et~al.}(2009)\citenamefont {Bais},
  \citenamefont {Slingerland},\ and\ \citenamefont {Haaker}}]{BSH0903}%
  \BibitemOpen
  \bibfield  {author} {\bibinfo {author} {\bibfnamefont {F.~A.}\ \bibnamefont
  {Bais}}, \bibinfo {author} {\bibfnamefont {J.~K.}\ \bibnamefont
  {Slingerland}}, \ and\ \bibinfo {author} {\bibfnamefont {S.~M.}\ \bibnamefont
  {Haaker}},\ }\href {\doibase 10.1103/PhysRevLett.102.220403} {\bibfield
  {journal} {\bibinfo  {journal} {Phys. Rev. Lett.}\ }\textbf {\bibinfo
  {volume} {102}},\ \bibinfo {pages} {220403} (\bibinfo {year} {2009})},\
  \Eprint {http://arxiv.org/abs/arXiv:0812.4596} {arXiv:0812.4596} \BibitemShut
  {NoStop}%
\bibitem [{\citenamefont {{Kapustin}}\ and\ \citenamefont
  {{Saulina}}(2011)}]{KS11}%
  \BibitemOpen
  \bibfield  {author} {\bibinfo {author} {\bibfnamefont {A.}~\bibnamefont
  {{Kapustin}}}\ and\ \bibinfo {author} {\bibfnamefont {N.}~\bibnamefont
  {{Saulina}}},\ }\href {\doibase 10.1016/j.nuclphysb.2010.12.017} {\bibfield
  {journal} {\bibinfo  {journal} {Nuclear Physics B}\ }\textbf {\bibinfo
  {volume} {845}},\ \bibinfo {pages} {393} (\bibinfo {year} {2011})},\ \Eprint
  {http://arxiv.org/abs/1008.0654} {arXiv:1008.0654 [hep-th]} \BibitemShut
  {NoStop}%
\bibitem [{\citenamefont {Wang}\ and\ \citenamefont {Wen}(2012)}]{WW1263}%
  \BibitemOpen
  \bibfield  {author} {\bibinfo {author} {\bibfnamefont {J.}~\bibnamefont
  {Wang}}\ and\ \bibinfo {author} {\bibfnamefont {X.-G.}\ \bibnamefont {Wen}},\
  }\href@noop {} {\  (\bibinfo {year} {2012})},\ \Eprint
  {http://arxiv.org/abs/arXiv:1212.4863} {arXiv:1212.4863} \BibitemShut
  {NoStop}%
\bibitem [{\citenamefont {{Levin}}(2013)}]{Lev13}%
  \BibitemOpen
  \bibfield  {author} {\bibinfo {author} {\bibfnamefont {M.}~\bibnamefont
  {{Levin}}},\ }\href {\doibase 10.1103/PhysRevX.3.021009} {\bibfield
  {journal} {\bibinfo  {journal} {Physical Review X}\ }\textbf {\bibinfo
  {volume} {3}},\ \bibinfo {eid} {021009} (\bibinfo {year} {2013})},\ \Eprint
  {http://arxiv.org/abs/1301.7355} {arXiv:1301.7355 [cond-mat.str-el]}
  \BibitemShut {NoStop}%
\bibitem [{\citenamefont {{Fuchs}}\ \emph {et~al.}(2013)\citenamefont
  {{Fuchs}}, \citenamefont {{Schweigert}},\ and\ \citenamefont
  {{Valentino}}}]{FSV13}%
  \BibitemOpen
  \bibfield  {author} {\bibinfo {author} {\bibfnamefont {J.}~\bibnamefont
  {{Fuchs}}}, \bibinfo {author} {\bibfnamefont {C.}~\bibnamefont
  {{Schweigert}}}, \ and\ \bibinfo {author} {\bibfnamefont {A.}~\bibnamefont
  {{Valentino}}},\ }\href {\doibase 10.1007/s00220-013-1723-0} {\bibfield
  {journal} {\bibinfo  {journal} {Communications in Mathematical Physics}\
  }\textbf {\bibinfo {volume} {321}},\ \bibinfo {pages} {543} (\bibinfo {year}
  {2013})},\ \Eprint {http://arxiv.org/abs/1203.4568} {arXiv:1203.4568
  [hep-th]} \BibitemShut {NoStop}%
\bibitem [{\citenamefont {Bais}\ and\ \citenamefont
  {Slingerland}(2009)}]{BS0916}%
  \BibitemOpen
  \bibfield  {author} {\bibinfo {author} {\bibfnamefont {F.~A.}\ \bibnamefont
  {Bais}}\ and\ \bibinfo {author} {\bibfnamefont {J.~K.}\ \bibnamefont
  {Slingerland}},\ }\href {\doibase 10.1103/PhysRevB.79.045316} {\bibfield
  {journal} {\bibinfo  {journal} {Phys. Rev. B}\ }\textbf {\bibinfo {volume}
  {79}},\ \bibinfo {pages} {045316} (\bibinfo {year} {2009})},\ \Eprint
  {http://arxiv.org/abs/arXiv:0808.0627} {arXiv:0808.0627} \BibitemShut
  {NoStop}%
\bibitem [{\citenamefont {{Kong}}(2014)}]{Kon14}%
  \BibitemOpen
  \bibfield  {author} {\bibinfo {author} {\bibfnamefont {L.}~\bibnamefont
  {{Kong}}},\ }\href {\doibase 10.1016/j.nuclphysb.2014.07.003} {\bibfield
  {journal} {\bibinfo  {journal} {Nuclear Physics B}\ }\textbf {\bibinfo
  {volume} {886}},\ \bibinfo {pages} {436} (\bibinfo {year} {2014})},\ \Eprint
  {http://arxiv.org/abs/1307.8244} {arXiv:1307.8244 [cond-mat.str-el]}
  \BibitemShut {NoStop}%
\bibitem [{\citenamefont {Eli{\"e}ns}\ \emph {et~al.}(2013)\citenamefont
  {Eli{\"e}ns}, \citenamefont {Romers},\ and\ \citenamefont {Bais}}]{ERB1301}%
  \BibitemOpen
  \bibfield  {author} {\bibinfo {author} {\bibfnamefont {I.~S.}\ \bibnamefont
  {Eli{\"e}ns}}, \bibinfo {author} {\bibfnamefont {J.~C.}\ \bibnamefont
  {Romers}}, \ and\ \bibinfo {author} {\bibfnamefont {F.~A.}\ \bibnamefont
  {Bais}},\ }\href {http://arxiv.org/abs/1310.6001} {\  (\bibinfo {year}
  {2013})},\ \Eprint {http://arxiv.org/abs/arXiv:1310.6001} {arXiv:1310.6001}
  \BibitemShut {NoStop}%
\bibitem [{\citenamefont {{Barkeshli}}\ \emph
  {et~al.}(2013{\natexlab{a}})\citenamefont {{Barkeshli}}, \citenamefont
  {{Jian}},\ and\ \citenamefont {{Qi}}}]{BJQ13}%
  \BibitemOpen
  \bibfield  {author} {\bibinfo {author} {\bibfnamefont {M.}~\bibnamefont
  {{Barkeshli}}}, \bibinfo {author} {\bibfnamefont {C.-M.}\ \bibnamefont
  {{Jian}}}, \ and\ \bibinfo {author} {\bibfnamefont {X.-L.}\ \bibnamefont
  {{Qi}}},\ }\href {\doibase 10.1103/PhysRevB.88.241103} {\bibfield  {journal}
  {\bibinfo  {journal} {\prb}\ }\textbf {\bibinfo {volume} {88}},\ \bibinfo
  {eid} {241103} (\bibinfo {year} {2013}{\natexlab{a}})},\ \Eprint
  {http://arxiv.org/abs/1304.7579} {arXiv:1304.7579 [cond-mat.str-el]}
  \BibitemShut {NoStop}%
\bibitem [{\citenamefont {{Barkeshli}}\ \emph
  {et~al.}(2013{\natexlab{b}})\citenamefont {{Barkeshli}}, \citenamefont
  {{Jian}},\ and\ \citenamefont {{Qi}}}]{BJQ13a}%
  \BibitemOpen
  \bibfield  {author} {\bibinfo {author} {\bibfnamefont {M.}~\bibnamefont
  {{Barkeshli}}}, \bibinfo {author} {\bibfnamefont {C.-M.}\ \bibnamefont
  {{Jian}}}, \ and\ \bibinfo {author} {\bibfnamefont {X.-L.}\ \bibnamefont
  {{Qi}}},\ }\href {\doibase 10.1103/PhysRevB.88.235103} {\bibfield  {journal}
  {\bibinfo  {journal} {\prb}\ }\textbf {\bibinfo {volume} {88}},\ \bibinfo
  {eid} {235103} (\bibinfo {year} {2013}{\natexlab{b}})},\ \Eprint
  {http://arxiv.org/abs/1305.7203} {arXiv:1305.7203 [cond-mat.str-el]}
  \BibitemShut {NoStop}%
\end{thebibliography}%
\end{document}